\documentclass[10pt,journal,compsoc]{IEEEtran}
\usepackage{Rev}
\usepackage{amsfonts} % if you want blackboard bold symbols e.g. for real numbers
\usepackage{graphicx} % if you want to include jpeg or pdf pictures and put it in a flexible way
\usepackage{amssymb}% http://ctan.org/pkg/amssymb
\usepackage{stfloats}
\usepackage{pifont}% http://ctan.org/pkg/pifont
\usepackage{caption}% http://ctan.org/pkg/caption
\usepackage{xcolor} % want to change the color(predefined) of the text or backgroung
\usepackage{colortbl} %color tables
\definecolor{LightCyan}{rgb}{0.7,1,1}
\usepackage{textcomp} % trademark is from here
\usepackage{multirow}% multirow in a tabular
\usepackage{eso-pic}% http://ctan.org/pkg/eso-pic
\usepackage{wrapfig} % placing figures to one side
\usepackage[ruled,vlined]{algorithm2e}
\usepackage{cite}
\usepackage{cryptocode}
\usepackage{threeparttable}
\usepackage{amsfonts} % if you want blackboard bold symbols e.g. for real numbers

\usepackage{multirow}% multirow in a tabular
\usepackage{eso-pic}% http://ctan.org/pkg/eso-pic
\usepackage{wrapfig} % placing figures to one side
\usepackage{algpseudocode} %pseudo code class
\usepackage[utf8]{inputenc}
\usepackage{enumitem}
\usepackage{amsmath}
\usepackage{url}
\usepackage{amsmath, cases}
\usepackage{amssymb}
\usepackage{comment}
\usepackage{amsthm}
\usepackage{hyperref}
\usepackage{mathtools, cuted} %cutted strip to display long equation
\usepackage{booktabs}%table rules
\usepackage{amsmath}
\usepackage{breqn}
\usepackage{pgf}
\usepackage{setspace}
\usepackage{soul}
\setuldepth{Berlin} % use tight underline
\usepackage{makecell} %use multi line in side table
\usepackage{stackengine} % optimize spacing in a table

\usepackage{bibentry}

\usepackage{tabularx,ragged2e,booktabs}
\newcolumntype{C}[1]{>{\Centering}m{#1}}

\newlength{\halfwidth}
\IEEEaftertitletext{\vspace{-3\baselineskip}}
\usepackage{setspace}

\usepackage{enumitem}
\setlist{nosep} % or \setlist{noitemsep} to leave space around whole list
\renewcommand{\baselinestretch}{0.98}

\usepackage{tikz}% draw network protocol
\usepackage{etoolbox}

\newcommand{\circled}[2][]{% use circled number
  \tikz[baseline=(char.base)]{%
    \node[anchor=text, shape=circle,draw, inner sep=0pt, minimum size=0.5em] (char){#1\strut};
    \node at (char.center) {\makebox[0pt][c]{#2}};}}
\robustify{\circled}

\usepackage{array}
\usepackage{arydshln}
\newcommand{\epc}{\textit{EPC Gen2}\xspace}
\newcommand{\wisecr}{\textit{Wisecr}\xspace}
\hfuzz=\maxdimen %optimise width layout.
\usepackage{amsmath,booktabs}
\usepackage{soul}

\begin{document}

\title{\wisecr: Secure Simultaneous Code Dissemination to Many Batteryless Computational RFID Devices}

\author{Yang Su, Michael Chesser, Yansong~Gao, Alanson P. Sample and Damith C.~Ranasinghe

\IEEEcompsocitemizethanks{\IEEEcompsocthanksitem Y. Su, M. Chesser and D. C. Ranasinghe are with Auto-ID Lab, School of Computer Science, The University of Adelaide, Australia. (yang.su01, michael.chesser, damith.ranasinghe)@adelaide.edu.au}

\IEEEcompsocitemizethanks{\IEEEcompsocthanksitem Y. Gao is with School of Computer Science and Engineering, Nanjing University of Science and Technology, China. yansong.gao@njust.edu.cn}

\IEEEcompsocitemizethanks{\IEEEcompsocthanksitem A. P. Sample is with Electrical Engineering and Computer Science, University of Michigan, USA. apsample@umich.edu}

\IEEEcompsocitemizethanks{\IEEEcompsocthanksitem We acknowledge support from the Australian Research Council Discovery Program (DP140103448). We also acknowledge support from the National Natural Science Foundation of China (62002167) and National Natural Science Foundation of JiangSu (BK20200461).}
}

\IEEEtitleabstractindextext{%
\begin{abstract}
Emerging ultra-low-power \textit{tiny scale} computing devices run on harvested energy, are intermittently powered, have limited computational capability, and perform sensing and actuation functions under the control of a dedicated firmware operating without the supervisory control of an operating system. Wirelessly updating or patching firmware of such devices is inevitable. We consider the challenging problem of \textit{simultaneous} and \textit{secure} firmware updates or patching for a typical class of such devices---Computational Radio Frequency Identification (CRFID) devices. We propose \wisecr, the first secure and simultaneous wireless code dissemination mechanism to multiple devices that prevents \textit{malicious code injection attacks} and \textit{intellectual property (IP) theft}, whilst enabling \textit{remote attestation of code installation}. 
Importantly, \wisecr is engineered to comply with existing ISO compliant communication protocol standards employed by CRFID devices and systems. We comprehensively evaluate \wisecr's overhead, demonstrate its implementation over standards compliant protocols, analyze its security, implement an end-to-end realization with popular CRFID devices and open-source the complete software package on GitHub.
\end{abstract}

}
\maketitle
\IEEEdisplaynontitleabstractindextext

\section{Introduction}\label{sec:intro}
The maturation of energy-harvesting technology and ultra-low-power computing systems is leading to the advent of intermittently-powered, batteryless devices that operate entirely on energy extracted from the ambient environment~\cite{lucia2017intermittent}. The batteryless, low cost simplicity and the maintenance free perpetual operational life of these 
\textit{tiny scale} computing platforms provide a compelling proposition for edge devices in the Internet of Things (IoT) and Cyber-Physical Systems.

Recent developments in \textit{tiny scale} computing devices such as \Revsource{rev:WISP}{\Rev{Wireless Identification and Sensing Platform (WISP)}}~\cite{sample2008design}, MOO~\cite{zhang2011moo} and Farsens Pyros~\cite{farsens2016}, or so called \textit{Computational Radio Frequency Identification} (CRFID) devices, are  highly resource limited, intermittently-powered, batteryless and operate on harvested RF (radio-frequency) energy. CRFID type devices are more preferable in scenarios that are challenging for traditional battery-powered sensors, for example, pacemaker control and implanted blood glucose monitoring~\cite{bandodkar2019battery} as well as domains where batteries are undesirable, for example, wearables for healthcare applications~\cite{RanasinghePlosOneWearableSensor}. While, significant industrial applications are exemplified by asset management, such as monitoring and maintenance in the aviation  sector~\cite{aantjes2017fast,marasova2020digitization,yang2018rfid,santonino2018modernizing}. 

Despite various embodiments, the fundamental architecture of those devices include: microcontrollers, sensors, transceivers, and, at times actuators, with the most significant component, the {\it code} or software imparting the devices with the ability to communicate and realize interactive tasks~\cite{li2017policy,noura2019interoperability}. Consequently, the update and patching of this firmware is inevitable. In the absence of standard protocols or system level support, firmware is typically updated using a wired programming interface~\cite{zhang2011moo,Wisp5}.
In practical applications, the wired interface results in a potential attack vector to tamper with the behavior of the devices~\cite{ning2019understanding} and a compromised device can be further hijacked to attack other networked entities~\cite{xu2019badbluetooth}. Disabling the wired interface after the initial programming phase at manufacture can prevent further access. But, leaves the wireless update option as the only method to alter the firmware or to re-purpose the devices post-manufacture. 

\begin{figure}[!ht]
    \centering
    \includegraphics[width=\linewidth]{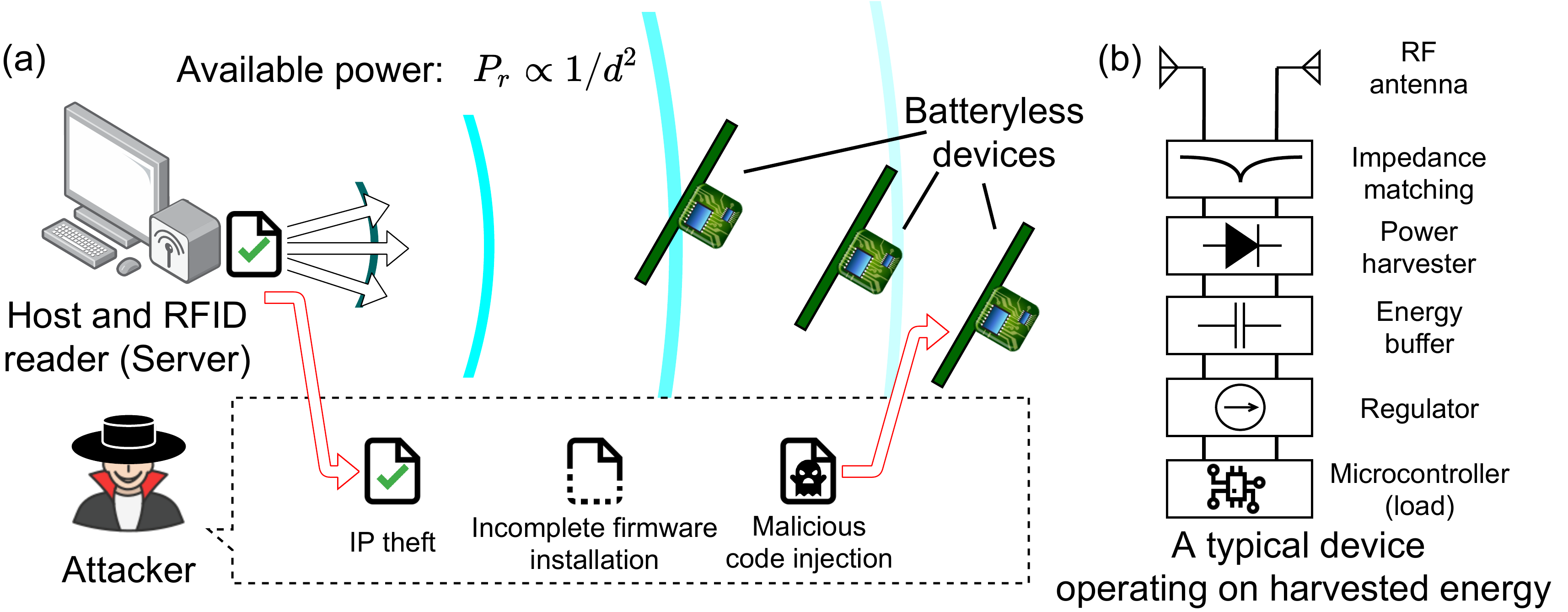}
    \caption{(a) An attacker may: i)~commit a malicious code injection attack such as alter code to inject bugs or load unauthorized code; ii)~attempt to prevent a new firmware installation and spoof the acknowledgment signal of a successful update; and iii) commit IP theft by exploiting the insecure wireless channel. (b) A typical device architecture where the amount of power available for harvesting decreases exponentially with distance $d$ from a powering source.}
    \label{fig:problem}
\end{figure}

However, \textit{point-to-point communications protocols over limited bandwidth communication channels} characteristic of RFID systems poses problems for \textit{rapid and secure} update of firmware over a wireless interface, post-manufacture. Notably, Federal Aviation Administration (FAA) in the United States has granted the installation of RFID tags and sensors on airplanes in 2018~\cite{rfid2018faa}, an increasing number of CRFID sensors are integrated with small aircraft and commercial airliners for maintenance history logging~\cite{aantjes2017fast} and aircraft health monitoring~\cite{yang2018rfid}. In such a scenario, a high-efficiency (simultaneous) and secure firmware update is desirable to ensure operational readiness and flight safety.

The recent \textit{Stork}~\cite{aantjes2017fast} protocol addressed the challenging problem of fast wireless firmware updates to multiple CRFID devices. But, as illustrated in \autoref{fig:problem},~\textit{Stork} allows any party, authorized or not, armed with a simple RFID reader to: i) mount malicious code injection attacks; ii) attempt to spoof the acknowledgment signal of a successful update to fool the Server; and iii) steal intellectual property (IP) by simply eavesdropping on the over-the-air communication channel. Although a recent protocol~\cite{su2019secucode} addressed the problem of malicious code injection attacks where by a single CRFID device is updated in turn, it neither supports simultaneous updates to many devices nor protects firmware IP and lacks a mechanism to validate the installation of code on a device.

\vspace{2mm}
\noindent\textbf{Why \textit{Secure} Wireless Firmware Update is Challenging?}

\vspace{2mm}
\noindent Constructing a secure wireless firmware update mechanism for ultra-low power and batteryless devices is non-trivial; designing a secure method is challenging. We elaborate on the challenges below.

\begin{figure}[!hb]
    \centering
    \includegraphics[width=0.8\linewidth]{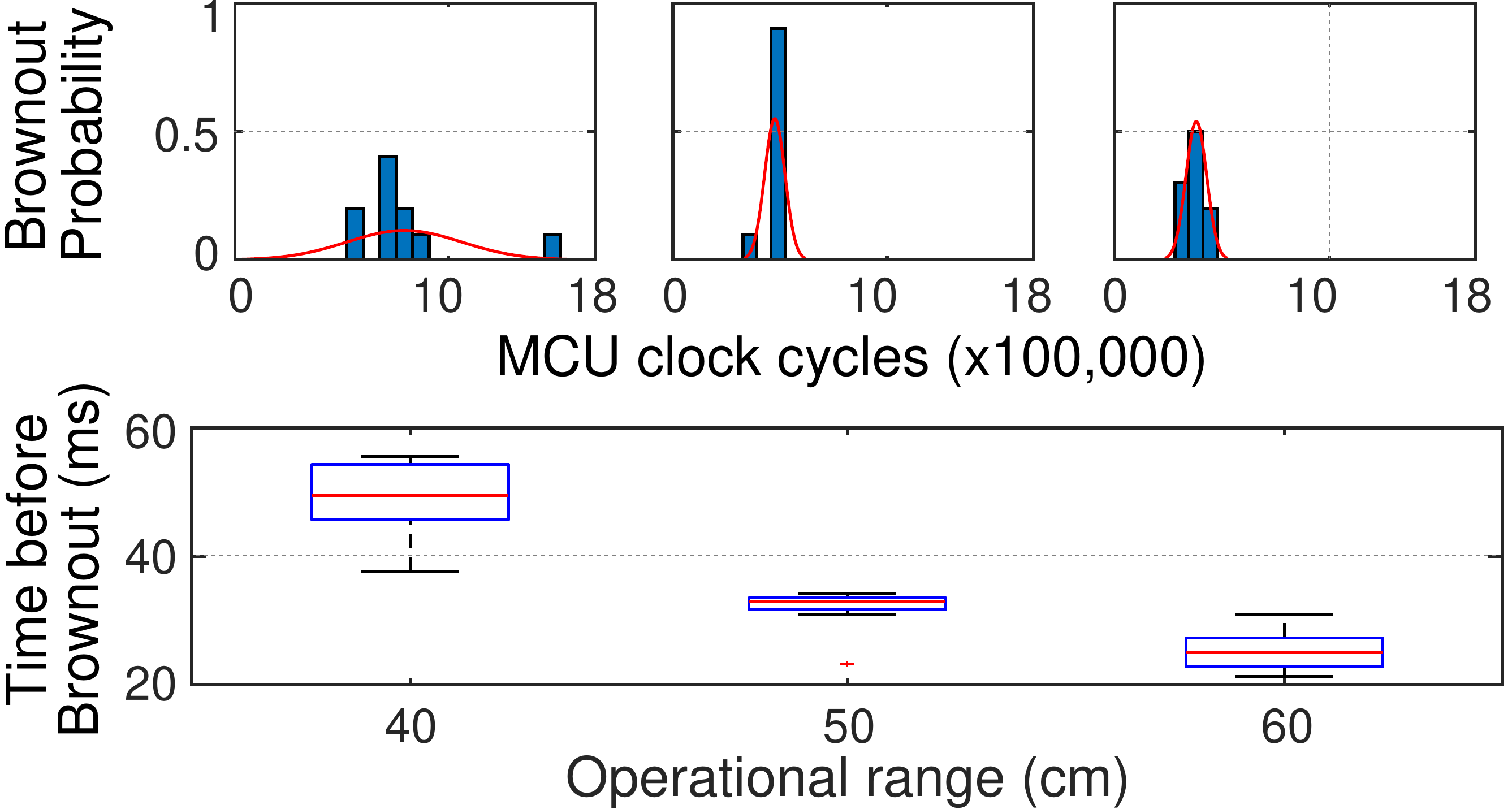}
    \caption{The distribution of available clock cycles (top) and corresponding time (bottom) before \textit{brownout}---exhaustion of available energy under computational load---for the open hardware and open software CRFID, WISP5.1LRG~\cite{Wisp5} built with a MSP430 microcontroller. We executed a message authentication code as the computational load and the device location is increased from 40~cm to 60~cm above a powering source---RFID reader antenna. We can observe that \textit{available MCU clock cycles and operational time periods before state loss are limited per charging-brownout cycle}.}
    \label{fig:mesh1}
\end{figure}

\noindent{\bf Limited and indeterminate powering.~}Energy-harvesting systems operate intermittently, only when energy is available from the environment. To operate, a device gradually buffers energy into a storage element (capacitor). Once sufficient energy is accumulated, the device begins operations. However, energy depletes more rapidly (e.g. milliseconds) during an operation compared to energy accumulation/charging (e.g. seconds). Further, energy accumulation in RF energy harvesting is sacrificed in backscatter communication links since a portion of the incident energy is reflected back during communications. Therefore, \textit{power failures are common and occur in millisecond time scales}~\cite{tan2016wisent,aantjes2017fast,dinu2019sia} as experimentally validated in~\autoref{fig:mesh1}.

Timing of power loss events across devices are ad-hoc. For example, harvested energy varies under differing distances; therefore, a computation task may only execute partially before power failure and it is hard to predict when it will occur. 

Further, saving and subsequently retrieving state at code execution checkpoints for a long-run application via an intermittent execution model~\cite{maeng2017alpaca} is not only: i)~costly in terms of computations and energy~\cite{wu2018r}---saving state in non-volatile memories such as Flash or \Revsource{rev:EEPROM}{\Rev{Electrically Erasable Programmable Read-Only Memory (EEPROM)}} consumes more energy than static RAM (SRAM) whilst reading state from \Revsource{rev:FRAM}{\Rev{Ferroelectric Random Access Memory (FRAM)}} consumes more energy than writing as we demonstrate in~\autoref{fig:TaMax}; but also: ii)~renders a device in a vulnerable state for an attacker to exploit when checkpoint state is stored on an off-chip memory, due to the lack of internal MCU memory, where non-invasive contact probes can be used to readout contents when the memory bus used is easily accessible~\cite{krishnan2018exploiting}. For example, common memory readout ports such as the I$^2$C (Inter-Integrated Circuit) bus used by a microcontroller for connecting to external memory devices are often exposed on the top layer of a printed circuit board (PCB). Hence, the content stored in an off-chip NVM can be easily read out using contact probes without any damage to the hardware~\cite{su2022leaving}.
 
These issues make the execution of long-run security algorithms, such as Elliptic Curve Diffie-Hellman (ECDH)\footnote{\scriptsize We are aware of optimized public key exchange implementations such as ECDH and Ring-LWE~\cite{liu2015efficient,liu2015efficientches}; however, they are impractical for passively powered resource-constrained devices. For example, ECDH~\cite{liu2015efficient}  with Curve25519 still requires 4.88 million clock cycles on a MSP430 MCU. In contrast, there are a very limited number of clock cycles available from harvested-energy before energy depletion~\cite{su2019secucode,tan2016wisent}---e.g. 400,000 clock cycles expected even at close proximity of 50~cm from an energy source, see \autoref{fig:mesh1}---and there is very limited time available for computations where a CRFID must reply before strict air interface protocol time-outs are breached.} key exchange, difficult to deploy securely.

Consequently, power must be carefully managed to avoid power loss and leaving the device in a potentially vulnerable state; and, we must seek computationally efficient security schemes.

\vspace{2mm}
\noindent{\bf Unavailability of hardware security support.~}Highly resource-constraint devices lack hardware security support. Thus, security features, including a trusted execution environment (TEE)---for example, ARM Trustzone~\cite{brasser2019sanctuary}---and dedicated memory to explicitly maintain the secrecy of a long term secret keys, as in~\cite{frassetto2018imix}, are unavailable. 

\vspace{2mm}
\noindent{\bf Constrained air interface protocols.~} The widely used wireless protocol for Ultra High Frequency (UHF) RFID communication only provides \textit{insecure unicast communication links} and supports no broadcast features or device-to-device communication possible in mesh networking typical of wireless sensor networks. 

\vspace{2mm}
\noindent{\bf Unavailability of supervisory control from an operating system.~} Unlike \textit{wireless sensor network nodes}, severely resource limited systems, such as CRFID, do not operate under the supervisory control of an operating system to provide security or installation support for a secure dissemination scheme.

\subsection{Our Study}
\noindent We consider the problem of secure and simultaneous code dissemination to multiple RF-powered CRFID devices operating under constrained protocols, device capability, and extreme on-device resource limitations---computing power, memory, and energy.

The scheme we developed overcomes the unique challenges, protects the firmware IP during the dissemination process, prevents malicious code injection attacks and enables remote attestation of code installation. More specifically, we address the following security threats: 
\begin{itemize}[leftmargin=15pt]
    \item \textit{Malicious code injection}: code alteration, loading unauthorized code, loading code onto an unauthorized device, and code downgrading. 
    \item \textit{Incomplete firmware installation}
    \item \textit{IP theft}: reverse engineering from plaintext binaries.
\end{itemize}

Consequently, we fulfil the \textit{urgent} and \textit{unmet} security needs in the existing state-of-the-art multiple CRFID wireless dissemination  protocol---Stork~\cite{aantjes2017fast}.

\subsection{Our Contributions and Results}
\noindent\textit{Contribution 1 (Section \ref{sec:protocol-design})}---\textbf{\wisecr is the first secure and simultaneous (\textit{fast}) firmware dissemination scheme to multiple batteryless CRFID devices.}
\noindent \wisecr provides three security functions for secure and fast updates: i)~preventing malicious code injection attacks; ii) IP theft; and iii) attestation of code installation. 
\wisecr achieves \textit{rapid} updates by supporting simultaneous update to multiple CRFID devices through a secure broadcasting of firmware over a {\it standard non-secure unicast air interface protocol}

\vspace{2mm}
\noindent\textit{Contribution 2 (Section~\ref{sec:sec-engineering} \& \ref{sec:exp-evals})}---\textbf{A holistic design trajectory, from a formal secure scheme design to an end-to-end implementation 
requiring only limited on-device resources.}
\noindent Ultra-low power operating conditions and on-device resource limitations demand both a secure and an efficient scheme. 
First, we built an efficient broadcast session key exchange exploiting commonly available hardware acceleration for crypto on microcontroller units (MCUs). 

Second, to avoid power loss and thus achieve uninterrupted execution of a firmware update session, we propose {\it new} methods: i) adaptive control of the execution model of devices using RF powering channel state information collected and reported by field deployed devices; 
ii) reducing disruptions to broadcast data synchronization across multiple devices by introducing the concept of a \textit{pilot tag} selection from participating devices in the update scheme to drive the protocol. These methods avoid the need for costly, secure checkpointing methods and leaving a device in a vulnerable state during power loss. 

Third, in the absence of an operating system, we develop an immutable bootloader to:~i)~supervise the control flow of the secure firmware update process; ii)~minimize the occurrence of power loss during an update session whilst abandoning a session in case an unpreventable power loss still occurs; and iii) manage the secure storage of secrets by exploiting commonly available on-chip memory protection units (MPUs) to realize an immutable, bootloader-only accessible, secrets. 

\vspace{2mm}
\noindent\textit{Contribution 3 (Section~\ref{sec:exp-evals} and Appendix~\ref{apd:Detailed-protocol}})---\textbf{ISO Standards compliant end-to-end \wisecr implementation.} \noindent We develop \wisecr from specification, component design to architecture on the device, and implementation. We evaluate \wisecr extensively, including comparisons with current non-secure methods, and validate our scheme by an end-to-end implementation. \wisecr is a standard compliant secure firmware broadcast mechanism with a demonstrable implementation using the widely adopted air interface protocol---ISO-18000-63 protocol albeit the protocol's lack of support for broadcasting or multi-casting---and using commodity devices from vendors. Hence, the \wisecr scheme can be adopted in currently deployed systems. We demonstrate the firmware dissemination process  here~\href{https://youtu.be/GgDHPJi3A5U}{\color{black}{\ul{https://youtu.be/GgDHPJi3A5U}}}.

\vspace{2mm}
\noindent\textit{Contribution 4}---\textbf{Open source code release.~} The tool sets and end-to-end implementation is open-sourced on GitHub:~\href{https://github.com/AdelaideAuto-IDLab/Wisecr}{\color{black}\ul{https://github.com/AdelaideAuto-IDLab/Wisecr}}. 

\vspace{2mm}
\noindent\textit{Paper Organization.}~Section~\ref{sec:protocol-design} presents the threat model, security requirements and \wisecr design; Section~\ref{sec:sec-engineering} details how the demanding security requirements are met under challenging settings;
Section~\ref{sec:exp-evals} discusses our end-to-end implementation, followed by performance and security evaluations; Section~\ref{sec:relatedwork} discusses related work with which \wisecr is compared. Section~\ref{sec:conclusion} concludes this work.

\section{\wisecr Design}\label{sec:protocol-design}
In this section, we describe the threat model including a summary of notations in \autoref{tab:notation}, followed by details of \wisecr design, and then proceed to identify the minimal hardware security requirements needed to implement the scheme. 

\begin{table}[htbp]\caption{Table of Notations}\label{tab:notation}
\footnotesize
\centering 
\begin{tabular}{p{0.12\linewidth} p{0.79\linewidth} }
\toprule
$\mathcal{S}$ & Server $\mathcal{S}$ is a single entity consisting of a host computer and a networked RFID reader.\\
$\mathcal{T}$ & Tokens $\mathcal{T}$ is a set of individual (CRFID) devices $T_i$.\\
\hline
$\textbf{DB}$  & Server's database, where each element is a three-tuple describing each CRFID token: i)~the unique and immutable identification number ${\bf id}_i$; ii)~the device specific secret key ${\bf k}$; and iii)~a flag denoting a device requiring an update ${\bf valid}$.\\

\hline
$\textbf{firmware}$  & The new firmware binary for the update.\\
$\textbf{firmware}^{(j)}$  & The $j_{\rm th}$ block of the \bf{firmware}.\\
$\textbf{nver}$  & The new version number (a monotonically increasing ordinal number with a one-to-one correspondence with each updated \textbf{firmware}).\\
$\textbf{ver}$  & A token's current version number.\\
${\bf sk}$  & The broadcast session key.\\
${\bf s}$  & The MAC tag computed by the Server.\\
${\bf c}$, ${\bf r}$  & Attestation challenge and response, respectively.\\

\hline
${\bf \Box}'$  &  An apostrophe denotes a value computed by a different entity, e.g., $\bf s'$ the MAC tag computed by a Token.\\
${\bf \Box}_i$  & A subscript $i$ denotes a specific entity, e.g.,  ${\bf k}_i$ is the device key of the $i_{\rm th}$ Token.\\
$\langle \Box \rangle$  & Encrypted data, e.g. $\langle {\bf sk}_i \rangle$  denotes the encrypted session key {\bf sk}, with the $i_{\rm th}$ token's key ${\bf k}_i$.\\

\hline
\textsf{RNG()}  & A cryptographically secure random number generator.\\
\textsf{SKP.Enc()}  & Symmetric Key Primitive encryption function described by ${\langle {\bf m}\rangle} \leftarrow \sf{SKP.Enc_{\bf sk}}({\bf m})$. Here, the plaintext $\bf m$ is encrypted with the $\bf sk$ to produce the ciphertext ${\langle {\bf m}\rangle}$.\\
\textsf{SKP.Dec()}  & Symmetric Key Primitive decryption function where ${\bf m} \leftarrow \sf{SKP.Dec_{\bf sk}}(\langle {\bf m}\rangle)$.\\
\textsf{MAC()}  & Message Authentication Code function.  By appending an authentication tag {\bf s} to the message {\bf m}, where ${\bf s} \leftarrow {\sf MAC_k(m)}$, a message authentication code (MAC) function can verify the integrity and authenticity of the message by using the symmetric key {\bf k}.\\
\textsf{SNIFF()} & \Revsource{rev:sniff_vt}{\Rev{The voltage ${\bf Vt}_i$ established by the power harvester within a fixed time $t$ from boot-up at token $i$, is measured by  \textsf{SNIFF}() described by ${\bf Vt}_i \leftarrow {\sf SNIFF}(t)$.}}\\
\textsf{PAM()} & Family of functions employed by the Power Aware Execution mode of operation proposed for the token to mitigate power-loss (brown-out) events.\\
\hline

\texttt{Query}  & We denote \epc commands using typewriter fonts.\\ 
\bottomrule
\end{tabular}
\label{tab:TableOfNotationForMyResearch}
\end{table}

\subsection{Threat Model}\label{sec:adversarymodel}
Communication with an RFID device operating in the UHF range is governed by the widely adopted ISO-18000-63:\Rev{2015}---also known as the EPCglobal Class 1 Generation 2 version 2 (C1G2v2)---air interface protocol or simply the \textit{EPC Gen2}.

\begin{figure}[!hb]
    \centering
    \includegraphics[width=\linewidth]{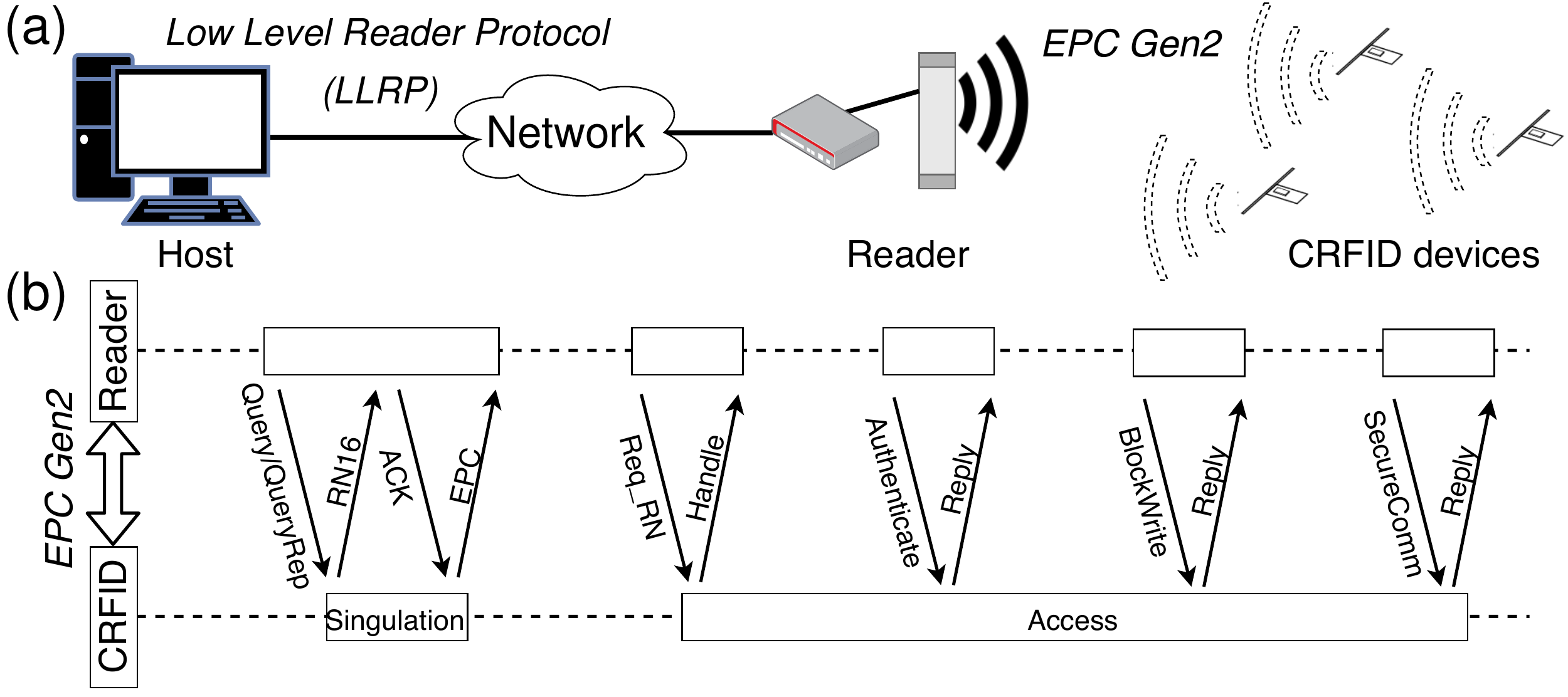}
    \caption{(a) System overview. (b) Control/data flow over the \epc air interface. Firmware update will take place once a device enters the \textit{Access} state and by implementing \wisecr over \textit{Access} command specifications such as \texttt{Authenticate}.}\label{fig:system-overview}
\end{figure}

\autoref{fig:system-overview} illustrates the communication channels in a networked RFID system. The RFID reader resides at the edge of the network and is typically connected to multiple antennas to power and communicate with RFID or CRFID devices energized by the antennas. The reader network interface is accessed by using a Low-Level Reader Protocol (LLRP) from a host machine. Communication with an RFID device operating in the UHF range is through the \textit{EPC Gen2} protocol. 

Our focus is on the insecure communication channel between the RFID reader connected antenna and the CRFID transponder or token $T_i \in \mathcal{T}$. Hence, we assume that the communication between a host and a reader is secured using standard cryptographic mechanisms~\cite{jin2017secure}. Therefore, a host computer and a reader are considered as a single entity, the Server, denoted as $\mathcal{S}$ (a detailed execution of an \epc protocol session to \textit{singulate} a single CRFID device from all visible devices in the field can be found in~\cite{tan2016wisent,aantjes2017fast}).

We assume a CRFID device can meet the pragmatic hardware security requirements, detailed in Section~\ref{sec:Hard_secu_requirement}. Further, after device provisioning, the wired interface for programming is disabled---using a common technique adopted to secure resource-constrained microcontroller based devices~\cite{dinu2019sia,su2019secucode}. Subsequently, both the trusted party and adversary $\mathcal{A}$ must use the wireless interface for installing new firmware on a token.

Building upon relevant adversary models related to wireless firmware update for low-end embedded devices~\cite{kohnhauser2016secure,feng2017secure}, we assume an adversary $\mathcal{A}$ has full control over the communication channel between the Server $\mathcal{S}$ and the tokens $\mathcal{T}$. Hence, the adversary $\mathcal{A}$ can eavesdrop, manipulate, record and replay all messages sent between  the Server $\mathcal{S}$ and the tokens $\mathcal{T}$. This type of attacker is referred to as an input/output attacker~\cite{piessens2016software}. 

We assume the firmware (application), potentially provided by a third party in the form of a stripped binary, may contain vulnerabilities or software bugs that can cause the program to deviate from the specified behavior with potential consequences being corruption of the bootloader and/or \Revsource{rev:NVM_define}{the non-volatile memory \Rev{(NVM)} contents.} Such an occurrence is possible when firmware is frequently written in unsafe languages such as C or C++~\cite{sammler2021refinedc,levy2017multiprogramming}. Hence, the firmware (application) cannot be trusted. In this context, similar to~\cite{kohnhauser2016secure,feng2017secure}, we also assume that the adversary cannot bypass any of the memory hardware protections (detailed in Section~\ref{sec:Hard_secu_requirement}) and an adversary cannot mount invasive physical attacks to extract the on-chip non-volatile memory contents. Such an assumption is practical, especially in deeply embedded applications such as pacemaker control~\cite{tan2016wisent} where wireless update is the only practical mechanism by which to alter the firmware and physical access is extremely difficult. 

As in~\cite{aysu2015end}, we also assume the adversary $\mathcal{A}$ cannot mount implementation attacks against the CRFID, or gain internal variables in registers, for example, using invasive attacks and side-channel analysis. We do not consider Denial of Service (DoS) attacks because it appears to be impossible to defend against such an attacker, for example, that disrupts or jams the wireless communication medium in practice~\cite{kohnhauser2016secure}. 

\begin{figureRevsourceS}[t!]
\centering
\resizebox{0.9\textwidth}{!}{% <------ Don't forget this %
    \fbox{$
        \Huge
        \begin{array}{lcl}
        {\text{Server}~\mathcal{S}} & & ~~~~~~~~~~~~~~~~~~~~~~~~~~~~~~~~~~~~~~~~~~~~~{\text{Tokens}~\mathcal{T}=\{T_1...T_m\}} \\
        {\bf DB} =  ({\bf id}_i,{\bf k}_i,{\bf ver}_i)~\textrm{for}~i=1...m~\text{Tokens} & &  ~~~~~~~~~~~~~~~~~~~~~~~~~~~~~~~~~~~~~~~~~~~~~~~~~~~~~~~~~~~~~~~~~~~~{\bf id}_i,{\bf k}_i\\
        {\bf firmware},{\bf nver} & & \\
        
        \cmidrule{1-3}
        \rowcolor{LightCyan}
        & \textbf{Prelude} & \\
        \cmidrule{1-3}
        {\bf sk} \leftarrow {\sf RNG()} & & \\
        {\bf firmware} = ({{\bf firmware}^{(1)}\|...\|{\bf firmware}^{(n)}}) & & \\
        \langle {\bf firmware}^{(j)} \rangle \leftarrow {\sf SKP.Enc}_{\bf sk}({\bf firmware}^{(j)}), \textrm{for}~j=1...n& & \\
        
        \cmidrule{1-3}
        \rowcolor{LightCyan}
        & \textbf{Security Association} & \\
        \cmidrule{1-3}
        \Rev{\textit{for} ~\text{each Token}~ T_i \in \mathcal{T}} & {\bf id}_i ~do,{\bf ver}_i, {\bf Vt}_i & \bf{Vt}_{\it i} \leftarrow \sf{SNIFF(t)}\\
        \quad \textit{if}~{\bf id}_i \notin {\bf DB} \textit{ then}\textrm{ reject and abort } & \xleftarrow{~~~~~~~~~~~~~~~~~~~~} & \\
        \quad \textit{else}~{\bf s}_i \leftarrow {\sf MAC}_{{\bf k}_i}({\bf firmware}\|{\bf ver}_i\|{\bf nver}) & & \\
        \quad \quad\langle {\bf sk}_i \rangle \leftarrow {\sf SKP.Enc}_{\bf k_{\it i}}({\bf sk})& {\bf \langle sk}_i \rangle,{\bf s}_i,{\bf nver},{\bf t_{{LPM}_{\rm i}}},{\bf t_{{active}_{\rm i}}} & \\
        \quad \quad({\bf t_{{LPM}_{\rm i}}},{\bf t_{{active}_{\rm i}}}) \leftarrow\sf{PAM.Get({\bf Vt_{\it i}})} & \xrightarrow{~~~~~~~~~~~~~~~~~~~~~~~~~~~~~~~~~~~~~~~~~~} & {\bf sk} \leftarrow {\sf SKP.Dec}_{\bf k_{\it i}}({\bf \langle sk}_i \rangle)\\

        \cmidrule{1-3}
        \rowcolor{LightCyan}
        & \textbf{Secure Broadcast} & \\
        \cmidrule{1-3}
        \Rev{\text{Simultaneously broadcast the new firmware}} & \langle {\bf  firmware}^{(1)} \rangle & \\
        \Rev{\text{to all Tokens } T_i \in \mathcal{T}}& \xrightarrow{~~~~~~~~~~~~~~~~~~~~~~~~~} & \\
        & \dots & \\
        & \langle {\bf  firmware}^{(n)} \rangle & \\
        & \xrightarrow{~~~~~~~~~~~~~~~~~~~~~~~~~}& \sf{PAM.Enter({\bf t_{{LPM}_{\rm i}}},{\bf t_{{active}_{\rm i}}}})\\
        & & {\bf firmware}^{(j)} \leftarrow {\sf SKP.Dec}_{\bf sk}({\bf \langle firmware}^{(j)} \rangle),\textrm{ for }j = 1,...,n  \\
        & & {\bf firmware} = ({{\bf firmware}^{(1)}\|...\|{\bf firmware}^{(n)}})\\
        
         \cmidrule{1-3}
         \rowcolor{LightCyan}
        & \textbf{Validation} & \\
        \cmidrule{1-3}
        \Rev{\textit{for} ~\text{each Token}~ T_i \in \mathcal{T}}& & {\bf s}_i' \leftarrow {\sf MAC}_{{\bf k}_i}({\bf firmware}\|{\bf ver}_i\|{\bf nver})\\
        & & \sf{PAM.Exit()}\\
        & & \textit{if}~{\bf s}_i == {\bf s}_i' \textit{ then}\textrm{ accept and apply update}\\
        & {\bf id}_i,{\bf ver}_i & \quad{\bf ver}_i \leftarrow {\bf nver}\\
        \quad \textit{if}~{\bf ver}_i == {\bf nver} \textit{ then} \textrm{ update successful} & \xleftarrow{~~~~~~~~~~~~~~~} & {\textit{else}} \textrm{ reject and abort}\\
        
        \cmidrule{1-3}
        \rowcolor{LightCyan}
        & \textbf{Remote Attestation} & \\
        \cmidrule{1-3}
        \Rev{\textit{for} ~\text{each Token}~ T_i \in \mathcal{T}}\\
        \quad {\bf sk} \leftarrow {\sf RNG()},~{\bf c} \leftarrow {\sf RNG()}, \langle {\bf sk}_i \rangle \leftarrow {\sf SKP.Enc}_{\bf k_{\it i}}({\bf sk}) & & \\
        \quad M \leftarrow \textrm{memory segment size to attest} & \langle {\bf sk}_i \rangle,~{\bf c},~\textit{elaborate},~M & \\
        \quad {\bf r} \leftarrow \textsf{MAC}_{{\bf sk}}({\bf c\|\textrm{selected segment}~(firmware)\| {\bf id}_i}\|{\bf ver}_i) & \xrightarrow{~~~~~~~~~~~~~~~~~~~~~~~~~~~~~~~~} & {\bf sk} \leftarrow {\sf SKP.Dec}_{\bf k_{\it i}}({\bf \langle sk}_i \rangle)\\
        \quad & {\bf r'} & {\bf r'} \leftarrow \textsf{MAC}_{{\bf sk}}({\bf c\|\textrm{selected segment}~(firmware)}\|{\bf id}_i\|{\bf ver}_i) \\
        \quad \textit{if}~{\bf r} == {\bf r'} \textit{ then} \textrm{ Attestation successful} \textit{ else} \textrm{ failure} & \xleftarrow{~~~~~~~~~~}  & \\
        \end{array}
    $}%\fbox
}% \resizebox
\caption{\textit \wisecr: The proposed wireless, secure and simultaneous code dissemination scheme to multiple tokens. Notably, the functions \textsf{SNIFF()} and \textsf{PAM()} facilitate the secure and uninterruptible execution of an update session. A single symmetric key cipher \textsf{SKP} can be exploited in practice to realise all of the primitives needed, including the \textsf{MAC()} function, to address the demands of a resource limited setting.}
\label{fig:protocol}
\end{figureRevsourceS}

\subsection{\wisecr Update Scheme}
\label{sec:protocol}
\wisecr enables the ability to securely distribute and update the firmware of multiple CRFID tokens, simultaneously. Given that a Server $\mathcal{S}$ must communicate with an RFID device $\mathcal{T}$ using \textit{\epc}, \wisecr is compatible with \epc by design. Generally, our scheme can be implemented after the execution of the anti-collision algorithm in the media access control layer of the \epc protocol, where a reader must first singulate a CRFID device and obtain a handle, $RN16$, to address and communicate with each specific device. \Revsource{rev:access_commands}{After singulating a device, \Rev{the server can employ commands such as:} \texttt{BlockWrite}, \texttt{Authenticate},  \texttt{SecureComm} and \texttt{TagPrivilege} \Rev{specified in the \epc access command set~\cite{epcglobal2015inc}}, to implement the \wisecr scheme.}
We describe the update scheme in  \autoref{fig:protocol} and defer the details of our scheme implemented over the \epc protocol to the Appendix~\ref{apd:Detailed-protocol}.

As described in \autoref{tab:notation}, the Server $\mathcal{S}$ maintains a database {\bf DB} of provisioned tokens, issues the new \textbf{firmware} and the corresponding version number ($\bf nver$). 

Each token $T_i$ has a secure storage area provisioned with: i)~an ${\bf id}_i$, the immutable identification number; ii)~${\bf k}_i$, the device specific secret key stored in NVM that is read-only accessible by the immutable bootloader. The {${\bf k}_i$ assigned to different devices are assumed to be \textit{independent and identically distributed} (i.i.d.)}; iii)~${\bf ver}_i$, the token's current firmware version number. The secure storage area is only accessible by the trusted and immutable bootloader provisioned on the device; this region is inaccessible to the firmware (application) and therefore cannot be modified by it.

We describe a dissemination session in four stages: i)~Prelude; ii)~Security Association; iii)~Secure Broadcast; and iv)~Validation. An update can be extended with an optional, v) Remote Attestation to verify the firmware installation.

\vspace{2mm}
\noindent\textbf{STAGE 1: Prelude (Offline).~}In this stage, the Server $\mathcal{S}$ undertakes setup tasks. The Server uses \textsf{RNG()} to generate a broadcast session key (${\bf sk}$).

The new $\textbf{firmware}$ is divided into segments---or block; each block $j$ is encrypted as $\langle {\bf firmware}^{(j)} \rangle$ $\leftarrow$ \textsf{SKP.Enc$_{\bf sk}$}(${\bf firmware}^{(j)}$), where $\langle {\bf firmware}^{(j)} \rangle$ denotes the encrypted firmware block $j$. The division of firmware is necessary as the narrow band communication channel and the \epc protocol does not allow {\it arbitrary} size payloads to be transmitted to a token.

\vspace{2mm}
\noindent\textbf{STAGE 2: Security Association.~} In this stage, the Server $\mathcal{S}$ distributes the broadcast session key (${\bf sk}$) to all tokens $\mathcal{T}$ and builds a secure broadcast channel over which to simultaneously distribute the firmware to multiple tokens. 

More specifically, each token in the energizing field of the Server responds with ${\bf id}_i$, \textbf{Vt}$_i$ and ${\bf ver}_i$. The token will not be included in the following update session if: i) the ${\bf id}_i$ of the responding token is not in Server's {\bf DB}; or ii) the token is not scheduled for an update (${\bf valid}_i==$~False). 
For tokens selected for an update, the Server computes a MAC tag ${\bf s}_i\leftarrow$ \textsf{MAC$_{{\bf k}_i}$}(${\bf firmware}$$\vert \vert$${\bf ver}_i$$\vert \vert$${\bf nver}$).
In practice, we cannot assume that each token is executing the same version of the firmware, therefore a token specific MAC tag is generated over the device specific key whilst the firmware is encrypted with the broadcast session key.

The Server establishes a shared session key with each token $T_i$ by sending $\langle {\bf sk}_i \rangle$, where $\langle {\bf sk}_i \rangle$ $\leftarrow$ \textsf{SKP.Enc$_{\bf k_{\it i}}$}$({\bf sk})$ and ${\bf k}_i$ is specific to the $i_{\rm th}$ token, and ${\bf nver}$. 
The $\langle {\bf sk}_i \rangle$ and ${\bf s}_i$ are transmitted to each token $T_i$. Each token decrypts the broadcast session key ${\bf sk} $ $\leftarrow$ \textsf{SKP.Dec$_{\bf k_{\it i}}$}$(\langle {\bf sk}_i \rangle)$---thus, all tokens selected for an update now possess the session key. 

Notably, as detailed in Section~\ref{sec:paem}, each token measures its \textit{powering channel state}  or its ability to harvest energy by measuring the voltage ${\bf Vt}_i$ established by the power harvester within a fixed time $t$ from boot-up, as ${\bf Vt}_i \leftarrow {\sf SNIFF}(t)$ to transmit to the Server.
There are two important reasons for measuring ${\bf Vt}_i$. First, to facilitate the power aware execution model (PAM) employed to mitigate power-loss at a given token. The Server uses the reported ${\bf Vt}_i$ to control the execution model of the $i_{\rm th}$ token. 
Specifically, the reported ${\bf Vt}_i$ is used by the Server to determine the length of time that a token dwells in low power mode (LPM) $t_{{\rm LMP}_i}$ and active mode $t_{{\rm active}_i}$ when executing computationally intensive tasks in the Secure Broadcast and Validation stages; here, the server determines $(t_{\rm LPM_i}, t_{\rm active_i}) \leftarrow \sf{PAM.Get}({\bf Vt}_i)$. 
Second, the Server uses the reported ${\bf Vt}_i$ to realize the \textit{Pilot-Observer Mode} of operation where one token is \textit{elected} based on its ${\bf Vt}_i$, termed the \textit{Pilot}, to control the flow in the Secure Broadcast stage by responding to server commands as detailed in Stage~3.

\vspace{2mm}
\noindent\textbf{STAGE 3: Secure Broadcast.~} In this stage, the encrypted firmware blocks $\langle {\bf firmware}^{(1..n)} \rangle$ are broadcasted; and each token stores the new encrypted firmware blocks in its application memory region (Segment $M$ defined in Section~\ref{sec:secure-storage}).
Once the broadcast is completed, each token starts firmware decryption and validation. The $\langle \bf firmware \rangle$ is decrypted using the session key $\bf sk$ as ${\bf firmware}^{(j)} \leftarrow {\sf SKP.Dec}_{\bf sk}(\langle {\bf firmware}^{(j)}\rangle)$. 

To realize a secure and power efficient logical broadcast channel under severely energy constrained settings, we use the \textit{Pilot-Observer mode}. Herein, all tokens, except the  \textit{Pilot} token elected by the Server, enters into an {\it observer mode}. The tokens in the {\it observer mode} silently listen and store encrypted data disseminated by the server; the \textit{Pilot} token performs the same operation whilst responding to the server commands. We employ two techniques within the Pilot-Observer Mode to mitigate power-loss and to achieve a secure broadcast to tokens: i) disabling energy consuming communication command reply from observers; and ii)~the concept of electing a \textit{Pilot} CRFID device to drive the update session as detailed in Section~\ref{sec:observerMode}.

\textit{Notably, the techniques described in Stage 2 and 3 form the foundation for the uninterrupted execution of the bootloader to ensure security and enhance the performance of the firmware dissemination under potential power-loss events.}

\vspace{2mm}
\noindent\textbf{STAGE 4: Validation.~}\label{sec:protocol-validation} In this stage, firmware is validated before installation. More precisely, a token specific MAC tag~${\bf {s}_{\it i}'}\leftarrow \textsf{MAC}_{{\bf k}_i}({\bf firmware} \vert \vert{\bf ver}_i \vert \vert {\bf nver})$ is computed by each token $T_i$. If the received MAC tag ${\bf s_{\it i}}$ matches the device computed ${\bf {s}_{\it i}'}$, the integrity of the firmware established and the issuing Server is authenticated by the token. Subsequently, the new firmware is updated and the new version number $\bf nver$ is stored as ${\bf ver}_i$. Otherwise, the firmware is discarded and the session is aborted. Notably, the \epc protocol provides a reliable transfer feature. Each broadcast payload is protected by a \Revsource{rev:CRC-16}{\Rev{16-bit Cyclic Redundancy Check (CRC-16)}} error detection method. Hence, the notification of a CRC failure to the Server results in the automatic re-transmission of the packet by the Server. Therefore, a MAC tag mismatch is more likely to be adversarial and discarding the firmware is a prudent action. At stage completion, each token switches from the \textit{observer} or \textit{Pilot} to the normal mode of operation after a software reset (reboot). Subsequently, all the temporary information such as session key ${\bf sk}$ and the token specific MAC tag ${\bf s}'_i$ in volatile memory will be erased. 

Once the firmware is installed, the Server is acknowledged with the status of each participating token in the session. This is achieved by performing an \textit{EPC Gen2} handshake after a reboot of the tokens, and comparing the version number ${\bf ver}_i$ reported from each token specified by ${\bf id}_i$ to the new firmware version number ${\bf nver}$ expected from the token. If the ${\bf ver}_i$ is up-to-date, the Server is acknowledged that the token $T_i$ has been successfully updated.

\vspace{1mm}
\noindent\textbf{Remark.~}It is theoretically possible to include a MAC tag in the acknowledgment message at the end of the Validation stage to authenticate the acknowledgment. But the implementation of this in practice is difficult under constrained protocols and limited resources typical of intermittently powered devices. This is indeed the case with the \epc air interface protocol and CRFID devices we employed. We discuss specific reasons in Appendix~\ref{apd:Detailed-protocol} where we detail implementation of the \wisecr Update Scheme over the \epc air interface protocol.

\vspace{2mm}
\noindent\textbf{STAGE 5 (Optional): Remote Attestation.~}\label{sec:protocol-attest}

The Server can elect to verify the firmware installation on a token by performing a remote attestation; a mechanism for the Server to verify the complete and correct software installation on a token. Considering the highly resource limited tokens, we propose a lightweight challenge response based mechanism re-using the \textsf{MAC()} function developed for \wisecr. The server sends a randomly generated challenge $c \leftarrow {\sf RNG()}$ and evaluates the corresponding response $r'$ to validate the installation. We provision a new session key ${\bf sk}$ to enable the remote attestation to proceed independent of the previous stages, whilst avoiding the derivation of a key on device to reduce the overhead of the attestation routine.

We propose two modes of attestation; a \textit{fast mode} and an \textit{elaborate mode} to trade-off veracity of the verification against computational and power costs. The \textit{fast} mode only examines the token serial number (${\bf id}_i$) and the version number ($\bf ver_i$). While the \textit{elaborate} mode traverses over an entire memory segment. The \textit{elaborate} mode is relatively more time consuming but allows the direct verification of the code installed on the target token $T_i$.

We illustrate (in \autoref{fig:protocol}) and demonstrate (in Section~\ref{sec:performance}) the \textit{elaborate} mode (more veracious and computationally intensive) where response ${\bf r'} \leftarrow $ $\textsf{MAC}_{{\bf sk}}$ $({\bf c\|}\textrm{selected segment}~{\bf (firmware)} \|{\bf id}_i\|{\bf ver}_i)$  attests the application memory segment containing the installed {\bf firmware}. In contrast, the \textit{fast} method computes the response as ${\bf r'} \leftarrow \textsf{MAC}_{{\bf sk}}({\bf c\|} {\bf id}_i\|{\bf ver}_i)$.

\subsection{Token Security Requirements and Functional Blocks}\label{sec:Hard_secu_requirement}

Our design is intentionally minimal and requires the following security blocks. 

\vspace{1mm}
\begin{description}
    \item [Immutable Bootloader] (Section~\ref{sec:secure-storage})~We require a static NVM sector $M_{\rm rx}$ that is write-protected to store the executable bootloader image to ensure the bootloader can be trusted post deployment where, for example, firmware (application) code vulnerabilities or software bugs do not lead to the corruption of the bootloader and the integrity of the bootloader can be maintained.
    
    \item [Secure Storage]
    (Section~\ref{sec:secure-storage})~To store a device specific secret, e.g., ${\bf k}_i$, we require an NVM sector $M_{\rm}$ read-\textit{only} accessible by the immutable bootloader in sector $M_{\rm rx}$. This ensures the integrity and security of non-volatile secrets post deployment since the firmware (application) code cannot be trusted, for example, due to potential vulnerabilities or software bugs that can lead to the corruption of non-volatile memory contents).
    
    \item [Uninterruptible Bootloader Execution] (Section~\ref{sec:unterruptbootloader})~During the execution of the bootloader stored in sector $M_{\rm rx}$, execution cannot be interrupted until the control flow is intentionally released by the bootloader. 
    
    \item [Efficient Security Primitives] (Section~\ref{sec:effi-security-prim})~The update scheme requires: i) a symmetric key primitive; and ii) a keyed hash primitive for the message authentication code that are both computationally and power efficient.
\end{description}

In Section~\ref{sec:sec-engineering} we discuss how the associated functional blocks are engineered on typical RF-powered devices built with ultra-low power commodity MCUs.

\section{On-device Security Function Engineering}\label{sec:sec-engineering}

To provide comprehensive evaluations and demonstrations, we selected the WISP5.1LRG~\cite{Wisp5} CRFID device with an open-hardware and software implementation for our token $\mathcal{T}$. This CRFID device uses the ultra-low power MCU MSP430FR5969 from Texas Instruments. Consequently, for a more concrete discussion, we will refer to the WISP5.1LRG CRFID and the MSP430FR5969 MCU in the following.

\subsection{Immutable Bootloader \& Secure Storage}\label{sec:secure-storage}
For resource limited MCUs, several mechanisms---detailed in the Appendix~\ref{apd:memory_management_compare}---exists for implementing secure storage: i) Isolated segments; ii) Volatile keys; iii) Execute only memory; and iv) Runtime access protections. 
We opt for achieving secure storage and bootloader immutability using Runtime Access Protection by exploiting the MCU's memory protection unit (MPU), which offers flexibility to the bootloader. In particular, the MPU allows \underline{r}ead/\underline{w}rite/e\underline{x}ecute permissions to be defined individually for memory segments at 
power-up---prior to any firmware (application) code execution. \wisecr requires the following segment permissions 
to be defined by the bootloader to prevent their subsequent modifications through application code by locking the MPU: 

\begin{comment}
\noindent\textbf{Segment 1}: (0x4400 to bootloader\_start - 1~KiB\footnote{\scriptsize Segment boundaries must be 1~KiB aligned, so secure storage area must be at least 1~KiB in size.}): Read + Write + Execute -- application code/data and application IVT.
\end{comment}

\begin{description}
\item [Segment $M$] is used as the secure storage area. During application execution, any access (reading/writing) to this segment results in an access violation, causing the device to restart in the bootloader.

\item [Segment $M_{\rm rx}$] contains the bootloader, device interrupt vector table (IVT), shared code (e.g., \epc implementation). During application execution, writing to this segment results in an access violation.

\item [Segment $M_{\rm rwx}$] covers the remaining memory, and is used for application IVT, code (\textbf{firmware}) and data. 

\end{description}

\subsection{Uninterruptible Bootloader Execution}\label{sec:unterruptbootloader}
The execution or control flow of the bootloader on the token must  be  uninterruptible  by  application  code  and  power-loss events  to  meet  our  security  objectives but dealing with brownout induced power-loss events is more challenging. Power loss leaves devices in vulnerable states for attackers to exploit; therefore, we focus on innovative, pragmatic and low-overhead power-loss prevention methods. Our approach deliberately mitigates the chances of power-loss. In case a rare power-loss still occurs, the token will discard all state---including security parameters such as the broadcast session key; subsequently, the Server will re-attempt to update the firmware by re-commencing a fresh update session with this token. We detail our solution below.

\subsubsection{Managing Application Layer Interruptions} 
The bootloader must be uninterruptible (by application code) for security considerations. For instance, the application code---due to an unintentional software bug or otherwise---could interrupt the bootloader while the device key is in a CPU register, so that the application code (exploited by an attacker) can copy the device key to a location under its control, or completely subvert access protections by overriding the MPU register before it is locked.

Recall, the memory segment $M_{\rm rx}$ (see Section~\ref{sec:secure-storage}) includes the memory region containing the IVT. This ensures that only the bootloader can modify the IVT. Since the IVT is under the bootloader's control, we can ensure that any non-maskable interrupt is unable to be directly configured by the application code, whereas all other interrupts are disabled during bootloader execution. Consequently, the interrupt configuration cannot be mutated by application code. 

\subsubsection{Managing Power-Loss Interruptions}\label{sec:managepower}
Frequent and inevitable power loss during the bootloader execution will not only interrupt the execution, degrading code dissemination performance but also compromise security. Although intermittent computing techniques relying on saving and retrieving state at check-points from NVM---such as Flash or EPPROM---is possible, these methods impose additional energy consumption and introduce security vulnerabilities revealed recently~\cite{krishnan2018exploiting}. 

Confronted with the complexity of designing and implementing an end-to-end scheme under extreme resource limitations, we propose an on-device Power Aware execution Model (PAM) to: i) avoid the overhead of intermittent computing techniques; and ii) enhance security without saving check-points to insecure NVM. 

We observe that only a limited number of clock cycles are available for computations per charge and discharge cycle (Intermittent Power Cycle or IPC) of a power harvester, as illustrated via comprehensive measurements in \autoref{fig:mesh1}. Further, the rate of energy consumption/depletion is faster than energy harvesting. We recognize that there are three main sources of power-loss: 
 i) (CPU) energy required for function computation exceeding the energy supply capability from the harvester;  ii) (FRAM.R/W) memory read/write access such as in executing \texttt{Blockwrite} commands; and iii) (RFID) power harvesting disruptions from communications---especially for backscattering data in response to \epc commands. 

\begin{figure}[!ht] 
    \centering
    \includegraphics[width=\linewidth]{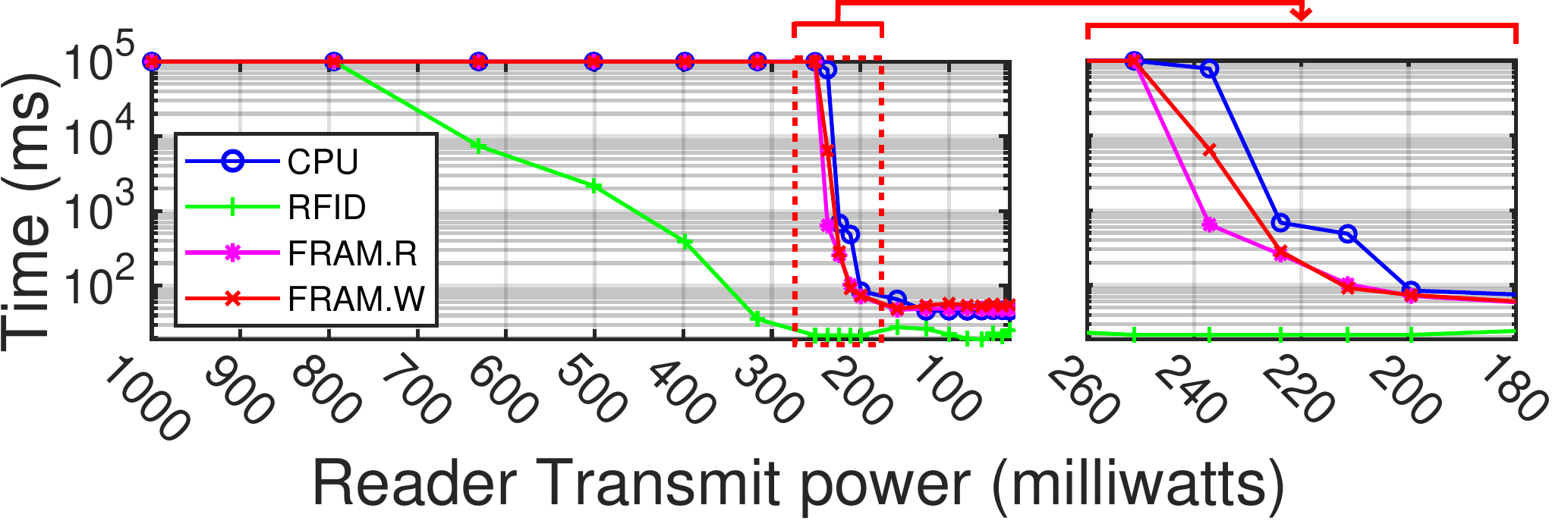}
    \caption{Impact of four key device tasks on power-loss: CPU (computations); FRAM.R/W (memory read/write accesses); and RFID (communications). The data above are collected by using special firmware. The power-loss event is captured by monitoring through a GPIO pin. We conduct 10 repeated measurements and report the mean time before power-loss. The plot provides a lateral comparison among four operations. The data is measured from a single CRFID device with oscilloscope probes attached to measure the device's internal state.}
    \label{fig:TaMax}
\end{figure}
To understand the severity of these four causes, we measure the maximum time duration before brownout/power-loss versus the harvested power level for each task---CPU, FRAM.R, FRAM.W and RFID. In the absence of a controlled RF environment (i.e., anechoic chamber), it is extremely difficult to maintain the same multipath reflection pattern. Especially when changing the distance between the radiating reader antenna and an instrumented CRFID device considering the multipath interference created by the probes, cables, and the nearby oscilloscope and researcher to monitor the device's internal state. To minimize the difficulty of conducting the experiments, we place the CRFID device at a fixed distance (20~cm) whilst keeping all of the equipment at fixed positions, and adjust the transmit power of the RFID reader through the software interface. According to the free-space path loss equation~\cite{friis1946free}, adjusting the transmit power of the RFID reader or changing the distance can be used to vary the available power at the CRFID device. We describe the detailed experimental settings in Appendix~\ref{apd:power_based_exp_setup}. For experiments without the requirement for monitoring the device's internal state, we still employ distance-based measurements as in previous studies~\cite{aantjes2017fast,wu2018r,su2019secucode}.

The results are detailed in \autoref{fig:TaMax}. For a transmit power greater than 800~mW, the CRFID transponder continuously operates without power failure within 100 seconds. If the reader transmit power is below 800~mW, the average operating time of the RFID task drops as the power level decreases. This is because the RFID communication process invokes shorting the antenna, during which the energy harvesting is interrupted. Notably, different from Flash memory, reading data from FRAM (FRAM.R) consumes more power than writing to it (FRAM.W) as a consequence of the destructive read and the compulsory write-back~\cite{thanigai2014msp430}. 
Consequently, we developed: 
\begin{itemize}[leftmargin=15pt]
    \item The {\it Pilot-Observer Mode} to reduce the occurrence of RFID tasks by enabling observing devices to listen to broadcast packets in silence whilst \textit{electing} a single \textit{Pilot} token to respond to the Server.
    \item The {\it Power Aware Execution Model (PAM)} to ensure memory access (FRAM tasks) and intensive computational blocks of the security protocol (CPU tasks) do not exceed the powering capability of the device.
\end{itemize}

\begin{figure}[!ht]
    \centering
    \includegraphics[width=\linewidth]{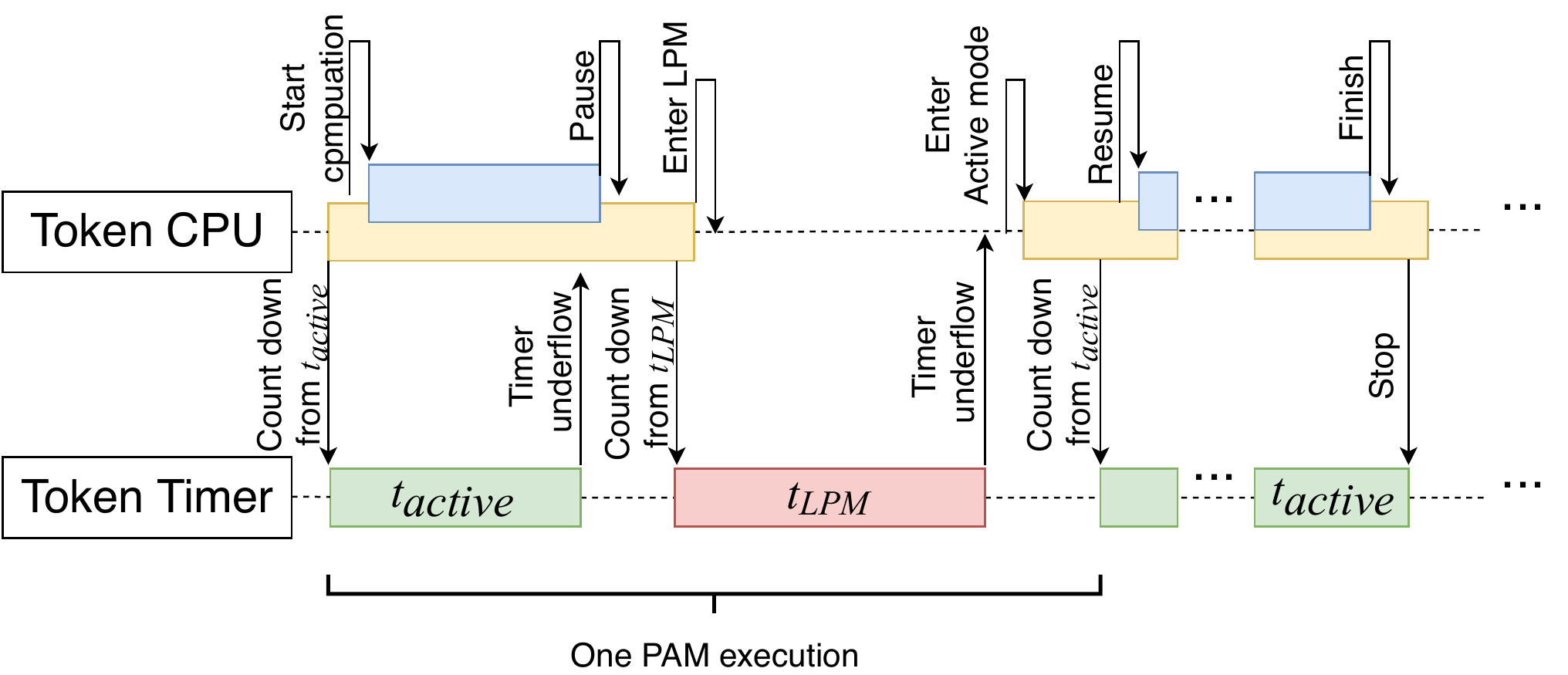}
    \caption{A sequence diagram describing our proposed PAM and illustrating the interaction between the Token's CPU execution and the hardware timer. One PAM execution cycle consists of an active mode followed by a low power mode (LPM), one complete task may involve multiple PAM cycles. Other operations, such as RFID communications and FRAM access, are coordinated by the token CPU. If the CPU is in the LPM state, the entire system will be halted to allow energy to accumulate.}
    \label{fig:PAM_sequence}
\end{figure}

\noindent\textbf{Power Aware Execution Model (PAM).~}\label{sec:paem} The execution mode enables a token to dynamically switch between \textit{active} power mode and lower power mode (LPM)---LPM preserves (SRAM) state and avoids power-loss while executing a task. PAM is illustrated in~\autoref{fig:PAM_sequence}. In active power mode, the token executes computations, and switches to LPM before power-loss to accumulate energy; subsequently, the token is awoken to active power mode to continue the previous computation after a period of $t_{\rm LPM}$. 

Our PAM model builds upon~\cite{buettner2011dewdrop,su2019secucode} in that, these are designed for execution scheduling to \textit{prevent power-loss from brownouts}. Compared to~\cite{su2019secucode}, we consider dynamic scheduling of tasks and in contrast to~\cite{buettner2011dewdrop} sampling  of  the  harvester  voltage (\textit{only possible in specific devices}) within the application code, we consider dynamic scheduling determined by the more resourceful Server $\mathcal{S}$ using a single voltage measurement reported by a token $\mathcal{T}$ (see Appendix~\ref{apd:PAM-methods} for detailed comparison). We outline the means of achieving our PAM model on a token below. 

During the Security Association stage shown in~\autoref{fig:protocol}, a token measures and reports the voltage ${\bf Vt}_i \leftarrow {\sf SNIFF}(t)$ to the Server. The ${\bf Vt}_i$ measurement indicates energy that can be harvested by token $T_i$ under the settings of the current firmware update session. 
According to ${\bf Vt}_i$, the Server determines the active time period $t_{\rm active}$ and LPM time period $t_{\rm LPM}$ for each CRFID device (detailed development of a model to estimate $t_{\rm active}$ and $t_{\rm LPM}$ from ${\bf Vt}$ is in Appendix~\ref{apd:paem}). \textit{Consequently, each CRFID device's execution model is configured by the Server with device specific LPM and active periods at run-time. Hence, an adaptive execution model customized to the available power that could be harvested by each CRFID device is realized. Notably, this execution scheduling task is outsourced to the resourceful Server}.

In this context, we assume the distances between the antenna and target CRFID devices are relatively constant during the \textit{short} duration of a firmware update. Firmware updates are generally a maintenance activity where CRFID integrated components are less likely to be mobile to ease maintenance, such as during the scheduled maintenance of an automated production line~\cite{fischer2017advancements}; night time updates in smart buildings when people are less likely to be present and facilities are inoperative~\cite{elmangoush2015application}; or the pre-flight maintenance or inspection of aircraft parts while parked on an apron~\cite{yang2018rfid}.  
 Further, it is desirable to maintain a stable powering channel in practice by ensuring a consistent distance during the maintenance or patching of devices for a short period. This is a more reasonable proposition than the wired programming of each device. Hence, the relatively fixed distance is a reasonable assumption in practice. Notably, given the challenging nature of the problem, previous non-secure firmware update methods, such as \textit{$R^3$}~\cite{wu2018r} and \textit{Stork}~\cite{aantjes2017fast}, were evaluated under the same assumption.

\begin{figure}[!ht]
    \centering
    \includegraphics[width=0.95\linewidth]{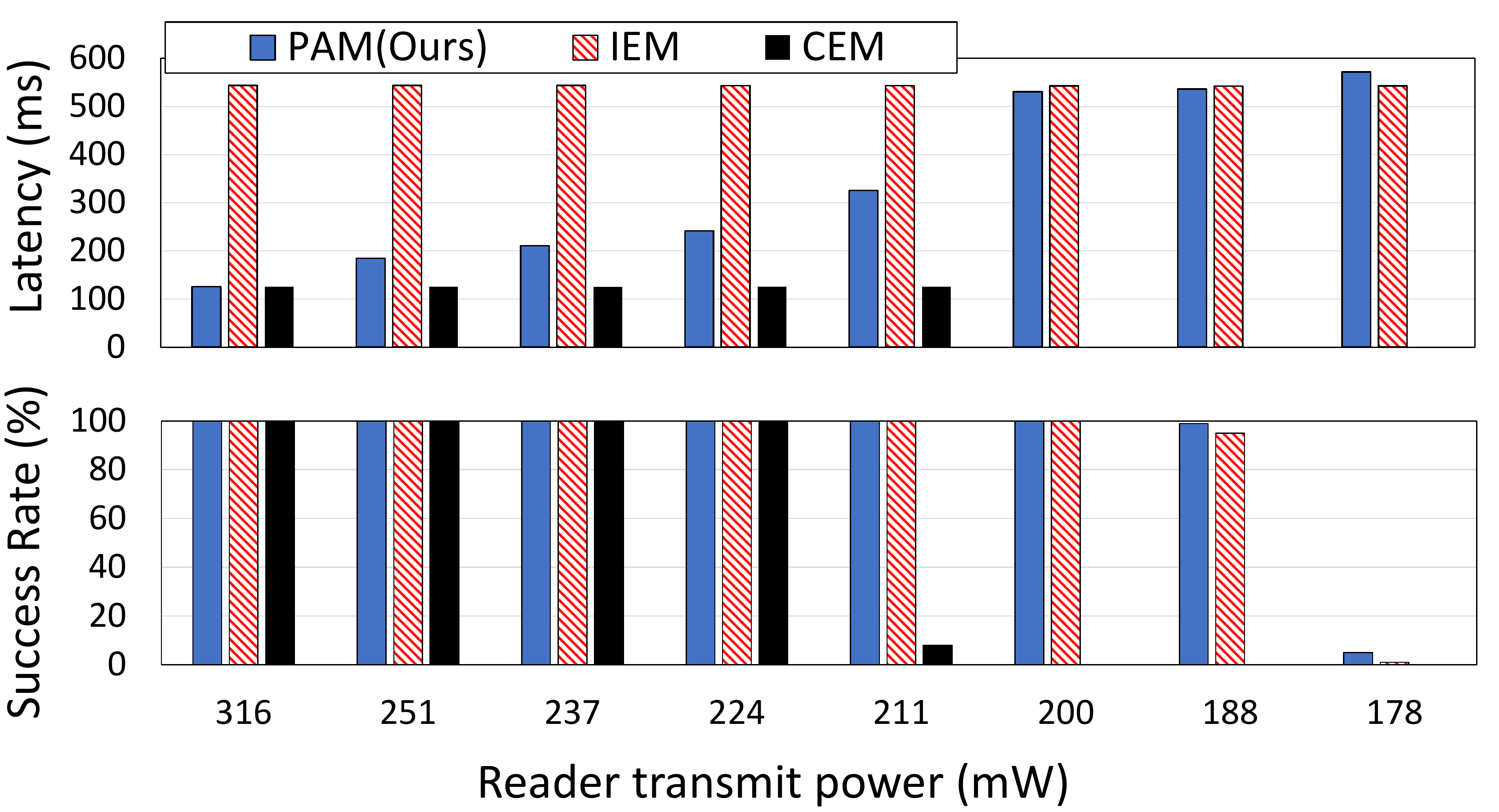}
    \caption{Experimental evaluation of the latency and success rate of our proposed power aware execution model (PAM) compared to the \Revsource{rev:IEM}{\Rev{Intermittent Execution Mode (IEM)}} method in~\cite{su2019secucode} and a typical continuous execution model (CEM). The long-run task used in this evaluation is a MAC computation over a 1,536-Byte random message typically requiring 125.5~ms to complete in CEM. Notably, CEM fails when the reader transmit power is below 200~mW. We conduct 100 repeated measurements and report the mean.}
    \label{fig:PAM-CMAC evaluation}
\end{figure}

\noindent\textbf{PAM Experimental Validation.} \label{sec:power-loss-meth}
To understand the effectiveness of our PAM method to reduce the impact of brownout, we execute a computation intensive module, \textsf{MAC()} using PAM, at power-up on a CRFID. We employ a few lines of code to toggle GPIO pins to indicate the successful completion of a routine. Notably, it is difficult to track a device's internal state without a debug tool attached to the device; however, if the debug interface is in use, it will either interfere with powering, or affect the timing by involving additional \Revsource{rev:JTAG}{\Rev{Joint Test Action Group (JTAG)}} service code. We measure the time taken to complete the \textsf{MAC()} execution (latency) and success rate (success over 10 repeated attempts under wireless powering conditions) with digital storage oscilloscope connected to GPIO pins indicating a successful execution before power loss. We compute a MAC over a 1,536-Byte randomly generated message to test the effectiveness of PAM in preventing a power loss event due to brown-out and compare performance with the execution method---denoted IEM---of a fixed LPM state (30~ms) \textit{programmatically} encoded during the device provisioning phase as in~\cite{su2019secucode}. 

The results in \autoref{fig:PAM-CMAC evaluation} show the effectiveness of PAM to mitigate the interruptions from power loss. This is evident when the success rate results without PAM---using the continuous execution model (CEM)---is compared with those of PAM at decreasing operating power levels. However, as expected, we can also observe that the dynamically adjusted execution model parameters ($t_{\rm LPM}$ and $t_{\rm active}$) of PAM at decreasing power levels to prevent power loss events and increase latency or the time to complete the routine. When PAM is compared with IEM, the dynamically adjusted execution model parameters allow PAM to demonstrate an improved capability to manage interruptions from power-loss at poor powering conditions; this is demonstrated by the higher success rates when the reader transmit power is at 188~mW and 178~mW. Since IEM encodes operating settings programmatically during the device provisioning phase~\cite{su2019secucode}, we can observe increased latency at better powering conditions when compared to PAM which allows a completion time similar to that obtained from CEM when power is ample. Thus, PAM provides a suitable compromise between latency and successful completion of a task at different powering conditions.

Notably, the RFID media access control (MAC) layer on a CRFID device is implemented in software as assembly code and executed at specific clock speeds to ensure strict signal timing requirements in the \epc air interface protocol. Additionally, protocol message timing requirements places strict limits on waiting periods for devices responses. Hence, we do not consider the direct control of the execution mode during communication sessions for managing power consumption and instead rely on the Pilot-Observer mode method we investigate next.

\begin{table}[ht!]
{
\centering
\caption{Execution overhead of pilot and observer tokens when receiving broadcast packets.}
\label{tab:Execution_overhead_RF_pilot_observer}
\resizebox{\columnwidth}{!}{%
\begin{tabular}{lcccc}
\toprule[1.5pt]
\multicolumn{1}{c}{\multirow{2}{*}{}} & \multicolumn{1}{c}{\multirow{2}{*}{Clock Cycles\footnotemark[1]}} & \multicolumn{2}{l}{Memory Usage (Bytes)} & \multicolumn{1}{c}{\multirow{2}{*}{Operation}} \\ 
\multicolumn{1}{c}{} & \multicolumn{1}{c}{} & ROM  & \multicolumn{1}{c}{RAM}  & \multicolumn{1}{c}{} \\ \hline
\addstackgap{Pilot token receiving} & 23,082 & 12 & 2 & \addstackgap{\makecell{apply decoding, \\prepare reply (e.g., compute CRC)}}\\\hline
\addstackgap{Pilot token replying} & 1,131 & 0 & 0 & communication (backscattering)\\\hline
\addstackgap{Observer token receiving} & 22,002 & 12 & 2 & apply decoding\\ %\hline
% \multicolumn{5}{l}{\bf Update with a 240-Byte firmware} \\ \hline
% Pilot & 2,905,560 & 12 & 2 & apply decoding and backscattering\\
% Observer & 2,640,240 & 12 & 2 & apply decoding\\
\bottomrule[1.5pt]
\end{tabular}
}\\
}
\begin{tablenotes}[flushleft]
\item \scriptsize{ \footnotemark[1]{Numbers are collected using a JTAG debugger where the reported values are averages over 100 repeated measurements. We provide a detailed discussion of the experiments and analysis process in Appendix~\ref{apd:overhead_receiving_broadcast_packets}}.}
\end{tablenotes}
\end{table}

\vspace{1mm}
\noindent\textbf{Pilot-Observer Mode.~}\label{sec:observerMode}
\begin{figureRevsourceS}[t]
    \centering
    \includegraphics[width=\mylinewidth]{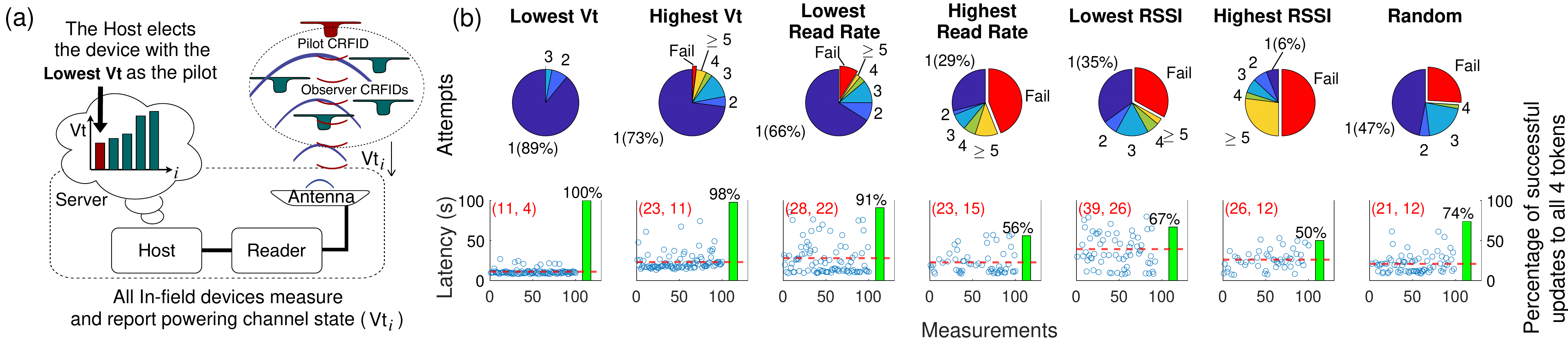}
    \caption{(a) Our proposed Pilot-Observer method. Only the elected \textit{Pilot} using the ${\rm Vt}_i$ based election method responds while \textit{observers} listen silently. (b) Evaluation of Pilot Token selection strategies. Measured Tokens are placed 40~cm above the reader antenna, where the powering condition is critical; we defer the reader to Appendix~\ref{apd:pilot_selection_distances} for further results from other operational distances. The number of attempts to have all 4 CRFIDs (Tokens) updated (pie chart) and the corresponding latency (scatter plot) over 100 repeated measurements. The mean number and the standard division for successful updates for all four tokens are also labeled in the scatter plot with (mean, std) given in seconds. }.\label{fig:pilot_selection_result}
\end{figureRevsourceS}
We observed and also confirmed in \autoref{fig:TaMax} that the task of responding to communication commands will likely cause power loss. During the Secure Broadcast stage shown in~\autoref{fig:protocol}, communication is dominated by repeated \texttt{SecureComm} commands with payloads of encrypted \textbf{firmware} and the data flow is uni-directional. Intuitively, we can disable responses from all the tokens to save energy. However, the \texttt{SecureComm} command under \epc requires a reply (ACK) from the token to serve as an acknowledgment~\cite{epcglobal2015inc}. An absent ACK within 20~ms will cause a protocol timeout and execution failure.

To address the issue, we propose the pilot-observer mode inspired by the method in~Stork~\cite{aantjes2017fast}. A critical and distinguishing feature of our approach is \textit{the intelligent election of a pilot token from all tokens to be updated}---we defer to Appendix~\ref{apd:ObsvsPilot} for a more detailed discussion on the differences.  
As illustrated in \autoref{fig:pilot_selection_result}(a), our approach places all in-field tokens to be updated into an \textit{observer} mode except one token \textit{elected by the Server} to drive the \epc protocol---this device is termed the \textit{Pilot} token. By doing so, observer tokens process all commands such as \texttt{SecureComm}---ignoring the handle designating the target device for the command---whilst remaining silent or muting replies to \textit{all} commands whilst in the \textit{observer} mode. Muting replies significantly reduce energy consumption and disruption to power harvesting of the tokens.

We propose \textit{electing} the token with the \textit{lowest} reported $\textbf{Vt}$ as the pilot based on the observation:~measuring powering channel state from the token, obtained from ${\bf Vt}_i \leftarrow {\sf SNIFF}(t)$, is the most reliable measure of power available to a given device (see the discussion in Appendix~\ref{apd:paem}). 

We recognize that it is very difficult to implement an explicit synchronization method to ensure the observer tokens stay synchronized with the pilot token in the secure broadcast update session. This is due to the level of complexity and overhead that an explicit method would bring and the consequence of such an overhead would be the increased occurrence of failures of a secure broadcast update session due to the additional task demands on tokens---we discuss the problem further in \autoref{sec:conclusion}. Although synchronization between the pilot and observer tokens are not explicit, the Pilot-Observer method implicitly enforces a degree of synchronicity. The selected pilot token---tasked with responding to the Host commands during the broadcast---has to spend more time than observer tokens to prepare the uplink packet and send a reply (ACK) to the Host (RFID reader) to meet the \epc specification requirements as summarized in~\autoref{tab:Execution_overhead_RF_pilot_observer}; hence the update process is controlled by the \textit{slowest} token. Further, since we elect the token with the lowest powering condition as the pilot, the update is controlled by the most energy-starved token. This strategy leads to other tokens having higher levels of available power and extra time to successfully process the broadcasted firmware and remain synchronized with the update process. Hence, \textit{if the pilot succeeds, observers are expected to succeed}.

\vspace{1mm}
\label{sec:pilot-election}
\noindent\textbf{Validation of Pilot Election Method.~}To demonstrate the effectiveness of our pilot token election method, we considered \textit{\textbf{seven}} different pilot token selection methods based on token based estimations of available power and indirect methods of estimating the power available to a token by the server: 1) Lowest $V_t$: this implies the pilot is selected based on the most conservative energy availability at token; 2) Highest $V_t$: the pilot has to reply to commands, hence we may expect higher power availability to prevent the failure of a broadcast session due to brownout at the pilot; 3) Lowest Read Rate: a slow pilot allows observers to gather more energy during the broadcast and remain synchronized with the protocol to prevent failure of the observer tokens; 4) Highest Read Rate: a faster pilot could reduce latency; 5) Lowest \Revsource{rev:RSSI}{\Rev{Received Signal Strength Indicator (RSSI):}} successful communication over the poorest channel may ensure all observers are on a better channel (RSSI acts as an indirect measure of the powering channel at a token); 6) Highest RSSI: prevent pilot failure due to a potentially poor powering channel at the pilot token; and 7) Random selection: \textit{monkey beats man on stock picks}.

We have extensively evaluated and compared all of the above seven pilot token selection methods. The experiment is conducted by placing 4 CRFID tokens at 20~cm, 30~cm, 40~cm and 50~cm 
above an reader antenna; each CRFID is placed in alignment with the four edges of the antenna (See \autoref{fig:vsSecuCode}.(d)). We repeated the code dissemination process {\it 100 times} at each distance test, for each of the 7 different pilot selection methods. We recorded two measures: i) latency (seconds); and ii) number of times the broadcast attempt updated all of the 4 in-field devices.As illustrated in \autoref{fig:pilot_selection_result}.(b), at 40~cm where the power conditions are more critical at a token, the selected pilot token with the lowest token voltage ({\bf Vt}) shows an enormous advantage, in terms of both latency and attempt number. Overall, our proposed approach (electing the lowest {\bf Vt}) ensures that most observer tokens are able to correctly obtain and validate the firmware in a given broadcast session on the first attempt in the critical powering regions of operation (we provide a theoretical justification in Appendix~\ref{apd:paem}).

We also evaluated the reduction in power consumption achieved from computational tasks, in addition to eliminating the communication task. While we provide a detailed discussion of the experiments and analysis process in Appendix~\ref{apd:overhead_receiving_broadcast_packets}, we summarize our experimental results in \autoref{tab:Execution_overhead_RF_pilot_observer}. The measurements show that the observer tokens, compared to the pilot token, consumes 2,211 less clock cycles \textit{per} firmware packet from the Host. This is a reduction of 9.13\% to process each firmware packet sent from the Host.

\vspace{1mm}
\noindent\textbf{Summary.~}\textit{Our pilot-election based method effectively reduces the chance of power failure as well as de-synchronization of the observer tokens during a firmware broadcast session. Our approach is able to ensure that more of the observer tokens are able to correctly obtain and validate the firmware during a single broadcast sessions}.

\subsection{Efficient Security Primitives}\label{sec:effi-security-prim}

\wisecr requires two cryptogrpahic primitives: i) symmetric key primitive \textsf{SKP()}; and ii) a keyed hash primitive for the message authentication code \textsf{MAC()}. When selecting corresponding primitives in the following, two key factors are necessary to consider:
\vspace{1mm}
\begin{enumerate}[leftmargin=*]
\item \textit{Computational cost}: The CPU clock cycles required for a primitives quantifies both the computation cost and energy consumption. Therefore, clock cycles required for executing the primitives need to be within energy and computational limits of the energy harvesting device. 
\item \textit{Memory needs}: On-chip memory is limited---only 2~KiB of SRAM on the target MCU---and must be shared with: i) the RFID protocol implementation; and ii) user code.
\end{enumerate}

\vspace{1mm}
\noindent\textbf{Symmetric Key Primitives.~}\label{sec:BlockCiphers} We fist compare block ciphers via software implementation benchmarks---clock cycle counts and memory usage---on our target microcontroller~\cite{buhrow2014block,dinu2015triathlon}. 
We also considered the increasingly available cryptographic co-processors in microcontrollers: our targeted device uses an MSP430FR5969 microcontroller and embeds an on-chip {\it hardware} \Revsource{rev:AES}{\Rev{Advanced Encryption Standard (AES)}} accelerator (we refer to as HW-AES)~\cite{instruments2014msp430fr58xx}. Based on the above selection considerations, HW-AES is confirmed to outperform others. Specifically, HW-AES to encrypt/decrypt a 128-bit block consumes 167/214 clock cycles with a power overhead of $21~\mu$A/MHz; therefore we opted for HW-AES for \textsf{SKP.Dec} function implementation on our target device. We configured AES to employ the Cipher-Block Chaining (CBC) mode\footnote{\scriptsize We are aware of CBC padding oracle attacks; however, in our implementation, the response an attacker can obtains is to the CMAC failure or success and not the success or failure of the decryption routine. Alternatively, the routines can be changed to employ authenticated encryption in the future, with an increase in overhead.} similar hardware AES resources can be found in a variety of microcontrollers, ASIC, FPGA IP core and smart cards. In the absence of HW-AES, a software AES implementation, such as tiny-AES-c can be an alternative.

\vspace{1mm}
\noindent\textbf{Message Authentication Code.~} MACs built upon BLAKE2s-256, BLAKE2s-128, HWAES-GMAC and HWAES-CMAC on our targeted MSP430FR5969 MCU were taken into consideration~\cite{su2019secucode}. 
We selected the 128-bit \textit{HWAES-CMAC}---Cipher-based Message Authentication Code\footnote{\scriptsize The implementation in NIST Special Publication 800-38B, \textit{Recommendation for Block Cipher Modes of Operation: the CMAC Mode for Authentication} is used here.}---based on AES since it yielded the lowest clock cycles per Byte by exploiting the MCU's AES accelerator (HW-AES).

\subsection{Bootloader Control Flow}\label{sec:control-flow}
\begin{figure*}[!ht]
    \centering
    \includegraphics[width=1.0\linewidth]{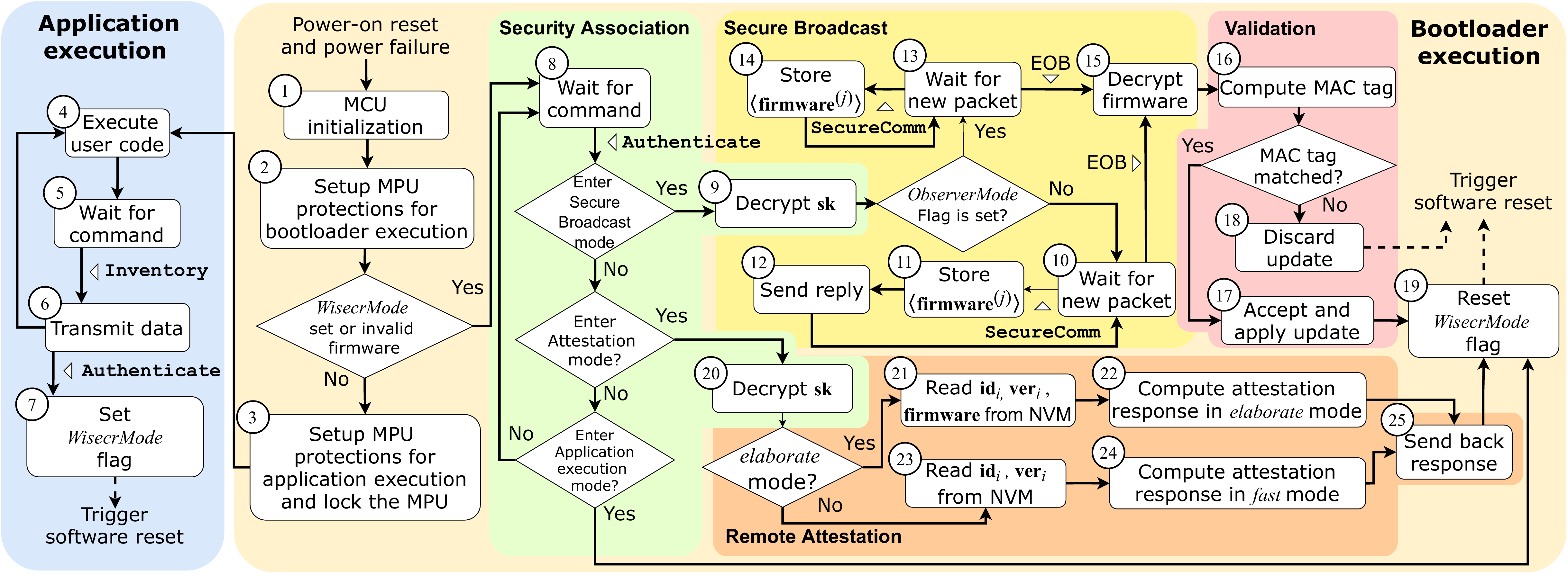}
     \caption{\wisecr Bootloader control flow. The Bootloader manages Security Association, Secure Broadcast, Validation and Remote Attestation stages.}
    \label{fig:ControlFlow}
\end{figure*}

We can now realize an uninterruptible control flow for an immutable bootloader built upon on the security properties identified and engineered to achieve our \wisecr scheme described in Section~\ref{sec:protocol}. We describe the bootloader control flow in \autoref{fig:ControlFlow}.

At power-up, \circled{1} the token performs MCU initialization routines, \circled{2} setup MPU protections for bootloader execution and carries out a self-test to determine whether a firmware update is required, or if there is a valid application installed. If no update is required and the user application is valid, \circled{3} the bootloader configures MPU protections for application execution, before handing over control to the application. During the application execution, \circled{4} the user code is executed, \circled{5} then the token $T$ waits for command from the Server $\mathcal{S}$, the Server $\mathcal{S}$ may send an \texttt{Inventory} commands to instruct the token $T$ to \circled{6} send back e.g., sensed data, or in \circled{7}~setup the \textit{WisecrMode} flag and trigger a software reset in preparation for an update. 

\vspace{1mm}
\noindent\textbf{Security Association.}~Upon a software reset, a set \textit{WisecrMode} flag directs the token $T$ to enter the Security Association stage. \circled{8} Subsequently, the token waits for further instructions from the Server. On reception of an \texttt{Authenticate} command, carrying an encrypted broadcast session key $\langle {\bf sk} \rangle_i$ and the MAC tag ${\bf s}_i$ of the new firmware, \circled{9} the token $T$ decrypts $\langle {\bf sk} \rangle_i$ with its device key ${\bf k}_i$ and acquires the session key ${\bf sk}$.

\vspace{1mm}
\noindent\textbf{Secure Broadcast.}~Recall, at this stage, all tokens selected for update will be switched into the observer mode except the pilot token that is set to respond to the Server.
The pilot token in state~\circled{10} receives a new chunk of encrypted firmware $\langle {\bf firmware}^{(j)} \rangle$, and \circled{11}~stores it into the specified memory location. \circled{12} The pilot token sends reply to the Server.
In \circled{13}, for tokens in observer mode, they silently listen to the communication traffic between the Server and the Pilot token without replying.
In other words, tokens under observer mode receives the encrypted firmware chunks, \circled{14} store chunks in memory but ignores unicast handle identifying the target device; this significantly saves observers' energy from replying as detailed in Section~\ref{sec:observerMode}. Once an End of Broadcast (\texttt{EOB}) message is received, all tokens stop waiting for new packets and start firmware decryption \circled{15}.

\vspace{1mm}
\noindent\textbf{Validation.}~The token computes a local MAC tag ${\bf s}_i'$, including the decrypted firmware, and compares it with the received MAC tag ${\bf s}_i$ \circled{16}. The firmware is accepted, applied to update \circled{17} the token, if ${\bf s}_i'$ and ${\bf s}_i$ are matched, and \circled{19} the \textit{WisecrMode} flag is reset; otherwise, \circled{18} the firmware is discarded and the update is aborted. Subsequently, each token performs a software reset to execute the new firmware or reinitialize in bootloader mode if the firmware update is unsuccessful.

\vspace{1mm}
\noindent\textbf{Remote Attestation.}~On reception of an \texttt{Authenticate} command with an instruction to perform remote attestation, \circled{20} the token first decrypts the session key ${\bf sk}$ and reads \circled{21} ${\bf id}_i$, ${\bf ver}_i$ and ${\bf firmware}$ from the NVM according to the specified memory address and size (firmware is only used in the \textit{elaborate} mode, in the \textit{fast} mode \circled{23} only ${\bf id}_i$ and ${\bf ver}_i$ are involved). The attestation response is then computed as ${\bf r} \leftarrow \textsf{MAC}_{{\bf sk}}({\bf c\|\textrm{selected segment}~(firmware)\| {\bf id}_i}\|{\bf ver}_i)$ if the \textit{elaborate} mode is selected \circled{22} or ${\bf r} \leftarrow \textsf{MAC}_{{\bf sk}}({\bf c\|{\bf id}_i}\|{\bf ver}_i)$ if using the \textit{fast} mode \circled{24}. Subsequently, the response ${\bf r}$ is sent back to the Server $\mathcal{S}$, \circled{25} the \textit{WisecrMode} flag is cleared and \circled{19} the device is reset.

Notably, a power-loss event
during the control flow will result in a reset and rebooting of the token $T$ and a transition out from the firmware update state. Most significantly, in such an occurrence, the immutable bootloader functionality is preserved. Although the loss of state implies that a Server $\mathcal{S}$ must reattempt the secure update, we prevent the token from being left in a vulnerable state to inherently mitigate security threats posed in the power \textit{off} state~\cite{krishnan2018exploiting}. 

\section{End-to-End Implementation and Experiments}\label{sec:exp-evals}
This section elaborates on software tools, the end-to-end implementation of \wisecr and extensive experiments conducted to systematically evaluate the performance of \wisecr. 

\subsection{End-to-End Implementation}\label{sec:implementation}
We implement a bootloader and a \textit{Server Toolkit} to wirelessly and simultaneously update multiple CRFID transponders using {\it commodity} networked RFID hardware and {\it standard} protocols. We realize the \wisecr scheme in~\autoref{fig:protocol} and the identified security property requirements. Specifically, we employ MSP430FR5969 MCU based WIPS5.1LRG CRFID devices. The MSP430FR5969 microcontroller has a 2~KiB SRAM and a 64~KiB FRAM internal memory. \textit{The implementation is a significant undertaking} and includes software development for the CRFID as well as the RFID reader and backend services for the server update mechanisms. Given that we have open sourced the project, we describe: i)~the \textit{bootloader} in Section~\ref{sec:securestorage}; and ii)~the Sever Toolkit and the protocols for Host-to-RFID-reader and Reader-to-CRFID-device communications for implementing \wisecr over \epc in Appendix~\ref{apd:Detailed-protocol}.  

\subsection{Bootloader and Memory Management}\label{sec:securestorage}

\noindent{\bf \wisecr Bootloader.~}The immutable bootloader supervises firmware execution while needing to be compatible with standard protocols, we developed our bootloader based on the Texas Instrument's recent framework for wireless firmware updates---\textit{MSP430FRBoot}~\cite{ryanbrownkatiepier2016}
---and the code base from recent work~\cite{su2019secucode} employing the framework. Such construction ensures a standard tool chain for compilation and update of new firmware whilst the usage of industry standards is likely to increase adoption in the future. We compile the firmware code into ELF (Executable and Linkable Format) made up of binary machine code and linker map specifying the target memory allocation---the memory arrangement is illustrated~\autoref{fig:segment-diagram}.

\renewcommand{\mylinewidth}{0.83\linewidth}
\begin{figureRevsource}[!ht]
    \centering
    \includegraphics[width=\mylinewidth]{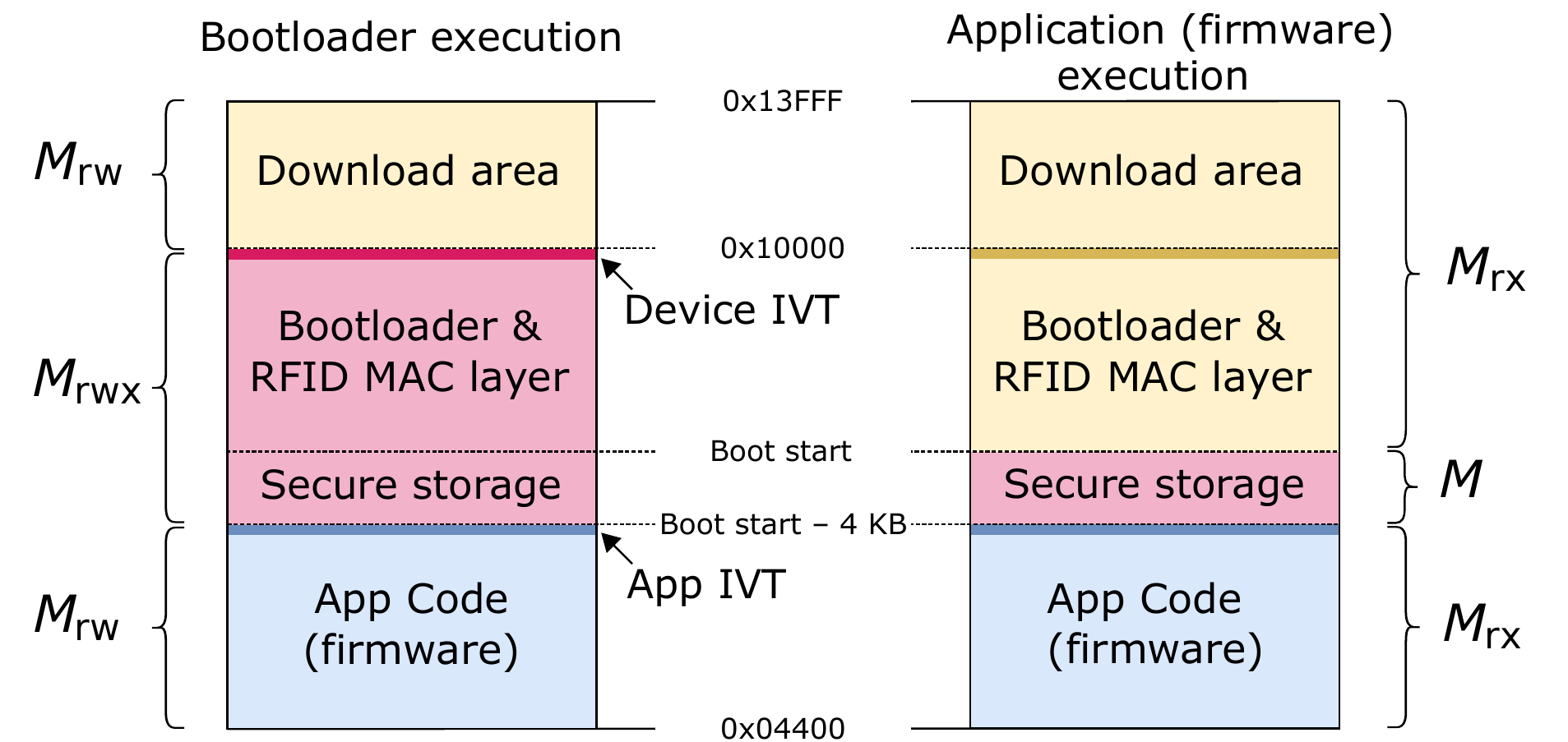}
    \caption{Memory Protection Unit (MPU) segmentation diagram and memory arrangement during bootloader execution (left) and application code execution (right). Due to hardware limitations, only 3 segments can be defined, and the secure storage area must be at least 1~KiB in size due to segment boundary alignment requirements~\cite{instrumentsmsp430framtechnology}}
    \label{fig:segment-diagram}
\end{figureRevsource}
\renewcommand{\mylinewidth}{\linewidth}

\noindent{\bf \wisecr Memory Management.} For implementing the Secure Storage component, several different mechanisms are available (as summarized in Appendix~\ref{apd:memory_management_compare}). In \wisecr, we select to use the
{\it Runtime access protection (e.g. using MPU segments at run-time)~} In this scheme, secure storage is available on device boot-up, but is locked (until next boot-up) by the bootloader before any application code is executed. We employed this method and the implemented memory arrangement is described in \autoref{fig:segment-diagram}. Switching to privilege mode is done by the bootloader on boot up, and the applications do not need to make any specific modifications. If an application attempts to access a privileged resource, e.g. overwriting the bootloader segment, a power-up clear (PUC) will be triggered by the  MPU to cause reboot, before any risky operation.

Importantly, any power-on reset, regardless of the cause, will result in a reboot and the execution of the bootloader before transferring control to application code---see Application execution stage in \autoref{fig:ControlFlow}. Prior to entering the application execution stage, the bootloader enforces the MPU policy for application execution (see \circled{3}). This step is necessary with the target MCU used in our implementation as it only allows defining three memory segments for protection. Thus, we are able to rely on the very limited protections provided by the MCU to achieve the security objectives of a trusted bootloader; recall, in our threat model, the bootloader is trusted, and we realize this in practice by ensuring that the bootloader provisioned is immutable where the secrets stored on device must remain inaccessible and immutable to the application firmware. Now, it is infeasible to revoke the policy enforced during the application execution stage, unless the device is power-cycled. But, the device will enter the bootloader stage after a reboot and MPU protections will be re-enabled prior to executing any application code. We summarize the memory protections developed using the limited MPU configurations to realize \wisecr security requirements in bootloader and application execution modes in Section~\ref{sec:unterruptbootloader}; therein, we also discuss alternative methods to realize such protections to ensure the generalization of \wisecr to other MCUs.

\begin{table*}[ht!]
{\centering
\caption{\wisecr Implementation and Execution Overheads}
\label{tab:building_block}
\resizebox{.83\linewidth}{!}{%
\begin{tabular}{lcccc}
\toprule[1.5pt]
\multicolumn{1}{c}{\multirow{2}{*}{}} & \multicolumn{1}{c}{\multirow{2}{*}{\quad Clock Cycles \quad}} & \multicolumn{2}{c}{Memory Usage (Bytes) \quad} & \multicolumn{1}{c}{\multirow{2}{*}{\quad HW Requirement \quad}} \\ 
\multicolumn{1}{c}{} & \multicolumn{1}{c}{} & \multicolumn{1}{c}{ROM}  & \multicolumn{1}{c}{RAM}  & \multicolumn{1}{c}{} \\ \hline
\multicolumn{5}{l}{\bf Functional blocks} \\
\hline
${\bf sk} \leftarrow \sf{SKP.Dec_{\bf k_{\mathnormal{i}}}}(\langle {\bf sk_{\mathnormal{i}}} \rangle)$ (For a 16-Byte block) & 772 & 706\footnotemark[2] & 62 & HW-AES \\
${\bf firmware}^{(j)} \leftarrow \sf{SKP.Dec_{\bf sk}}(\langle {\bf firmware}^{(j)} \rangle)$ (For a 16-Byte block) & 772 & 690 & 62 & HW-AES \\
${\bf s}_i$ $\leftarrow$ $\sf{MAC_{\bf sk}}$(${\bf firmware||ver_{\mathnormal{i}}||nver}$) & 26,698\footnotemark[1] & 2,374 & 72 & HW-AES \\
$ \bf Vt_{{\mathnormal{i}}} \leftarrow \textsf{SNIFF(}\mathnormal{t}\textsf{)}$ & 1,530 & 276 & 12 & ADC \\ 
\textsf{PAM.Enter(}{\bf $ \rm t_{LPM_{{\mathnormal{i}}}}$, $ \rm t_{active_{{\mathnormal{i}}}}$}\textsf{)} and \textsf{PAM.Exit()}\footnotemark[3] & 51 & 86 & 6 & Timer\\
Setup MPU for bootloader execution & 76 & 58 & 0 & MPU\\
Setup MPU for application execution & 91 & 60 & 0 & MPU\\
\Rev{Secure Storage (used for ${\bf k}_i$, ${\bf id}_i$ and ${\bf ver}_i$)} & \Rev{0} & \Rev{19} & \Rev{0} & \Rev{MPU}
\\ \hline 
\multicolumn{5}{l}{\bf Stages}\\
\hline
Security Association & 2,302 & 982 & 74 & ADC, HW-AES\\
Secure Broadcast & 11,604\footnotemark[1] & 760 & 68 & HW-AES, Timer\\
Validation & 26,724\footnotemark[1] & 2,390 & 77 & HW-AES, Timer\\
Remote Attestation (\textit{elaborate} mode) & 18,360\footnotemark[4] & 3,172 & 80 & HW-AES \\
Remote Attestation (\textit{fast} mode) & 5,574 & 3,172 & 80 & HW-AES\\
\hline
\multicolumn{5}{l}{\bf Total} \\
\hline
\wisecr (excluding Remote Attestation, for the \textbf{pilot} Token) & 40,630\footnotemark[1] & 3442 & 154 & All of the above \\
SecuCode (using fixed key, updates a single token) & 27,092\footnotemark[1] & 2442 & 78 & HW-AES, Timer \\
\bottomrule[1.5pt]
\end{tabular}
}\\
}
\begin{tablenotes}[flushleft]
 \item \scriptsize{\footnotemark[1]{For a typical 240-Byte firmware}} \quad
 \scriptsize{\footnotemark[2]{16 Bytes of this value incorporates the device key ${\bf k}_i$ in secure storage}} \quad
 \scriptsize{\footnotemark[3]{A single PAM execution, as defined in~\autoref{fig:PAM_sequence}}} \quad
 \item \scriptsize{\footnotemark[4]4,397 clock cycles for the setup routine--a constant overhead for any block size, 1,060 clock cycles for each 16-Byte block and $n+2$ clock cycles for an $n$-Byte padding to form a 16-Byte block, if required.}
\end{tablenotes}
\end{table*}

\begin{figure}[!ht]
  \includegraphics[width=\linewidth]{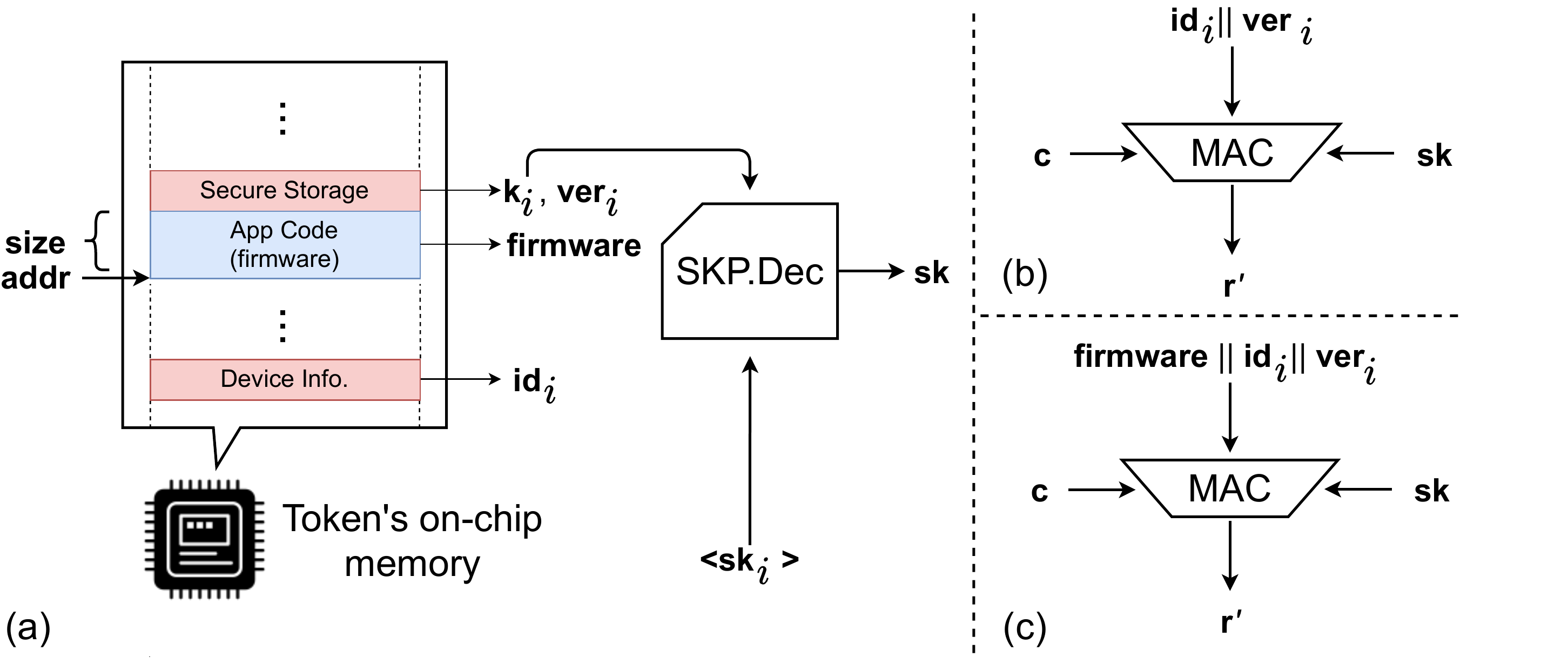}
  \caption{(a) Token memory and allocations, (b) \textit{fast} mode, (c) \textit{elaborate} mode.}
  \label{fig:boat1}
\end{figure}

\noindent{\bf Remote Attestation.}~As shown in \autoref{fig:boat1}~(a) Remote Attestation routine is an integral part of the immutable bootloader, it has access to all memory segments: Token device key ${\bf k}_i$, version number ${\bf ver}_i$, tag serial number ${\bf id}_i$ and the installed user application code \textbf{firmware}. A challenge \textbf{c} is a one-time use random string that is generated by the Server $\mathcal{S}$. The ${\bf r}'$ is the report to be transferred back to the Server. On receiving an encrypted session key ${\bf \langle sk_{\it i} \rangle}$, the Token decrypts it with \textsf{SKP.Dec} and obtains the session key ${\bf sk}$.   In the \textit{fast} mode illustrated in \autoref{fig:boat1}~(b), the response ${\bf r}'$ is calculated based on ${\bf id}_i$ and ${\bf ver}_i$. In contrast, the response ${\bf r}'$ is computed over an entire memory segment, such as the \textbf{firmware}, in the \textit{elaborate} mode as depicted in \autoref{fig:boat1}~(c). 

\subsection{\wisecr Implementation Overheads}
\label{sec:implementation_footprint}
We first evaluate the execution and implementation overhead of each \wisecr\space {\it security functional block}; results\footnote{\scriptsize Tested under the CCS 9.0.1.0004 development environment, with compiler TI v18.12.1.LTS Optimization settings: -O = 3; -opt\_for\_speed = 5.} are summarized in \autoref{tab:building_block}. The execution overhead is measured in terms of clock cycles for installing a firmware of 240 Bytes. The implementation overhead is measured in memory usage, consisting of ROM for code and constants and RAM for run-time state. Secondly, we assess the necessary hardware modules such as hardware AES accelerator (HW-AES), analog-to-digital converter (ADC) and Timer. All functional blocks are implemented in platform independent C source files. We summarize the implementation costs for each stage; the most significant implementation and execution overhead is from the Validation stage because of the computationally heavy \textsf{MAC()} function---recall Remote Attestation is an optional stage. 

In \autoref{tab:building_block}, we also compare with SecuCode\footnote{\scriptsize In~\cite{su2019secucode}, the device key is derived from SRAM PUF responses. In our comparison, we opt for a device key in the protected NVM instead to make a fair overhead comparison. Further, the key derivation module only adds a fixed overhead at start-up.}; presently, the only CRFID firmware update scheme considering security---notably, the scheme only prevents malicious code injection attacks and is designed to update a single device at a time as highlighted in the method comparison \autoref{tab:comparison}.  In summary, \textit{Wisecr} consumes 49.97\% more clock cycles, 40.95\% more ROM and 97.43\% more RAM space than the SecuCode. However, \textit{Wisecr} can update multiple devices simultaneously to attain a performance advantage as we demonstrate in Section~\ref{sec:implementation_footprint} because SecuCode only updates one device at a time. In addition, Wisecr provides IP protection (encryption of the firmware transmitted over the insecure wireless channel) and validates firmware installation. As expected, the performance improvements and additional security features are achieved at an increased implementation overhead.

\subsection{Experiment Designs and Evaluations}
\label{sec:exp-designs}
We employed four WIPS5.1LRG CRFID devices and an Impinj R420 RFID reader with a 9~dBic gain antenna to communicate with and energize the CRFID devices. All of the experiments are conducted with the RFID reader feeding the antenna with 30 dBm output power. Given that, the recent \epc V2.0~\cite{epcglobal2015inc} security commands such as the \texttt{SecureComm} are not yet widely supported by commercial RFID readers, like the Impinj Speedway R420, we map the unsupported \epc command to the \texttt{BlockWrite} command as in~\cite{su2019secucode}. We summarize our evaluations below\footnote{\scriptsize Notably, our extensive experiments to evaluate techniques and modules we developed are detailed in the Appendices with references in the main article}:

\begin{itemize}[leftmargin=15pt]
    \item We evaluate performance---latency and throughput in Section~\ref{sec:performance}---and compare with: i) SecuCode~\cite{su2019secucode}, the scheme only prevents malicious code injection attacks and is designed to update a single device at a time as shown in the method comparison Table~\ref{tab:comparison}; ii) \wisecr intentionally set to sequential mode---broadcasting/updating one token at a time---termed \wisecr(Seq), to assess the gains from \wisecr broadcasting firmware to multiple devices simultaneously; and iii) non-secure multi-CRFID dissemination scheme, Stork~\cite{aantjes2017fast} (see Section~\ref{sec:stork-compare})
    
    \item We evaluate the efficacy of our Power Aware Execution Model (PAM) and Pilot Token selection method (results in the Section. \ref{sec:pilot-election}).
    
    \item We demonstrate \wisecr in an end-to-end implementation (see Section~\ref{sec:demo} and demonstration video at~\href{https://youtu.be/GgDHPJi3A5U}{\color{black}\ul{https://youtu.be/GgDHPJi3A5U}})
\end{itemize}

\begin{figure*}[!tp]
    \centering
    \begin{tabular}{@{}p{.80\textwidth}p{.19\textwidth}@{}}
    \vspace{-15pt}
    \raisebox{-\height}{\includegraphics[width=\linewidth]{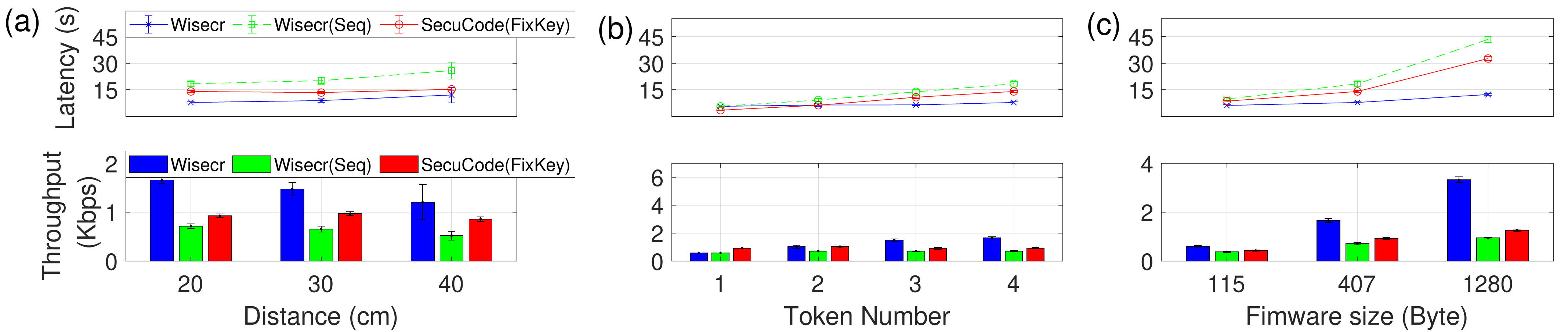}}&
    \raisebox{-\height}{\includegraphics[width=\linewidth]{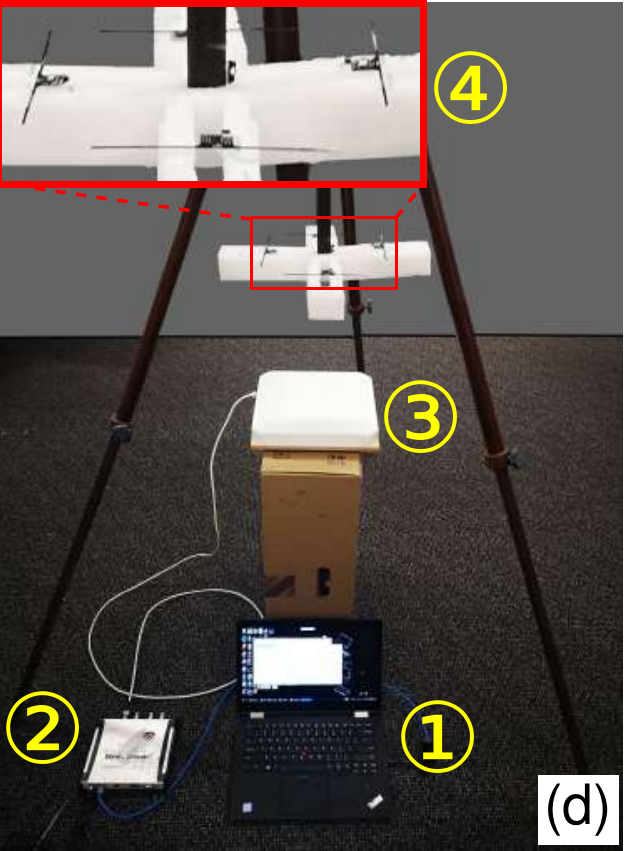}}\\
    \vspace{-55pt}
    \caption{Comparing \wisecr, \wisecr(Seq)---\wisecr operating in sequential mode, SecuCode~\cite{su2019secucode} \textit{with no secure broadcast} support under increasing: (a) Distance (4 Tokens with a 407 Byte firmware), (b) Number of Tokens (at 20~cm with 407 Byte firmware) and (c) Firmware size (at 20~cm with 4 Tokens). The experimental setup illustrated in (d) shows \circled{1} a host PC, \circled{2} a RFID reader, \circled{3} a reader antenna and \circled{4} four CRFID devices mounted on a foam frame, supported by a wooden tripod.} 
    \label{fig:vsSecuCode}
    \end{tabular}
    \vspace{-15pt}
\end{figure*}

\subsection{Performance Evaluation and Results}\label{sec:performance}
We use two performance metrics: i)~end-to-end~\textit{latency}; and ii)~\textit{throughput} (bits per second). Latency is the time elapsed from the Server Toolkit transmitting the firmware to an RFID reader using LLRP commands to the time the Server Toolkit confirms the acknowledgment of successfully updating {\it all chosen tokens}.  Throughput is the total data transferred over the latency, where $Throughput~(bps) = (\left |\textbf{firmware}|~\times~\textit{Num. of Tokens})\right/Latency$. Our evaluation is carried out under three different controlled variables: i)~distance; ii)~number of tokens; and iii) firmware size. In each experiment, only one variable is changed  
and our results are collected over 100 repeated measurements, with outliers outside of 1.5 times upper and lower quartiles removed.

First, we transfer the same firmware code (a 407 Byte accelerometer service) to four target devices located at distances from 20~cm to 40~cm, covering good to poor powering channel states, to evaluate the \textit{stability} of the \wisecr scheme. As expected, we can observe in \autoref{fig:vsSecuCode}(a) that the performance of all three schemes downgrades with increasing distance. Notably, \wisecr and SecuCode outperforms the \wisecr(Seq). Because, \wisecr(Seq) incurs a larger overhead through the need for executing a Security Association stage to setup the broadcast security parameters, for each device. Further, SecuCode (without IP protection) does not need to execute the power intensive firmware decryption operations, and thus is less likely to encounter power-loss and update session failure or require the extra time necessary by \wisecr to decrpyt the firmware. As expected, \wisecr outperforms albeit the added security functions provided in comparison to SecuCode.

Second, following the method and metrics in Stork~\cite{aantjes2017fast}, we tested the performance of the schemes by increasing the number of CRFID devices to be updated. We can see in \autoref{fig:vsSecuCode}(b) that the latency of both \wisecr(Seq) and SecuCode increases linearly while \wisecr remains relatively steady regardless of the number of devices to update. Consequently, \wisecr exhibits significantly improved throughput as the number of device increases. Further, we examine performance under increasing firmware size: 115 Bytes; 407 Bytes (a sensor service code); and 1280 Bytes (a computation code firmware). Efficacy of broadcasting to multiple devices simultaneously is validated by latency and throughput performance detailed in \autoref{fig:vsSecuCode}(c). %visible as the firmware gets larger.

Third, all three candidates show improved efficiency as larger firmware sizes are transmitted; \wisecr's significantly better efficiency can be attributed to the broadcast method.

We also conducted an evaluation of the execution overhead (latency) introduced by the remote attestation routines. Since remote attestation is performed on one singulated token at an instance, we only examine its impact on one individual device. The latency can be aggregated for multiple devices. As illustrated in \autoref{fig:vsSecuCode}, the remote attestation in \textit{fast} mode always completes in 1.1~ms, regardless of the firmware size. While the time required to complete the \textit{elaborate} mode scales from 1.3~ms to 6.5~ms with respect to firmware sizes from 115~Bytes to 1280~Bytes.
\begin{figure}[!ht]
    \centering
    \includegraphics[width=0.9\linewidth]{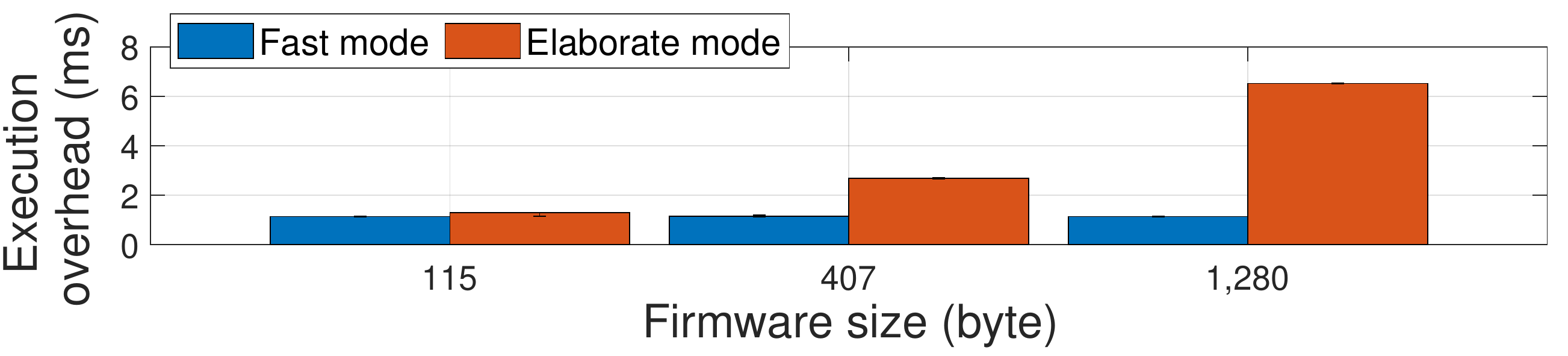}
     \caption{The execution overhead of the attestation function with respect to different firmware sizes in an end-to-end experiment setting. Mean values over 100 repeated measurements are plotted.} 
    \label{fig:attestation_performance_overhead}
\end{figure}

\begin{figure}[!ht]
    \centering
    \includegraphics[width=\linewidth]{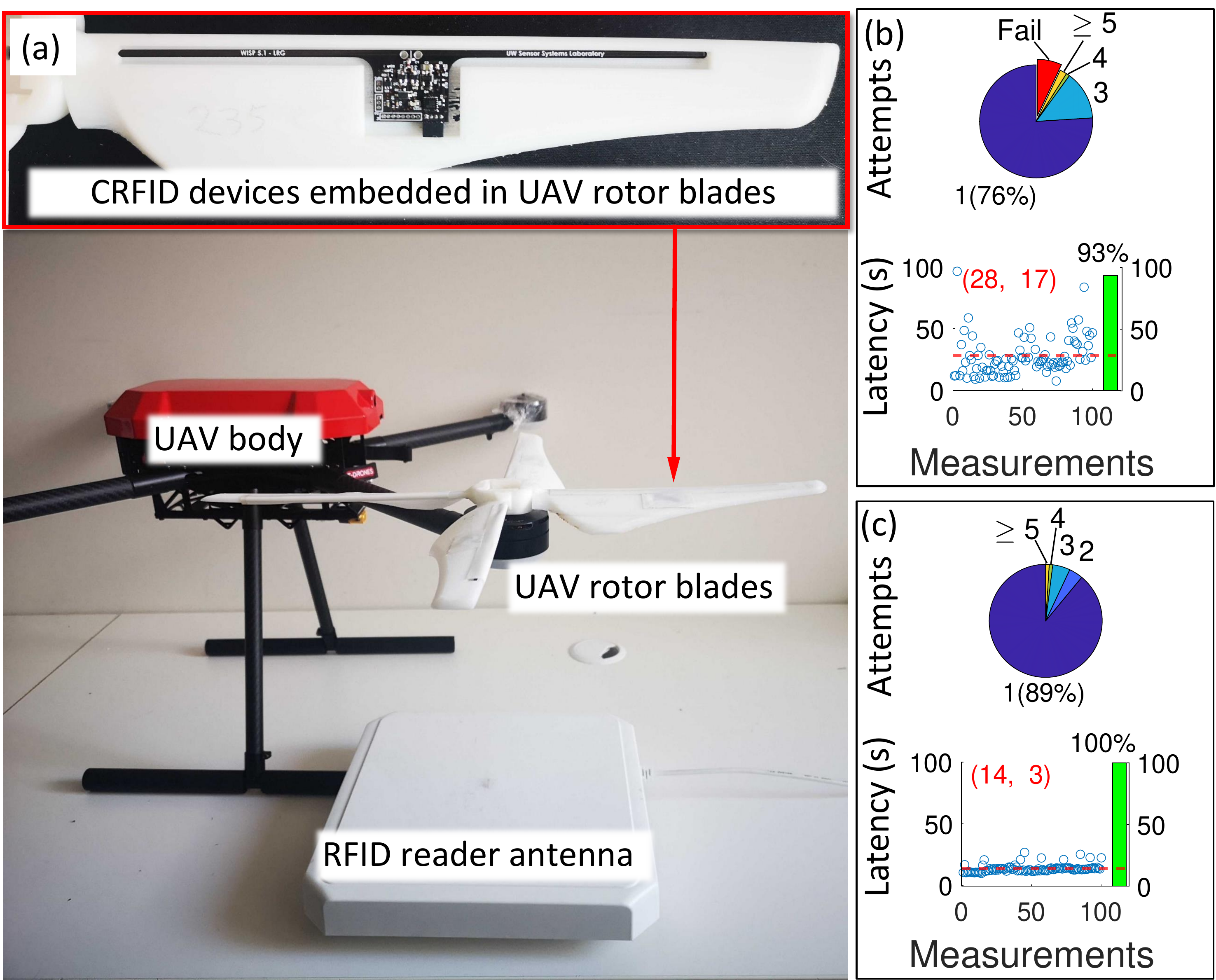}
    \caption{(a) Application scenario: wirelessly updating the firmware of four CRFID sensor devices embedded in an unmanned aerial vehicle (UAV) rotor prototype\protect\footnotemark. With \wisecr, we simultaneously, securely and wirelessly updated the four devices with a 391-Byte code. Experimental results when: (b) the rotor assembly is attached to the UAV; and (c) when the rotor assembly is free standing over the antenna. The green bar graph denotes the percentage of successful updates to all 4 devices.}
    \label{fig:UAV_rotor}
\end{figure}
\footnotetext{\scriptsize The UAV rotor aerodynamic model is adopted and modified from: https://grabcad.com/library/4-bladed-propeller-experimental-1}
\vspace{-2mm}

\subsection{Case Study}\label{sec:demo}

% \vspace{1mm}
% \noindent{\bf End-to-End Demonstration.~}\label{sec:demo}
As shown in \autoref{fig:UAV_rotor}, we employ four CRFID devices embedded in a 3D printed unmanned aerial vehicle (UAV) rotor prototype. \textit{The CRFID devices are factory programmed with firmware for monitoring the centrifugal load and the wired programming interface is disabled prior to embedding and deployment. All four CRFID devices are required to be updated wirelessly and securely with code for monitor the blade flapping vibration in a wind tunnel test.} 

We employed \wisecr to simultaneously, securely and wirelessly update the four CRFID sensor devices with the 391-Byte firmware code. The experimental results summarized in \autoref{fig:UAV_rotor} show successful updates for 100 repeated attempts in two settings: i)~when the rotor is attached to the UAV \wisecr updated all of the devices in the first attempt in 76\% of the time with an average latency of 28.12~seconds; and ii)~when the rotor assembly is free standing, \wisecr updated all of the devices in the first attempt in 89\% of the time with an average latency of 14.27 seconds\footnote{\scriptsize Although previous studies have performed experiments with static tags, we also attempted to update the tokens while the rotor blades are mobile by rotating them at approximately 1~RPM. We were able to update tokens over 50\% of the time. As expected, the changing powering conditions dramatically reduced the update rate (see details in Appendix~\ref{apd:update_moving_tokens})}.

A demonstration of the secure firmware update process in using our end-to-end implementation illustrated in \autoref{fig:UAV_rotor} is available at \href{https://youtu.be/GgDHPJi3A5U}{\color{black}\ul{https://youtu.be/GgDHPJi3A5U}}. We open source the complete source code for  \wisecr and related tools at~\href{https://github.com/AdelaideAuto-IDLab/Wisecr}{\color{black}\ul{https://github.com/AdelaideAuto-IDLab/Wisecr}}.

\subsection{Security Analysis}\label{sec:securityanalysis}
Under the threat model in Section~\ref{sec:adversarymodel}, we analyze \wisecr security against: i) IP theft (code reverse engineering); ii) malicious code injection attacks under code alteration, loading unauthorized code, loading code onto an unauthorized device, and code downgrading; 
and iii) incomplete code installation.

\vspace{1mm}
\noindent{\bf IP Theft or Code Reverse Engineering.~}The wirelessly broadcasted firmware is encrypted using a one-time 128-bit broadcast session key $\textbf{sk}_i$. Without the knowledge of the broadcast session key, the adversary $\mathcal{A}$ is unable to obtain the plaintext \textbf{firmware} through passive eavesdropping. Once the firmware is decrypted on the token, the binaries cannot be accessed outside the immutable bootloader. Therefore, the probability of a successful IP theft is now determined by the security provided by the size of the selected keys---we employ 128-bit keys in the \wisecr implementation. 

\vspace{1mm}
\noindent{\bf Loading Unauthorized Code.~}The security property of a MAC tag ${\bf s}_i$ $\leftarrow$ \textsf{MAC$_{{\bf k}_i}$}(${\bf firmware}$$\vert \vert$${\bf ver}_i$$\vert \vert$${\bf nver}$) ensures that the adversary $\mathcal{A}$ cannot commit a malicious firmware injection attack without the knowledge of the session key $\textbf{sk}$
---recall that the sharing of $\textbf{sk}$ is protected through a secure channel established with a token specific private key $\textbf{k}_i$. Hence, only the Server with the session key $\textbf{sk}$ is able to produce a valid MAC tag to a token $T$. Hence, the probability of launching a successful attack by the adversary $\mathcal{A}$ is limited to the probability of successfully obtaining the device key $\textbf{k}_i$ and/or the session key $\textbf{sk}$ in a man-in-the-middle attack to successfully construct a MAC tag and disseminate a malicious firmware that will be correctly validated by the target tokens. Therefore, without knowledge of ${\bf k}_i$ or $\textbf{sk}$, the probability of fooling a token $T$ to inject a malicious firmware by $\mathcal A$ is no more than the brute-force attack probability $2^{-128}$ determined by the key size $|{\bf k}_i|=128$.

\vspace{1mm}
\noindent{\bf Code Alteration.~} For each firmware update, a MAC tag ${\bf s}_i\leftarrow$ \textsf{MAC$_{{\bf k}_i}$}(${\bf firmware}$$\vert \vert$${\bf ver}_i$$\vert \vert$${\bf nver}$) is used to verify code integrity. Therefore, without knowledge of the session key $\bf sk$, the adversary $\mathcal{A}$ can not generate a valid MAC tag for an altered firmware. Any attempted changes to the firmware during the code dissemination will be detected and, thus, firmware  discarded by the token.

\vspace{1mm}
\noindent{\bf Loading Code onto an Unauthorized Device.~} Each authorized device $i$ maintains a unique and secret device specific key ${\bf k}_i$ only known to the Server $\mathcal{S}$. Therefore, an unauthorized device cannot decrypt the session key ${\bf sk}_i$ to install a recorded firmware encrypted with ${\bf sk}_i$ of an authorized device.

\vspace{1mm}
\noindent{\bf Code Downgrading.~}The adversary $\mathcal{A}$ can attempt to downgrade the firmware to an older version to facilitate exploitation of potential vulnerabilities in a previous distribution. This can be attempted through a replay attack by impersonating the Server. However, for each firmware update, 
a specific MAC tag: ${\bf s}_i$ $\leftarrow$ \textsf{MAC$_{{\bf k}_i}$}(${\bf firmware}$$\vert \vert$${\bf ver}_i$$\vert \vert$${\bf nver}$) is generated based on two monotonically increasing version numbers: ${\bf ver}_i$ and ${\bf nver}$. Without direct access to modify the Secure storage contents in region $M$ to re-write the current version number of a device to an older version, the attacker cannot replay a recorded MAC tag from a previous session to force a token to accept an older firmware. By virtue of the monotonically increasing version numbers, and the secure MAC primitive protected by the session key ${\bf sk}$, a successful software downgrade using a replay attack cannot be mounted by the adversary $\mathcal{A}$.  

\vspace{1mm}
\Revsource{rev:incomplete_firm_installation}{\noindent{\bf Incomplete Firmware Installation.~} \Rev{In a man-in-the-middle attack,} the adversary $\mathcal{A}$ may attempt to prevent a new firmware installation and spoof a device response with an updated version number during an interrogation---the version number is public information and sent in plaintext as shown in \autoref{fig:protocol}. Hence, a token may be left executing an older firmware version with potential firmware vulnerabilities. The Server can verify that a specific disseminated firmware deployed by the token is the same as the one issued by the Server by performing a remote attestation. In our \wisecr scheme, we have considered an optional remote attestation stage (detailed in Section~\ref{sec:protocol}) to allow the Server to verify the firmware installation on a given token. In general, both code installation and attestation are bounded to the device key ${\bf k}_i$, thus, it is infeasible to fool the Server without knowledge of the device key.}

\vspace{1mm}
\Revsource{rev:evasdrop_key}{\noindent{\bf Related-key Differential Cryptanalysis Attack.~} In our security association stage, the session key ${\bf sk}$ is encrypted multiple times with different device keys ${\bf k}_i$. A related-key differential cryptanalysis of round-reduced AES-128 is demonstrated in~\cite{fouque2013structural}. The prerequisite to mount such an attack is that the attacker knows or is able to  choose a relation between several secret keys and gain access to both plaintext and ciphertext~\cite{biryukov2009related}. In our approach, as mentioned in Section \ref{sec:adversarymodel}, ${\bf k}_i$ chosen by the server are \textit{i.i.d.}, and only the ciphertext is publicly available where the attacker is not able to force the server to encrypt a plaintext of their choosing. Therefore related-key attacks cannot be mounted under our threat model. Even if such an attack is possible, to the best of our knowledge, there is no successful full-round related-key differential cryptanalysis attack against  AES-128, which serve as the \textsf{SKP} function in our implementation.}

\section{Related Work and Discussion}\label{sec:relatedwork}
 
Here, we discuss related works in the area of \textit{wireless code dissemination to CRFID devices}. Notably, the topic of wireless code updating has been intensively studied for battery powered Wireless Sensor Network (WSN) nodes. For example, a recent study by Florian {\it et al.}~\cite{kohnhauser2016secure} proposed updating the firmware and peer-to-peer attestation of multiple mesh networked IoT devices. Physical unclonable functions~\cite{su2019secucode} and blockchains~\cite{lee2017blockchain} have also been utilized in code dissemination

But, such schemes~\cite{kohnhauser2016secure,lee2017blockchain} are designed for {\it battery powered} WSN nodes featuring {\it device-to-device communication capability}, and {\it supervisory control of operating systems} (e.g TinyOS), {\it absent} in batteryless CRFID devices (see challenges in Section~\ref{sec:intro}).
Further, batteryless devices operate under extreme energy and computational capability limitations. In particular, harvested energy is intermittent and limited and thus devices face difficulties supporting security functions such as key exchange using Elliptic Curve Diffie-Hellman used in~\cite{kohnhauser2016secure}. Therefore, we focus on research into wireless dissemination schemes for CRFID like devices in our discussions. Further, we also discuss and compare our work with existing wireless methods developed for CRFID to highlight the aim of our scheme to fulfill unmet security objectives for simultaneous wireless firmware update to multiple CRFIDs.

\vspace{1mm}
\noindent\textbf{Wireless Code Dissemination to CRFID.~}Given the recent emergence of the technology, there exists only a few studies on code dissemination to CRFID devices. \textit{Wisent} from Tan {\it et al.}~\cite{tan2016wisent}, and \textit{$R^3$} from Wu {\it et al.}~\cite{wu2018r} demonstrate a wireless firmware update method for CRFIDs. Subsequently, Aantjes {\it et al.}~\cite{aantjes2017fast} proposed \textit{Stork}, a fast multi-CRFID wireless firmware transfer protocol. Stork forces devices to ignore the RN16 handle in down-link packets---a form of promiscuous listening---to realize a logical broadcast channel in the absence of \epc support for a broadcast capability; RN16 specifies the designated packet receiver. This technique enables Stork to simultaneously program multiple CRFIDs to achieve \textit{fast} firmware dissemination. Our Pilot-Observer mode (Section~\ref{sec:observerMode}) is based on a similar concept, but instead of always selecting the first seen device, as in Stork, we strategically elect the token with the lowest $\rm V_t$ as the \textit{Pilot} to achieve higher broadcast success. Brown and Pier from Texas Instrument (TI) developed~\textit{MSPBoot}~\cite{luisreynoso2013} in late 2016. They demonstrate an update using a UART or SPI bus to interconnect a MSP430 16-bit RISC microcontroller and a CC1101 sub-1GHz RF transceiver. 
However, none of the schemes, \textit{Wisent}, \textit{$R^3$}, Stork and the work in~\cite{luisreynoso2013}, considers security when wirelessly updating firmware. SecuCode~\cite{su2019secucode} took the first step to prevent malicious code injection attacks where a single CRFID device is updated. But SecuCode lacks scalability and performs device by device updates, cannot protect the firmware IP, and provides no validation of firmware installation. 

\vspace{1mm}
\noindent\textbf{Remote Attestation.~} Attestation enables the Server to establish trust with the token's hardware and software configuration. Existing mainstream remote attestation methods are unsuitable for a practicable realization in the context of intermittently powered, energy harvesting platforms we focus on. For example, peer-to-peer attestation adopted in~\cite{kohnhauser2016secure} is impractical in the CRFID context due to the lack of a device-to-device communication channel. In boot attestation~\cite{schulz2017boot}, a public-key based scheme is proposed to fit ownership and third party attestation, however public-key schemes are too computationally intensive for the batteryless devices we consider. Recently, a publish/subscribe mechanism based asynchronous attestation for large scale WSN is presented in SARA~\cite{dushku2020sara}, however, such a publish/subscribe paradigm is yet to be supported over a standard \epc protocol. In contrast, our lightweight-remote-attestation mechanisms (see Section~\ref{sec:protocol}) are inspired from the recent study in~\cite{dinu2019sia} where methods are designed for resource limited platforms and require no additional building blocks besides those necessary for \wisecr.

\begin{table}[!ht]
	\centering 
	\caption{Comparison with Related Studies.}
\resizebox{0.48\textwidth}{!}{
\begin{tabular}{lccccccc}
\toprule[1.5pt]
\multicolumn{1}{c}{\multirow{5}{*}{Work}} & \multicolumn{1}{c}{\multirow{5}{*}{\begin{tabular}[c]{@{}c@{}}Multiple\\ devices\end{tabular}}} & \multicolumn{2}{c}{Power-loss mitigation} & 
\multicolumn{3}{|c}{\textbf{{Security}}} & \multicolumn{1}{c}{\multirow{5}{*}{\begin{tabular}[c]{@{}c@{}}Public\\ source\\ code\\release\end{tabular}}} \\ \cline{3-4} \cline{5-7}

\multicolumn{1}{c}{}& 
\multicolumn{1}{c}{}                                                                            & \multicolumn{1}{c}{\begin{tabular}[c]{@{}c@{}}Commun.\\ Energy \\ Reduction\end{tabular}} & \multicolumn{1}{c}{\begin{tabular}[c]{@{}c@{}}Adaptive\\ IEM\end{tabular}} & \multicolumn{1}{|c}{\begin{tabular}[c]{@{}c@{}}Prevent\\ Malicious\\ Code\\ Injection\end{tabular}} & \multicolumn{1}{c}{\begin{tabular}[c]{@{}c@{}}Prevent\\ IP\\ Theft\end{tabular}} & \multicolumn{1}{c}{\begin{tabular}[c]{@{}c@{}}Code\\ instal.\\ attest.\end{tabular}} & \multicolumn{1}{c}{}\\ \hline \hline
        Wisent~\cite{tan2016wisent} & \ding{56} & \ding{56} & \ding{56} & \ding{56} & \ding{56} & \ding{56} & \ding{51} \\ 
        R$^3$~\cite{wu2018r} & \ding{56} & \ding{56} & \ding{56} & \ding{56} & \ding{56} & \ding{56} &  \ding{56} \\ 
        Stork~\cite{aantjes2017fast} & \ding{51} & \ding{51} & \ding{56} & \ding{56} & \ding{56} & \ding{56} & \ding{51}\\ 
        SecuCode~\cite{su2019secucode} & \ding{56} & \ding{56} & \ding{56} & \ding{51} & \ding{56} & \ding{56}  & \ding{51} \\ 
	     \begin{tabular}{@{}c@{}} \textbf{\wisecr} (Ours) \end{tabular} & \ding{51} & \ding{51}  & \ding{51} & \ding{51} & \ding{51} & \ding{51} & \ding{51}\\ \bottomrule[1.5pt]
\end{tabular}
}
\label{tab:comparison}
\end{table}

\vspace{1mm}
 \noindent\textbf{Comparisons.~} \autoref{tab:comparison} summarizes a comparison of \wisecr with wireless code dissemination studies {\it specific to passive CRFID devices}. \wisecr is the only {\it secure} scheme (prevent IP theft and malicious code injection, and provide attestation of firmware installation) for \textit{simultaneous} update of passively powered devices. We have also extensively compared the performance of \wisecr with the \textit{non-secure} code update method of \textit{Stork} to provide an appreciation for the security overhead in an end-to-end implementation below.
 
 \begin{figure*}[!ht]
\centering
  \includegraphics[width=\linewidth]{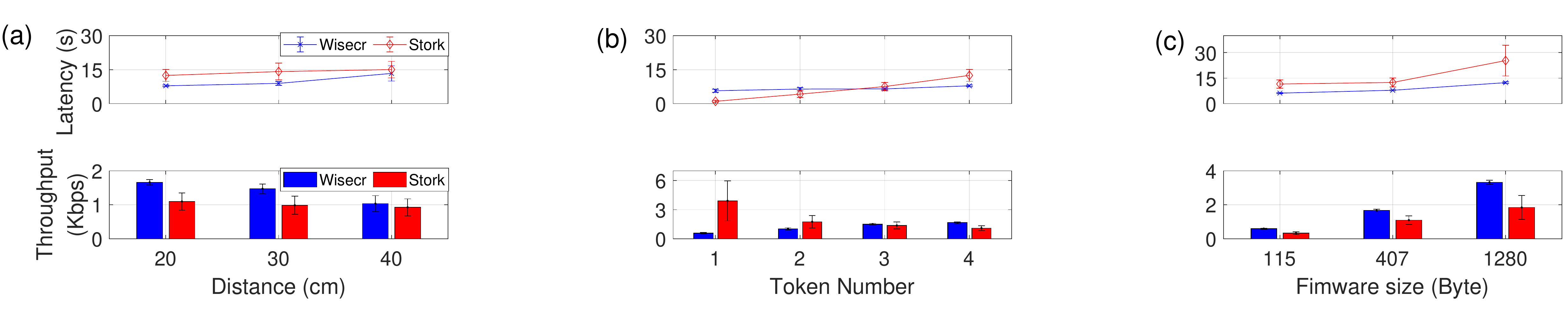}
  \caption{We repeated the firmware update process for \wisecr (ours) and Stork to compare between Stork (providing \textit{no security properties}) and our \wisecr scheme under three different test settings: (a) increasing operational range; (b) increasing number of tokens, and (c) payload sizes.}
  \label{fig:vsStork}
\end{figure*}

\subsection{Comparison with Stork (Insecure Method)}\label{sec:stork-compare}
We extensively compared the performance of our secure \wisecr scheme with Stork~\cite{aantjes2017fast}. Both methods aim to offer multiple CRFID wireless firmware updates, but security objectives are not considered by Stork. Comparisons are carried out under three different test settings:
\vspace{1mm}
\begin{itemize}[leftmargin=15pt]
    \item Four different operating ranges from 20~to 50~cm. 
    \item Updating 1 to 4 tags concurrently in-field to evaluate reduced latency to update (or improvements to scalability) (as in Stork, tags are at 20~cm from the antenna).
    \item Consider three different firmware sizes (as in Stork, tags are at 20~cm from the antenna).
\end{itemize}
We measured two performance metrics: i) latency; and ii) throughput as defined in Section~\ref{sec:performance} for each test setting. For both Stork and \wisecr, we used the same binary files generated from the same compilation. The settings in the \wisecr Server Toolkit are: 10 attempts per broadcast run; lowest voltage $\bf Vt_{i}$ pilot token selection method; Send Mode is set to Broadcast. Meanwhile, the settings in Stork are: BWPayload is set to throttle; Update Only is set to True; Reprogramming Mode is set to Broadcast; Compression is Disabled (as compression simply reduces the size of the firmware, we did not enable this option). Results are mean values from 100 repeated protocol update runs. Comparison results from our experiments are detailed in \autoref{fig:vsStork}:

\begin{itemize}[leftmargin=15pt]
\item From \autoref{fig:vsStork} (a): when powering channel state is reasonable i.e. from 20~cm to 40~cm, \wisecr outperforms Stork, as \wisecr does not need per-block checking and relies on the pilot token selection method to improve broadcast performance. However, Stork performs better than \wisecr at 50~cm when the powering channel condition is very poor because Stork is able to continue to transmit the firmware across power failures since the stork scheme does not need to meet any security requirements and therefore is able to send firmware payload in plaintext from the last correctly received payload. In contrast \wisecr gracefully fails when a power interruption cannot be prevented and all states (such as the session key) are lost and a new broadcast attempt must be made by the Server Toolkit.

\item From \autoref{fig:vsStork} (b): \wisecr exhibits significantly better latency and throughput as the number of tokens increases. 

\item From \autoref{fig:vsStork} (c): both protocols show improved efficiency as larger payload/firmware sizes are  transmitted. Notably, \wisecr exhibits much better efficiency attributed to the significantly reduced latency experienced during the broadcast method. 
\end{itemize}

The gains in performance and faster updates to many devices achieved by \wisecr can be attributed to: i) our pilot token selection method to drive the protocol where a significantly large proportion of in-field tokens are updated in a single attempt; and ii) our validation method of complexity $O(n \cdot 1)$  where $n$ is the number of tokens---although computationally intensive---is more lightweight than the read back mechanism of Stork, $O(n \cdot k)$ where $k$ is the code size, relying on the narrow band communication channel employed by CRFID devices. 

\vspace{-2mm}
\section{Conclusion and Future Work}
\label{sec:conclusion}
Our study proposed and implemented the first secure and simultaneous wireless firmware update to many RF-powered devices with remote attestation of code installation. We explored highly limited resources and innovated to resolve security engineering challenges to implement \wisecr. The scheme prevents malicious code injection, IP theft, and incomplete code installation threats whilst being compliant with standard hardware and protocols. \wisecr performance and comparisons with state-of-the-art through an extensive experiment regime have validated the efficacy and practicality of the design, whilst the end-to-end implementation source code is released to facilitate further improvements by practitioners and the academic community.

\vspace{0.1cm}
\noindent\textbf{Limitations and Future Work:~}Our study is not without limitations. Although the \wisecr reasonably assumes (as in~\textit{$R^3$}~\cite{wu2018r} and \textit{Stork}~\cite{aantjes2017fast}) that the devices are at relatively constant distances from an RFID reader antenna during the short duration firmware update process (approximately 16~s in the application demonstration), occasionally this assumption may not hold. For example, when re-programming devices on a conveyor belt, it is desirable to not delay the movement of the tags and to re-program whilst distances are changing. Development of a solution to the problem in the context of resource limited devices operating over a highly constrained air interface protocol is a challenging problem and an interesting direction for future work.

Additionally, every \texttt{BlockWrite} command used to broadcast the firmware contains a CRC-16 field for error detection~\cite{epcglobal2015inc}. In our Pilot-Observer mode, only the pilot-detected CRC errors are indicated to the Host; future work could consider the problem of broadcast reliability further. Notably, developing a method is non-trivial given the limitations of the \epc air interface protocol.

\Revsource{rev:sync_limitation}{In our secure update method, the \textit{pilot} token and \textit{observer} tokens do not explicitly synchronize progress due to the constrained air interface protocol not supporting token-to-token communications as well as the extremely limited available power. \Rev{This also makes it difficult for the server to immediately, i.e. during the \textit{Secure Broadcast} stage, to detect an observer token that has de-synchronized.} Any synchronization would need to involve the Host performing this function, such as reading back of memory contents from a token's download region and re-transmitting missing packets, at set intervals~\cite{aantjes2017fast}. Hence, developing an explicit synchronization method in the context of a secure broadcast method forms an interesting direction for future work.}

Further, it will be interesting to consider alternative mechanisms for validating the installation of firmware, such as the exploitation of the reply signal to the final packet delivering firmware to tokens. We leave this interesting direction to explore for future work.

%%%%%%%%%%%%%%%%%%%%%%%%%%%%%%%%%%% References %%%%%%%%%%%%%%%%%%%%%%%%%%%%%%%%%%%%%%%%%%%%
\vspace{-2mm}
\bibliographystyle{IEEEtran}
\bibliography{main}

\vspace*{-1.4cm}
\begin{IEEEbiography}[{\includegraphics[width=0.8in,height=1.0in,clip]{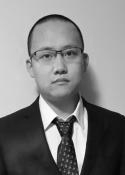}}]{Yang Su}
(S'14) Yang Su received the B.Eng. degree with first class Honours in Electrical and Electronic Engineering from The University of Adelaide, Australia in 2015. He worked as a Research Associate at The University of Adelaide from 2015-2016 and he is currently pursuing the Ph.D. degree. His research interests include hardware security, physical cryptography, embedded systems and computer security. 
\end{IEEEbiography}

\vspace*{-1.4cm}
\begin{IEEEbiography}[{\includegraphics[width=0.8in,height=1.0in,clip]{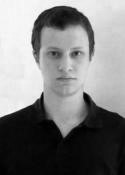}}]
{Michael Chesser} received his B.Sc Advanced degree in 2016 and his Honours (First Class) degree in Computer Science in 2017 from The University of Adelaide, Australia. Michael has worked as a consultant at Chamonix Consulting and, more recently, at the School of Computer Science, The University of Adelaide as a Research Associate. His research interests are in compilers, embedded systems, system security and virtualization.
\end{IEEEbiography}

\vspace*{-1.6cm}
\begin{IEEEbiography}[{\includegraphics[width=0.8in,height=1.0in,clip]{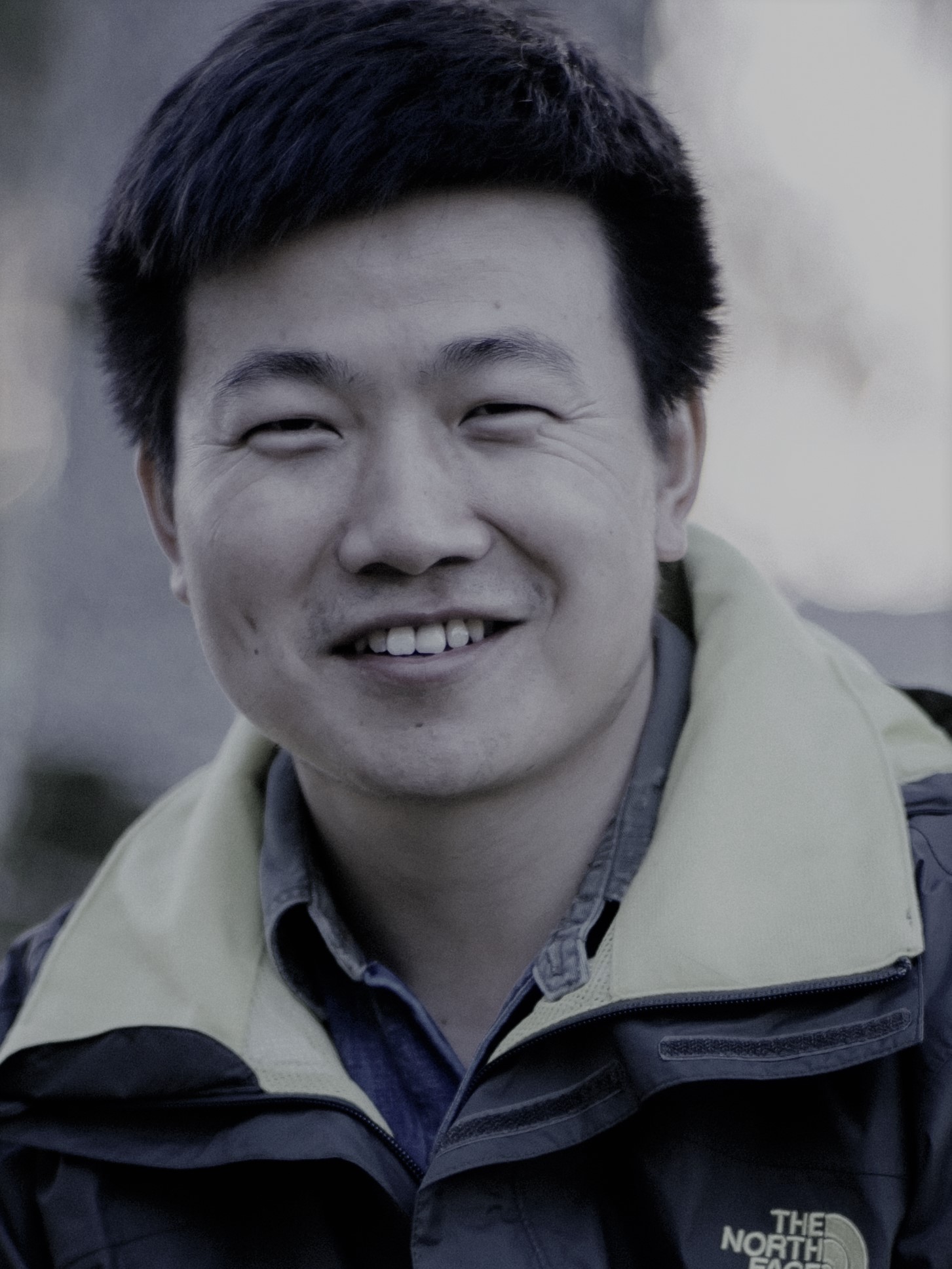}}]{Yansong Gao}
received his M.Sc degree from University of Electronic Science and Technology of China in 2013 and Ph.D degree from the School of Electrical and Electronic Engineering in the University of Adelaide, Australia, in 2017. He is now with School of Computer Science and Engineering, NanJing University of Science and Technology, China. His current research interests are AI security and privacy, hardware security, and system security.
\end{IEEEbiography}
 \vspace*{-1.6cm}

\begin{IEEEbiography}[{\includegraphics[width=0.8in,height=1.0in,clip]{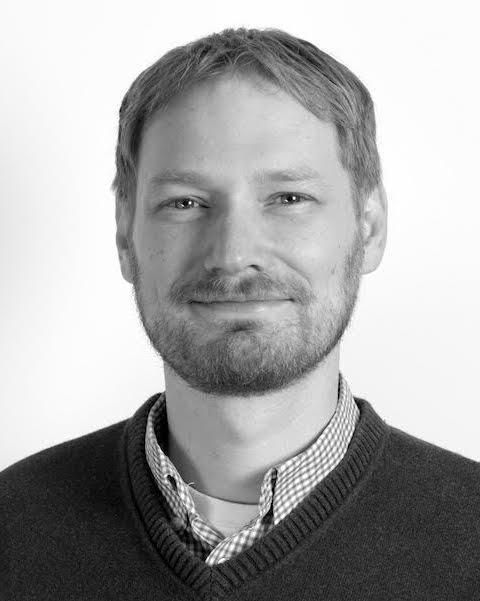}}]{Alanson Sample} received his Ph.D. in electrical engineering from the University of Washington in 2011. He is currently an Associate Professor in the department of Electrical Engineering and Computer Science at the University of Michigan. Prior to returning to academia, he spent the majority of his career working in academic minded industry research labs.  Most recently Alanson was the Executive Lab Director of Disney Research in Los Angeles. Prior to joining Disney, he worked as a Research Scientist at Intel Labs in Hillsboro. He also held a postdoctoral research position at the University of Washington. Dr. Sample's research interests lie broadly in the areas of Human-Computer Interaction, wireless technology, and embedded systems. 
\end{IEEEbiography}

\vspace*{-1.2cm}

\Urlmuskip=0mu plus 1mu\relax

\begin{IEEEbiography}[{\includegraphics[width=0.8in,height=1.01in,clip]{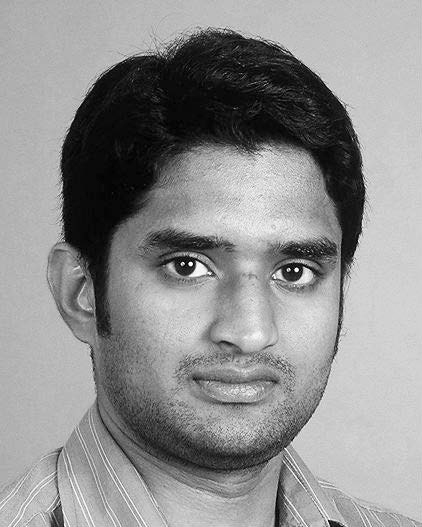}}]{Damith C. Ranasinghe}
received the Ph.D. degree in Electrical and Electronic Engineering from The University of Adelaide, Australia. From 2005 to 2006, he was a Visiting Scholar with the Massachusetts Institute of Technology and a Post-Doctoral Research Fellow with the University of Cambridge from 2007 to 2009. He joined The University of Adelaide in 2010, and is currently an Associate Professor with the School of Computer Science. His research interests include embedded systems, system security, autonomous systems, and deep learning.
\end{IEEEbiography}

\clearpage

\appendices

\section{Memory Management Comparison}\label{apd:memory_management_compare}
\Revsource{rev:appA}{For implementing the Secure Storage component, several different mechanisms can be used:
\begin{enumerate}
    \item \textbf{Isolated segments (e.g. using IP encapsulation hardware).~}
    \textit{Requirements and limitations}: It requires hardware features (such as IP encapsulation) to implement.
    
    \item \textbf{Volatile secret keys} (see, for example, SRAM PUF~\cite{holcomb2008power}). 
    \textit{Requirements and limitations}: It might be computationally expensive to derive and erase the secret key, need a mechanism to `restore' the volatile area on device restart, and a different mechanism is required to ensure the bootloader is immutable.
    
    \item \textbf{Execute only memory (e.g. using MPU segments at compile time)~}In this scheme secure memory is encoded as instructions (e.g. MOV) in execute only memory (XOM) (see~\cite{kohnhauser2016secure}). \textit{Requirements and limitations}: Implementation can be complex, since any code in this region must ensure that it does not leak data (e.g. in CPU registers or volatile memory), including from arbitrary jumps into the code. Additionally, since the bootloader is also unable to write to this region, it cannot be used to protect dynamic secrets (e.g. the broadcast session key).
    
    \item \textbf{Runtime access protection (e.g. using MPU segments at runtime)~} In this scheme, secure storage is available on device boot-up, but is locked (until next boot-up) by the bootloader before any application code is executed. We selected this method in \wisecr. \textit{Requirements and limitations}: Method requires MPU hardware with runtime configuration and locking, and the implementation must ensure that the MPU is always configured correctly before any application code is executed.
\end{enumerate}}

\begin{comment}
\section{RFID Protocol Overview}\label{apd:protocol-session}

The following steps are used to communicate with a CRFID device, or in general any \textit{EPC Gen2} protocol compliant transponder:

\begin{itemize}[leftmargin=15pt]
\item {\bf Host to Reader.} An application on the host machine constructs LLRP commands to build an \texttt{ROSpec} and \texttt{AccessSpec} to control the reader and transmits these specifications to a reader. 

\item {\bf Reader to CRFID.} As part of the anti-collision algorithm in the media access control layer, the reader must first singulate a CRFID device and obtain a handle, $RN16$. After singulation, a device can be placed in Access mode to conduct the firmware update. We illustrate the protocol flow in \autoref{fig:EPC_G2_sessions}. 
\end{itemize}

\begin{figure}[!ht]
    \centering
    \includegraphics[width=0.8\linewidth]{./pic/SecuCode-Batch_overview2.pdf}
    \caption{An \textit{EPC Gen2} protocol session.}
    \label{fig:EPC_G2_sessions}
\end{figure}
\end{comment}

\section{Detailed \wisecr Update Scheme}\label{apd:Detailed-protocol}

\noindent{\bf Server Toolkit.} Our Server Toolkit can be executed on a host with network connectivity to an RFID reader. The App loads in the ELF file generated from the compilation process, parses and slices it into MSPBoot specified 128-bit-long commands suitable for the RFID reader. The App then uses LLRP commands to construct \texttt{AccessSpecs} and \texttt{ROSpecs}. These encode \textit{EPC C1Gen2} protocol commands such as \texttt{BlockWrite} and \texttt{SecureComm} commands to discover, engage and configure a networked RFID reader to execute the \wisecr protocol.
Notably, we follow the same protocols for Host-to-RFID-reader and Reader-to-CRFID-device communications as in~\cite{aantjes2017fast,su2019secucode} and detail implementation of \wisecr over \epc in the \autoref{fig:system-overview}.

\begin{figure}[!ht]
    \centering
    \includegraphics[width=1.0\linewidth]{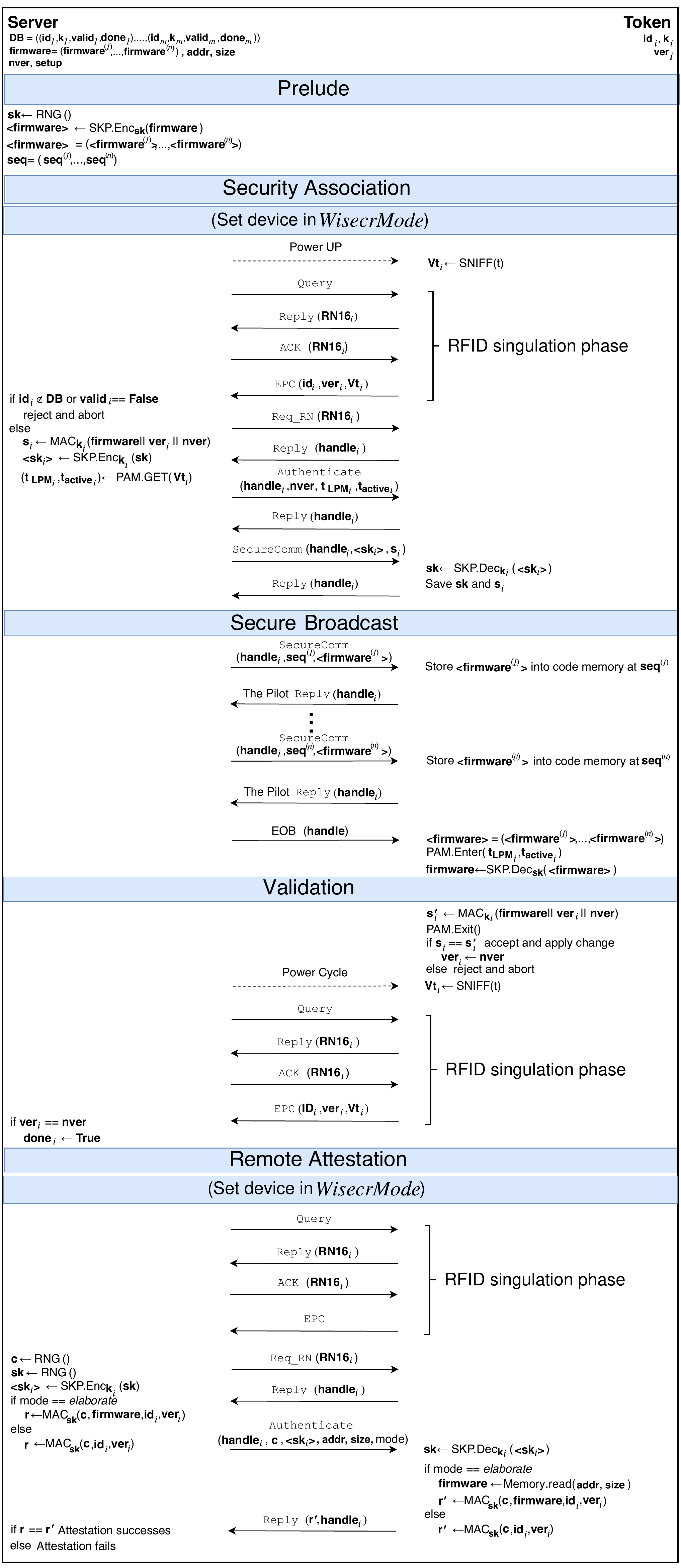}
    \caption{\wisecr protocol implemented over the \epc protocol. To reduce complexity, we only provide the protocol sessions after a device is placed in the \textit{WisecrMode}---as described in \autoref{fig:ControlFlow}---and highlighted here.} 
    \label{fig:Detailed_protocol}
\end{figure}
\noindent{\bf \wisecr Update Scheme Over \epc.}
As described in  \autoref{fig:protocol}, \wisecr enables the ability to distribute and update firmware of multiple CRFID tokens, simultaneously. Given that a Server $\mathcal{S}$ communicates with an RFID device using \textit{\epc}, a practicable, scalable and secure code dissemination scheme must be implemented over \epc. This requires communication between three separate entities: i)~the host  machine; ii)~the reader; and iii)~CRFID transponders---see main text \autoref{fig:system-overview}. Our scheme description here focuses on the communication between CRFID devices---tokens $\mathcal{T}$---and the RFID reader---the Server $\mathcal{S}$ over the \epc protocol as illustrated in \autoref{fig:Detailed_protocol}---our open source code base provides a complete description (\href{https://github.com/AdelaideAuto-IDLab/Wisecr}{\color{black}\ul{https://github.com/AdelaideAuto-IDLab/Wisecr}}).

\noindent{\bf Authenticating the Acknowledgment in the Validation Stage.}~It is theoretically possible to compute and include a MAC tag in an acknowledgment message at the end of the Validation stage but the implementation of this in practice is not possible. Our current acknowledgment signal is based on exploiting the last message in a singulation session---see RFID singulation phase in the Validation stage in \autoref{fig:Detailed_protocol}. This last message returns the the unique device identifier or EPC (electronic product code). We piggy back the current version number ${\bf ver}_i$ in the data field part of this message as our \textit{soft} validation signal. This is possible because at power-up (whether it be software or hardware reset) the device executes the bootloader and during its execution copies the ${\bf id}_i$ and ${\bf ver}_i$ to a global SRAM memory region for use after the MPU (memory protection unit) prevents access to this content stored in the secure storage area (see memory protection unit segmentation diagram in \autoref{fig:segment-diagram}.

Notably, even if a MAC tag was computed by the bootloader after power cycling (reboot) in the Validation stage, and stored in global memory for access after MPU protections to be used during the singulation phase that follows, the data field in the last EPC message is limited to 96 bits. Thus, there is inadequate room to piggyback a 128-bit MAC tag. Further, this would be an added overhead since the MAC computation would need to occur each time a CRFID device is powered (booted), irrespective of whether the MAC is needed or not.
      
Thus, it is better to compute a MAC tag after reboot, on demand and when necessary by: i)~singulating a token first; ii)~instructing the token to compute the MAC tag; and iii)~requesting the MAC tag. This is essentially the method we employ in the \textbf{Remote Attestation} stage that follows the \textbf{Validation} stage.

\section{Low Overhead Execution Scheduling}\label{apd:PAM-methods}

Buettner,  {\it et al.} proposed Dewdrop~\cite{buettner2011dewdrop} execution model to \textit{prevent} a brownout event under unreliable powering by executing tasks only when they are likely to succeed by monitoring the available harvested power. Dewdrop explores a dynamic on-device task scheduling method, however, requires the overhead of collecting samples of the harvester voltage and task scheduling by the device's application code. 

In addition, Dewdrop is only suitable for CRFID devices equipped with a passive charge pump, such as WISP4.1~\cite{sample2008design} since Dewdrop requires directly measuring the charging rate of the reservoir capacitor---charge storage element. In the follow-up, WISP version 5.1, the passive charge pump is replaced with a S-882z active charge pump and the reservoir capacitor is only connected to the load when $V_{\rm cap}$ developed across the capacitor exceeds the reference voltage of 2.4~V ($V_{\rm ref}$). Consequently, in WISP5.1LRG the voltage delivered to the microcontroller is a sharp step-up, rather than a ramp-up function related to harvested power. Therefore, the charging-up process cannot be directly monitored by the technique in Dewdrop.  

\section{Powering Channel State Measurement and Power Aware Execution Model (PAM)}\label{apd:paem}
\vspace{1mm}
\noindent\textbf{Observation.~}Generally, increasing distance of the token from a powering source lowers the harvester output power, RSSI and Read Rate of a given token.

\vspace{1mm}
\noindent\textbf{Proposition.~}Measure powering channel state from the token is the most reliable measure of power available at a device. 

\begin{comment}
\begin{equation}
    \label{eqn:PowerReceived}
    P_{r}=\frac{P_{t} G_{t} G_{R} \lambda^{2}}{(4 \pi R)^{2}}
\end{equation}
\end{comment}

\vspace{1mm}
\noindent\textbf{Validation.~} 
\begin{equation}
    \label{eqn:RSSI}
    R S S I=P_{t} G_{t}^{2} G_{p a t h}^{2} K ~\textrm{where}~ G_{p a t h}=\left(\frac{\lambda}{4 \pi d_{o}}\right)^{2}|H|^{2}
\end{equation}

\begin{equation}
    \label{eqn:H}
    H=1+\sum_{i=1}^{N} g_{t}^{i} g^{i} \Gamma_{i} \frac{d_{o}}{d_{i}} e^{-j k\left(d_{i}-d_{e}\right)}
\end{equation}

Although RSSI or received message rate (Read Rate)~\cite{tan2016wisent} measured by an RFID reader (Server) could provide a simple method to measure the powering channel state at a CRFID transponder, we observed these measures to be highly unreliable. This can be mostly understood by considering the complexity of the signal propagation model~\cite{nikitin2010phase}---see equations \eqref{eqn:RSSI} to \eqref{eqn:H} for details. We can see that RSSI depends not only on the transmit power $P_t$, the transmitter antenna gain $G_t$ and backscatter coefficient $K$, but also the path gain $G_{path}$. However, $G_{path}$ depends on signal wavelength $\lambda$, line of sight distance $d_0$ and the multi-path factor $H$; where $H$ is a complex function of angle alignment of the transmitter $g_t^i$ and the device $g^i$, angle-dependent reflection coefficient $\Gamma_{i}$ of $i$-th object, $i$-th path length for a total of $N$ multi-paths---and the influences from the random access nature\footnote{\scriptsize Notably, the RFID air interface relies on a slotted ALOHA media access control protocol.} of the media access control protocol used by the RFID air interface affecting RSSI and read rate measurements. 

\vspace{1mm}
\begin{center}
\noindent\textit{Therefore, the powering channel state is best estimated by the field deployed CRFID device harvesting available power at the device.}
\end{center}
\vspace{1mm}

Thus, we introduce a power aware execution model where:
\begin{itemize}[leftmargin=15pt]
    \item The powering channel state at a CRFID tag is measured by the device using a single measurement of harvested power at boot-up (called a \textit{Power Sniff} denoted as $ {\bf Vt}_i \leftarrow \textsf{SNIFF(t)}$ in the update scheme). The voltage measure, ${\bf Vt}_i$, is used to estimate the power that can be harvested by a given CRFID device in the field; and
    \item The workload of scheduling execution of on-device tasks is determined by the resourceful Server (RFID reader and network infrastructure) as opposed to the resource limited CRFID device based on the channel state measurements from a CRFID. 
\end{itemize} 

We consider a harvesting device operating under the commonly used charge-burst mode where charge is first accumulated in a storage element---often a capacitor---and charge is released for useful work when an adequate mount of charge (measured in terms of the voltage) is reached at the charge storage element.
Our power aware execution model (PAM) requires the Server (RFID reader and/or host) involved in the update to derive two parameters: i)~time estimation to charge to a set voltage ($t_c$); ii)~time estimation to a brown-out ($t_b$) based on the CRFID reported measurement ${\bf Vt}_i$. {\it Here, we reason that the distance between the antenna and the given CRFID device does not change during a given update session. Therefore, the power estimated at the begining of the session is valid during the entire session}.

\begin{figure}[ht]
    \centering
    \includegraphics[width=\linewidth]{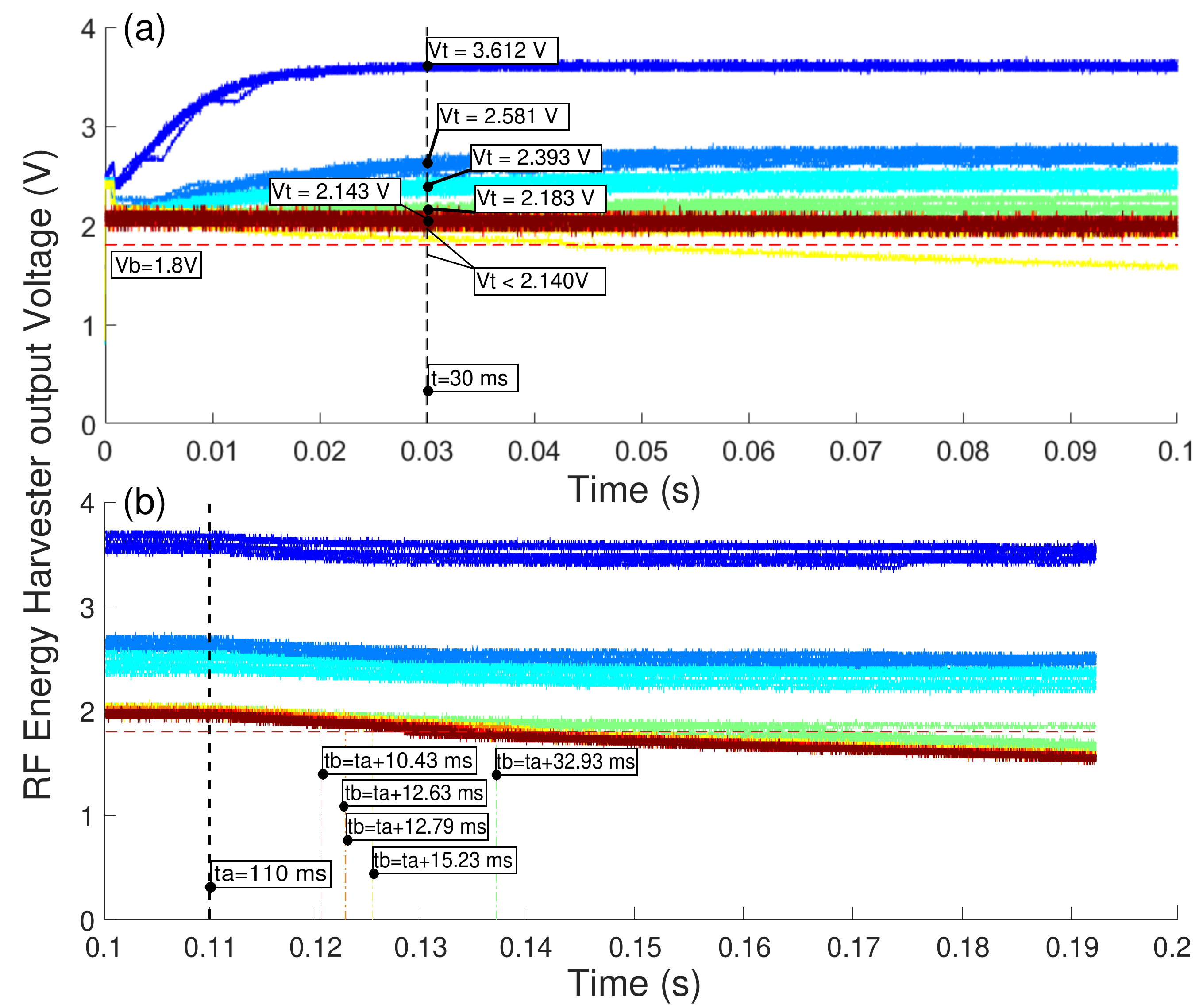}
    \caption{Power measurements. (a) Charging while the MCU is in LPM mode and (b) discharging when the MCU is in active mode (we employed a MAC computation for the task). The charge and discharge experiments at each fixed distance were repeated 10 times to obtain a mean response.}
    \label{fig:test_touch_measure}
\end{figure}

Unfortunately, the RF energy harvester on a CRFID device is a non-linear circuit component whose output voltage changes over time as input RF power level varies. Therefore, it is non-trivial to model the RF energy harvester~\cite{buettner2011dewdrop}. 
Therefore, instead of relying on an analytical model, we adopt an experimental method to derive the parameters for PAM empirically. We measure, in repeated measurements, the output voltage generated by the power-harvesting network on the WISP CRFID. We also measure the expected workload we can obtain from a CRFID device, before a brownout causes power loss. We extrapolate from these measurements to construct an empirical model for the \textsf{PAM()} function employed by the Server to determine the best set of execution parameters from the reported voltage ${Vt}$ at runtime.

As depicted in \autoref{fig:test_touch_measure} (a), the output voltage from the charge pump grows at different rates and as a function of received power. The traces do not  intersect, therefore, if we measure the charge pump output voltage at an early time, e.g, 30~ms as labeled in the picture, we can obtain a voltage across the reservoir capacitor $V_{\rm cap}$, with which we can predict the time required for the reservoir capacitor to be charged to a certain voltage level. The charging rate of a capacitor becomes slower following a logarithm trend. It is too conservative to use the time to charge to nearly 100\% saturated voltage; we empirically determined an adequate LPM time. We use the time to charge to approximately 63\% of the saturated voltage, then compared to waiting for a nearly fully charged capacitor, we can reduce charge time by 75\% and still accumulate  adequate energy to execute the chunk of task under the active time period $t_{active}$.

The time before brownout is also a function of received power. We can observe in \autoref{fig:test_touch_measure} (b), when the powering condition is good, the CRFID device can execute the MAC computation we employed for the load, continuously; starting from $ta = 110$~ms without power failure. However, the device starts to fail or brown-out causing a power loss at $t$ = 142.92~ms, giving a time to brown out of $t_b = t_a +$ 32.93~ms for ${Vt}$ below 2.183~V. As expected, we can see that $t_b$ decreases as the received power decreases.

\vspace{1mm}
\noindent{\textbf{\textsf{PAM} function formulation.}~}From our results, we can conclude: i) for $Vt \geq $ 2.393~V, the CRFID token may continuously operate with no power loss. In such cases, $t_{LPM}$ = 0, and $t_{active}$ = $\infty$ is used (execution does not need to be altered; ii) for $Vt$ within 2.183~V and 2.393~V, we can employ $t_{LPM}$ = 10~ms and $t_{active}$ = 29~ms (90\% the measured $t_b$ ) for a conservative approach to prevent power failures; iii) for $Vt$ within 2.143~V and 2.183~V, use $t_{LPM}$ = 15~ms and $t_{active}$ = 14~ms; iv) for $Vt$ within 2.140~V and 2.143~V, use $t_{LPM}$ = 25~ms and $t_{active}$ = 11~ms; and v) if $Vt \leq$  2.140~V, then the CRFID device cannot accumulate adequate energy under such a condition to complete a computation intensive task, and we strongly suggest to not perform a code update in this case. We describe \textsf{$(t_{active},t_{LPM})\leftarrow$PAM.Get($Vt$)} function in equation \eqref{eqn:pam-get-tweek}.

\begin{equation}
\label{eqn:pam-get-tweek}
\centering
\resizebox{.9\hsize}{!}{$
{
 \textsf{PAM.Get}(Vt) =  \left\{\begin{matrix}
 [\infty, 0] &,& Vt \geq 2.393~V\\ 
 [29~ms,10~ms] &,& 2.393~V > Vt \geq 2.183~V\\ 
 [14~ms,15~ms] &,& 2.183~V > Vt \geq 2.143~V\\ 
 [11~ms,25~ms] &,& 2.143~V > Vt \geq 2.140~V\\ 
 [9~ms,30~ms] &,& Vt < 2.140~V
\end{matrix}\right.}$
}
\end{equation}

\section{Transponder Modes \& Broadcast Channel}\label{apd:ObsvsPilot}
Stork~\cite{aantjes2017fast} proposed exploiting promiscuous listening by choosing one in-field token (CRFID) to stay in the non-overhearing mode and the rest in overhearing (observer or promiscuous listening) mode to create a logical broadcast channel. This method overcomes the unicast media access layer protocol to facilitate wireless code dissemination to multiple CRFID devices. Building upon Stork, our Pilot-Observer mode (Section~\ref{sec:managepower}) makes a key improvement. In contrast to Stork~\cite{aantjes2017fast}, for the token under Non-overhearing/Pilot mode, instead of selecting the first device responding to an interrogation signal from the Server, we elect the the device with the lowest reported voltage ($V_t$) as the pilot token. The similarties and differences between our work and the Stork are summarized in~Table~\ref{tab:TransponderRoles}.

This method forces the broadcast session to be  driven by the device with the lowest available energy, and the highest probability to brownout; thus, increasing the chance of all non responding tags remaining in synchronicity with the the processing of the broadcasted firmware, simply because these devices are able to harvest more power than the pilot token whilst also not having to respond to the Server.

\begin{table}[t]
    \centering
    \caption{Comparing the Settings of Tokens}
    \label{tab:TransponderRoles}
    \resizebox{\linewidth}{!}{
        \begin{tabular}{ccccc} 
            \toprule[1.5pt]
            Method & \begin{tabular}[c]{@{}c@{}}Pilot/Non-Overhearing \\ role selection\end{tabular} & Transponder mode & \begin{tabular}{@{}c@{}} Ignore handle in \\ \texttt{SecureComm}/ \\ \texttt{BlockWrite} \end{tabular}    & \begin{tabular}{@{}c@{}} \texttt{SecureComm}/ \\ \texttt{BlockWrite} \\ response \end{tabular}  \\ \hline \hline
            
            \multicolumn{1}{c}{\multirow{2}{*}{\begin{tabular}[c]{@{}c@{}}Stork~\cite{aantjes2017fast} \\ (insecure protocol)\end{tabular}}}& \multirow{2}{*}{First seen by Host} & Overhearing& \ding{51}& \ding{56} \\ \cdashline{3-5}
            
            & & non-Overhearing & \ding{56}& \ding{51} \\ \hdashline
            \multicolumn{1}{c}{\multirow{2}{*}{\begin{tabular}[c]{@{}c@{}}\wisecr \\ (Our secure protocol)\end{tabular}}} & \multirow{2}{*}{The \textit{lowest Vt} Token} & Observer& \ding{51} & \ding{56} \\ \cdashline{3-5}
            & & Pilot& \ding{56}& \ding{51}\\ \bottomrule[1.5pt]
        \end{tabular}
    }
\end{table}

\section{Pilot Election Experiments}\label{apd:pilot_selection_distances}

We have collected experiments for 20~cm, 30~cm, 40~cm and 50~cm as shown in~\autoref{fig:Pilot_selection_20_to_50}. Generally, at 20~cm and 30~cm, all methods can succeed, with little differences in terms of the number of attempts and latency. In comparison, all methods failed to update all four tokens in 10 attempts at 50~cm; \textit{although several devices were often updated, no attempt resulted in all four tokens being updated at this powering level}, hence the success rate is reported as zero. 
However, in the regions where devices are likely to operate at the threshold of powering, seen at 40~cm in our experiments, our proposed pilot election method performs best. Further, the proposed method is also seen to perform more consistently; indicated by the consistent success rate across different powering conditions achieved with different distances.

\begin{figureRevsourceS}[!ht]
    \centering
    \includegraphics[width=\mylinewidth]{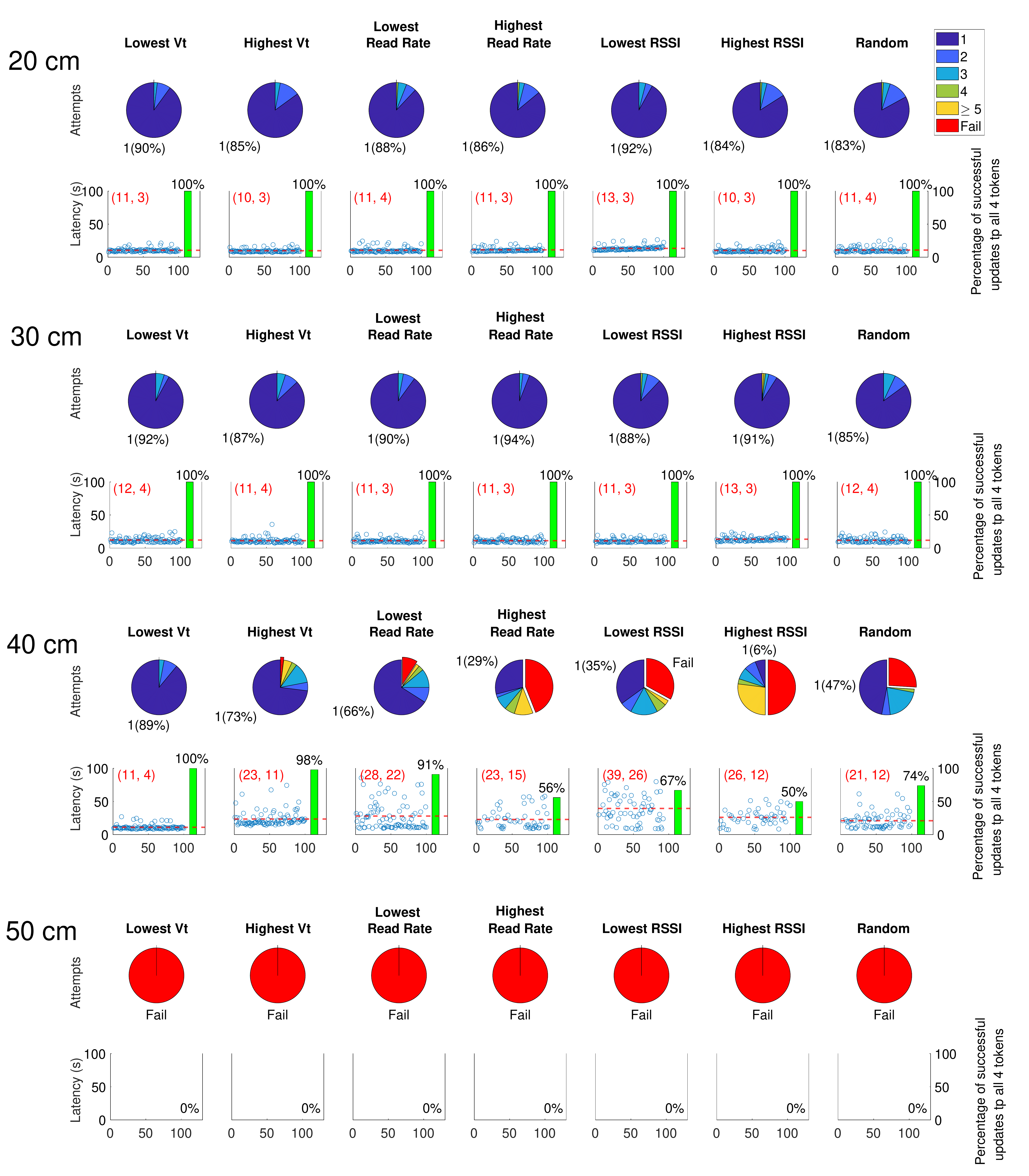}
     \caption{Evaluation of the pilot token selection strategies we propose. Four tokens were placed at 20~cm, 30~cm, 40~cm and 50~cm above a reader antenna. The pie chart shows the number of attempts to have all 4 CRFID devices or tokens updated. The scatter plot shows the corresponding latency for successful updates. The results are obtained over 100 repeated measurements, where each measurement included 10 attempts to update all four CRFID devices. The bar graph denotes the number of successful updates---defined as all four tokens being updated over the 100 repeated measurements. Here (mean, standard deviation) latency statistics are given in red text.} 
    \label{fig:Pilot_selection_20_to_50}
\end{figureRevsourceS}

\section{Execution overhead for receiving broadcast packets}\label{apd:overhead_receiving_broadcast_packets}

It is non-trivial to analyze the clock cycles for receiving a packet and sending a reply, as opposed to other tasks, as they are executed under a constant clock speed. Notably, the CRFID tokens do not have a hardware implementation of the wireless communication protocol. The RFID communication protocol is implemented in software. Since RFID communications require strict timing requirements, the CPU clock is dynamically configured and the relevant source code is written in assembly language to meet the strict timing requirements. For example, when the device is receiving a packet, the CPU works at 16~MHz. When the device is sending a reply by modulating its antenna impedance, the CPU works at 12~MHz. So it is difficult to employ our previous method to predict clock cycles by monitoring GPIO pins.

\begin{figure}[!ht]
    \centering
    \includegraphics[width=1.0\linewidth]{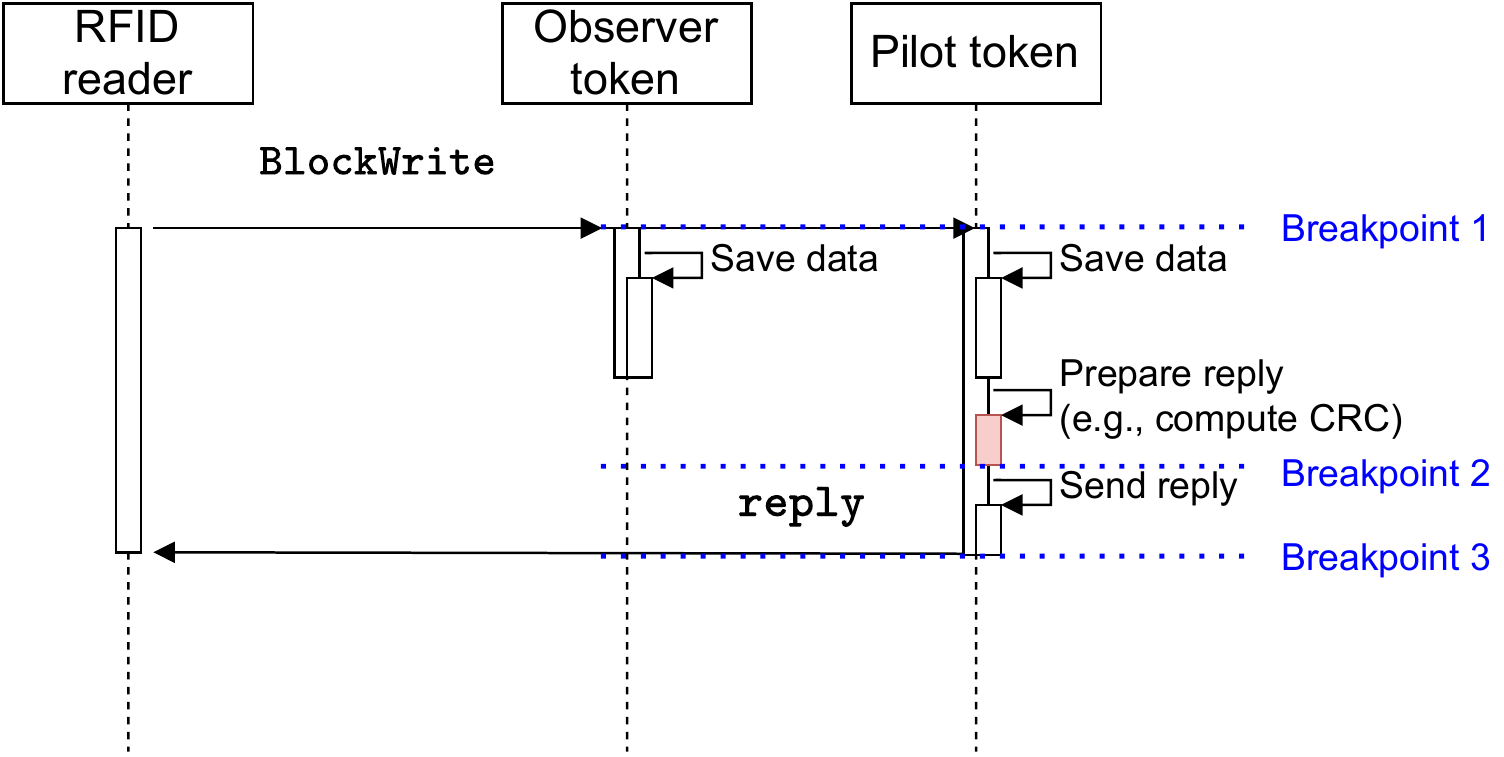}
     \caption{Method to measure clock cycles for receiving a packet and acknowledging a received packet.}
    \label{fig:Wisecr_breakpoints}
\end{figure}

To overcome the above challenge, we used debugger tools to read the CRFID device's internal states by inserting three breakpoints: Breakpoint 1: before the receiving routine is called; Breakpoint 2: in between the receiving routine and the reply routine; Breakpoint 3: after the reply routine (as illustrated in \autoref{fig:Wisecr_breakpoints}). Clock cycles for receiving and replying can be acquired by looking at the clock profile counter at each breakpoint.

The \texttt{SecureComm} command in the \epc specification is yet to be widely supported in commercial RFID hardware. Therefore, as we mentioned in Section~\ref{sec:managepower}, we implement this command over the existing \texttt{BlockWrite} specified to support variable payload size as described in~\cite{epcglobal2015inc}. However, the impinj R420 RFID reader we used in experiments implements it in a distinct manner. Irrespective of the specified payload size, the Impinj R420 reader (software version 5.12.3.240) always splits the payload into multiple \texttt{BlockWrite} commands, each command carrying a payload of only 2 Bytes~\cite{tan2015robust}. Further, busy waiting is used while receiving a command from the RFID reader (for example, at \url{https://git.io/J1lO8}). Hence, clock cycle results can vary from measurement to measurement. Hence, in our experimental results, we report the average clock cycles for such \texttt{BlockWrite} commands over 100 repeated measurements. We summarize the results below:

\begin{itemize}
    \item Pilot token's cycles to receive a \texttt{BlockWrite} packet (with 2 Byte payload): 23,082
    \item Pilot token's cycles to send a reply to the \texttt{BlockWrite} packet: 1,131
    \item Observer token's cycles to receive a \texttt{BlockWrite} packet (with 2 Byte payload): 22,002
\end{itemize}

The clock cycles for an observer token to receive a \texttt{BlockWrite} packet is 1,080 smaller than the pilot token (this number is obtained by calculating the difference between the pilot and observer tokens' clock cycle counter at Breakpoint 2, over the average obtained from 100 measurements). This is because the observer does not need to prepare the reply packet (ACK), which requires a CRC-6 calculation, as illustrated in \autoref{fig:Wisecr_breakpoints}.

The total number of packets required in an update can be computed with:

$$N_{packets} = \frac{\text{firmware size (in Bytes)}}{\text{2 Bytes per {\tt BlockWrite} command}}$$

The number predicted using the above equation is in an ideal case, the actual number may be higher considering retransmissions, for example, due to communications errors identified using the CRC.

\begin{figure}[!ht]
    \centering
    \includegraphics[width=1.0\linewidth,angle=0,origin=c]{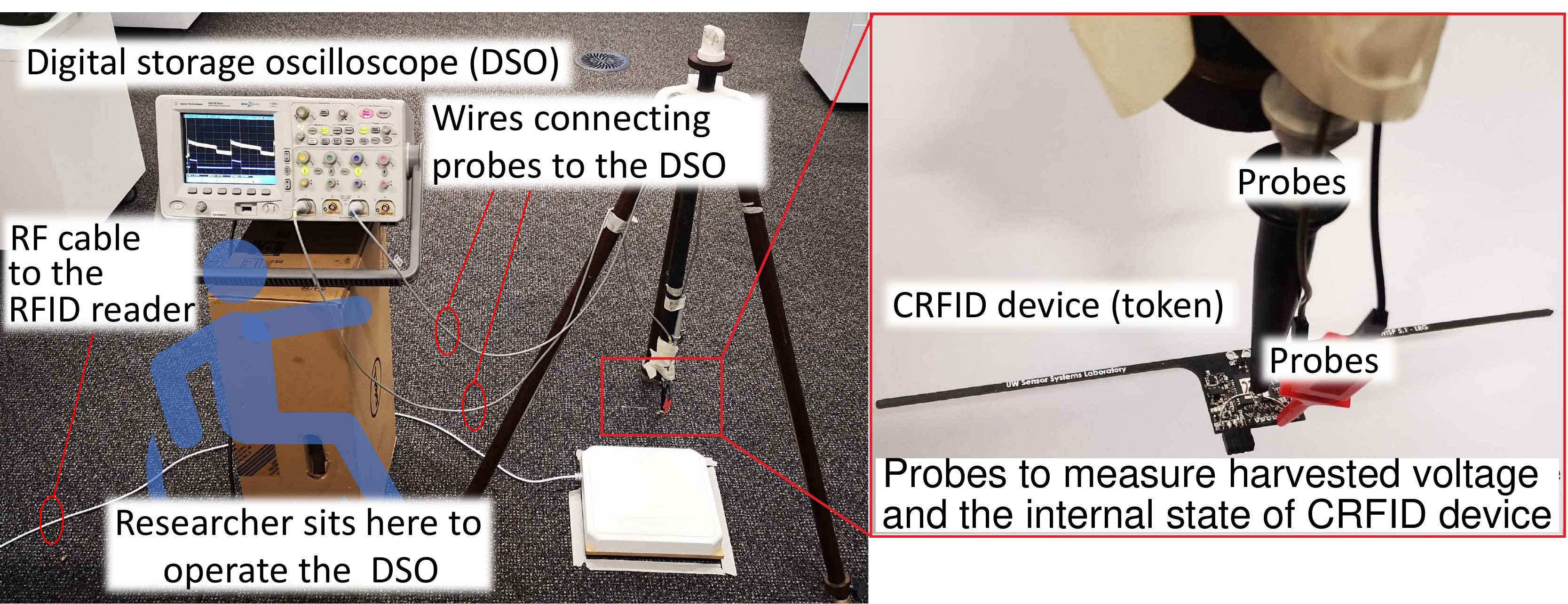}
     \caption{Experiment setup and instrumented CRFID device used for the experiments summarized in \autoref{fig:TaMax} and \autoref{fig:PAM-CMAC evaluation}. Notably conducting the experiments require a researcher to be seated next to the DSO. The probes, the changing positions of the wires leading from the probes, especially at different distances, and the orientation of the researcher significantly impacted these measurements.}
    \label{fig:WISP_instrumented}
\end{figure}

\section{Experiment Setup to Access Token's Internal State} \label{apd:power_based_exp_setup}
Some of our experiments require to measure the internal states of the token. For example the impact of four key device tasks on power-loss and the evaluation of our
proposed power PAM method in Section~\ref{sec:managepower}.  However, we do not have a precisely controlled RF environment (i.e., anechoic chamber) to remove the impact of the multipath signals constructively and destructively interfering with the RF powering of a device.

The measurement processes for \autoref{fig:TaMax} and \autoref{fig:PAM-CMAC evaluation} were different and complicated by the probes and wires that we need to attach to the device, the Digital Storage Oscilloscope (DSO) we need to keep near the setup as well as the close proximal presence of the researcher to operate the DSO to enable taking the  measurement as shown in \autoref{fig:WISP_instrumented}.

To try to mitigate the influence of factors discussed above, we conducted these experiments following the method we describe below:

\begin{itemize}
    \item Ensure we used the same CRFID device for both experiments. 
    
    \item Try our best to keep the multipath environment the same across the two experiments.  Hence, instead of adjusting the distance (which suffers from different multipath reflection and interference) and the changing positions of the probes and wires, we have the CRFID device fixed at 20~cm above the reader antenna, and \textit{adjusted the transmit power of the RFID reader following the method in~\cite{su2015investigating}.}
    
    We can employ this method because, according to the free-space path loss equation~\cite{friis1946free} given below, adjusting the transmit power $P_t$ of the RFID reader has a similar impact on the received power $P_r$ as changing the distance $d$. Because, the RF wavelength $\lambda$ is relatively constant (notably RFID employs frequency hopping regulations) whilst the RFID reader antenna gain $G_t$ and CRFID device antenna gain $G_r$ are fixed.
    $$ P_{r}=P_{t} G_{t} G_{r}\left(\frac{\lambda}{4 \pi d}\right)^{2} $$ 
\end{itemize}

Notably, in the evaluation of PAM, what we are interested in is the available harvested power, distance is just one factor we can use to control the available power at the device in our experiments. Adjusting the transmit power endows us with a more accurate form of control over the available power as that can be done using the RFID reader software, programmatically, and without interfering with the devices setup (such as changing distances and the arrangement of the probe wires and the multi-path environment).

\section{Firmware Update to Mobile Tokens}\label{apd:update_moving_tokens}
 
\begin{figure}[!ht]
    \centering
    \includegraphics[width=\linewidth]{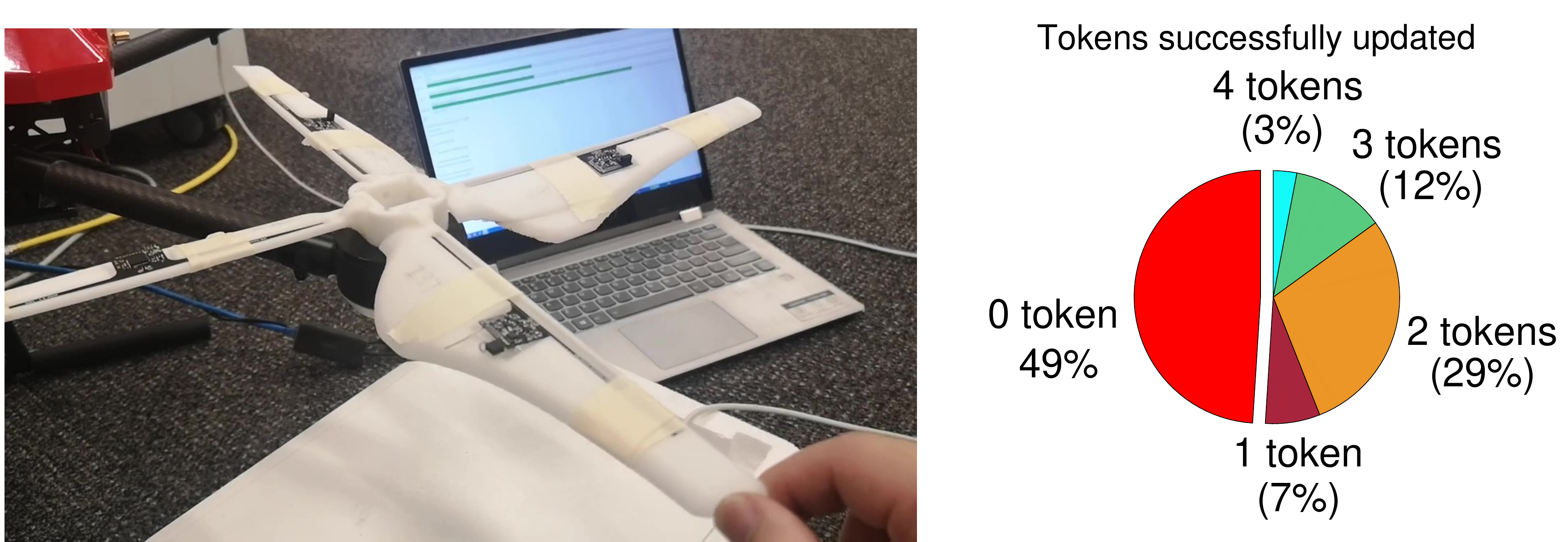}
     \caption{Attempt to update firmware while the UAV rotor is rotated by hand at approximately 1~RPM. The pie chart shows the number of CRFID devices updated in the 100 repeated firmware update executions.} 
    \label{fig:spinning_rotor_image}
\end{figure}

We tested firmware updates to mobile CRFID devices. We rotated the rotor by hand, with a rotation speed of approximately 1~RPM (revolutions per minute) and conducted 100 repeated firmware updates to the four CRFID devices in the rotor blades (as illustrated in the video \url{https://youtu.be/AVrf0rNM0z8}). Here, as in other experiments, for each firmware update, the Host makes 10 attempts to update all 4 devices. The result in~\autoref{fig:spinning_rotor_image} shows that 3 updates resulted in disseminating firmware to all four tokens, and in 51 updates, at least one token is updated. We can conclude that updating firmware to mobile CRFID devices is possible but the success rate can be expected to reduce dramatically.

\end{document}